\documentclass[11pt,preprint,prd,aps,superscriptaddress,nofootinbib,showpacs,a4paper]{revtex4}
\usepackage{graphicx}
\usepackage{epsfig}
\usepackage{axodraw}
\usepackage{amsmath}
\usepackage{amssymb}
\newcommand{\Kvec}{$K^*(892)$ \hspace{-0.07cm}}
\newcommand{\Kscal}{$K^*_0(1430)$ \hspace{-0.07cm}}
\newcommand{\mKpi}{$m_{K\pi}$ \hspace{-0.07cm}}

\newcommand{\lapprox}{%
\mathrel{%
\setbox0=\hbox{$<$}\raise0.6ex\copy0\kern-\wd0\lower0.65ex\hbox{$\sim$}}}
\newcommand{\lbl}[1]{\label{eq:#1}}
\newcommand{\braque}[1]{{\langle #1 \rangle}}
\newcommand{\rf}[1]{(\ref{eq:#1})}

\newcommand{\fpid} {f^2_\pi}
\newcommand{\fpiq} {f^4_\pi}
\newcommand{\mkd} {m^2_K}
\newcommand{\mpid} {m^2_\pi}
\newcommand{\mpiq} {m^4_\pi}
\newcommand{\metap}{m_{\eta'}}
\newcommand{\metapd} {m^2_{\eta'}}
\newcommand{\metaxd} {m^2_{\eta'}}
\newcommand{\mkq} {m^4_K}
\newcommand{\be}{\begin{equation}}
\newcommand{\en}{\end{equation}}
\def\im{ {\rm Im}\,}
\def\tr{ {\rm tr}\,}

\begin{document}

\title{\mbox{\boldmath $CP$} violation and kaon-pion interactions in \mbox{\boldmath $B\to K \pi^+\pi^-$} decays}

\author{B.~El-Bennich\footnote{Present address: ANL, Argonne, USA.}}
\affiliation{Laboratoire de Physique Nucl\'eaire et de Hautes \'Energies (IN2P3--CNRS--Universit\'es Paris 6 et 7), Groupe Th\'eorie, 
                  Universit\'e Pierre et Marie Curie, 4 place Jussieu, 75252 Paris, France}
\affiliation{Physics Division, Argonne National Laboratory, Argonne, Illinois, 60439, USA}

\author{A.~Furman}
\affiliation{ul. Bronowicka 85/26, 30-091 Krak\'ow, Poland}

\author{R.~Kami\'nski} 
\affiliation{Division of Theoretical Physics, The Henryk Niewodnicza\'nski Institute of Nuclear Physics,
                  Polish Academy of Sciences, 31-342 Krak\'ow, Poland}

\author{L.~Le\'sniak}
\affiliation{Division of Theoretical Physics, The Henryk Niewodnicza\'nski Institute of Nuclear Physics,
                  Polish Academy of Sciences, 31-342 Krak\'ow, Poland}

\author{B.~Loiseau}
\affiliation{Laboratoire de Physique Nucl\'eaire et de Hautes \'Energies (IN2P3--CNRS--Universit\'es Paris 6 et 7), Groupe Th\'eorie,
                  Universit\'e Pierre et Marie Curie, 4 place Jussieu, 75252 Paris, France}

\author{B.~Moussallam}  
\affiliation{Groupe de Physique Th\'eorique, Institut de Physique Nucl\'eaire (IN2P3--CNRS),
Universit\'e Paris-Sud 11, 91406 Orsay Cedex, France}
               
\date{\today }

\begin{abstract}
We study $CP$ violation and the contribution of the strong kaon-pion interactions in the three body $B\to K \pi^+ \pi^-$ decays.
We extend our recent work on the effect of the two-pion $S$- and $P$-wave interactions to that of the corresponding kaon-pion ones.
The weak amplitudes have a first term derived in QCD factorization and a second one as a phenomenological contribution added to the QCD penguin amplitudes.
The effective QCD coefficients include the leading order contributions plus next-to-leading order vertex and penguins corrections. 
The matrix elements of the transition to the vacuum of the kaon-pion pairs, appearing naturally in the factorization formulation, are described by the strange $K \pi$ scalar ($S$-wave) and vector  ($P$-wave) form factors.
These  are determined from Muskhelishvili-Omn\`es coupled channel equations 
using experimental kaon-pion $T$-matrix elements, 
together with chiral symmetry and asymptotic QCD constraints. 
From the scalar form factor study, the modulus of the $K^*_0(1430)$ decay
constant is found to be $(32\pm 5$) MeV. 
The additional phenomenological amplitudes are fitted to reproduce the $K \pi$
effective mass and helicity angle distributions, the $B \to K^*(892) \pi $
branching ratios and the $CP$ asymmetries of the recent data from Belle and
BaBar collaborations. We use also the new measurement by the  BaBar group of the  phase difference between the $B^0$ and $\bar B^0$ decay amplitudes to $K^*(892) \pi $.
Our predicted $B^\pm \to \ K^*_0(1430)\pi^\pm$,  $K^*_0(1430) \to K^\pm\pi^\mp$ branching fraction, equal to $(11.6 \pm 0.6)\times 10^{-6}$, is smaller than  the result of the analyzes of both collaborations. For the neutral $B^0$ decays, the predicted value is   $(11.1 \pm 0.5)\times 10^{-6}$.
In order to reduce the large systematic uncertainties in the experimental determination of the $B \to \ K^*_0(1430)\pi$ branching fractions, a new parametrization is proposed.
It is based on the $K\pi$ scalar form factor,  well constrained by theory and experiments other than those of $B$ decays.

\end{abstract}

\maketitle

\section{Introduction}
Rare two-body and quasi two-body charmless hadronic decays of $B$ mesons~\cite{Biesiada2007} are a rich field for tests of the standard model and QCD~\cite{Beneke2006}.
Furthermore three-body charmless hadronic $B$ decays provide an interesting ground, not only for searches on $CP$ violations but also to study hadronic physics~\cite{3bodyworkshop}.
Strong interaction effects, in particular through the presence of two-body resonances and their interferences, can influence weak decay observables.
Strong interaction phases are necessary for the occurrence of $CP$ violation and it is essential to have a description, as reliable as possible, of the interactions between the detected hadrons. 
 Dalitz-plot analyzes allow to extract effective mass and angular 
distributions of the produced meson pairs. 
 If one pair is created in two (or more) different states, one can see 
specific interference effects leading to additional and interesting $CP$ asymmetries. 
 These arise from the variations of the strong phases with energy, such variations are absent in the two-body $B$ decays where the energy is fixed.
 The meson-meson final state interactions must be addressed using theoretical 
constraints, such as unitarity, analyticity and chiral symmetry, and 
experimental data from processes other than $B$ decays.
 Then, for a given $B$-decay, the comparison between the theoretical model and experimental results will determine the strong phases needed to generate the measured direct $CP$ asymmetries.

Recently, BaBar and Belle Collaborations have performed detailed Dalitz plot analyzes for different $B\to K\pi^+\pi^- $ 
decays~\cite{AubertPRD72,Abe2005,Garmash:2005rv,AubertPRD73,Abe0509047,Garmash2007,AubertLP2007,Aubert:2008bj,Aubert:2007bs}.
One observes an accumulation of events for $\pi\pi$ or $K\pi $ effective 
masses lower than 2 GeV and in particular the presence of the scalar mesons
 $f_0(980)$, $K_0^*(1430)$  (and to a lesser extent $f_0(1500)$) and of the
  vector mesons  $\rho(770)^0$, $K^*(892)$.
The event distributions of the Dalitz plots are usually studied using the isobar model in which the decay amplitudes are parametrized by sums of Breit-Wigner terms and a background.
In Refs.~\cite{AubertPRD72,AubertPRD73,AubertLP2007,Aubert:2008bj,Aubert:2007bs}, an effective range nonresonant component has been used in the $K\pi$ $S$-wave amplitude.
 The aim was to ameliorate the description of the low $K\pi$ effective mass spectrum. 
 
An important breakthrough in the theory of $B$ decays recently achieved 
is the confirmation of the validity of factorization, as a leading
order  approximation in an expansion in inverse powers of the $b$ quark mass $m_b$~\cite{bene03}. 
This concerns the $B$ decays into two mesons and was
later reformulated using the soft collinear effective theory approach
to QCD~\cite{BauerFPCP06}. 
Detailed comparisons between theory, based on QCD factorization (QCDF), and 
experiment were made in the case of decays into two pseudo-scalar mesons and one pseudo-scalar and one vector meson~\cite{bene03,2bodyQCDF}. 
Agreement is generally quite fair.  
However, some phenomenological parameters are introduced.

 In this paper, we study the decays of the $B$ into three mesons, 
$B\to K\pi^+\pi^-$, for which, to our knowledge, no proof of factorization has 
been given.
 However, we restrict ourselves to specific kinematical configurations in 
which the three mesons are quasi aligned in the rest frame of the $B$. 
 This condition is met, in particular, in the low effective $K\pi$ mass region ($\lesssim$ 2 GeV) of the Dalitz plot where most of the $K\pi$ resonant structures are seen.
 We will denote such processes as $B\to (K\pi)\pi$ where the mesons of the  $K\pi$ pair move, more or less, in the same direction.
Three-body interactions are expected to be suppressed in such conditions. 
Then, it seems reasonable to assume the validity of factorization for this quasi two-body $B$ decay~\cite{Beneke3body} where we assume that the ($K \pi$) pair originates from a quark-antiquark state.

In a previous work~\cite{fkll}, the decays $B\to (\pi^+\pi^-)_S\  K$ as well as $B\to
(\bar{K} K)_S\  K$, where the two comoving mesons in the $(\pi^+ \pi^-)$ and $(K\bar K)$ pairs are in $S$-states, were studied using an approximate construction of relevant scalar form factors proposed in Ref.~\cite{Meissner:2000bc}. 
The decays $B\to (\pi^+\pi^-)_P\  K$, with the two pions in a $P$ state, were subsequently studied in Ref.~\cite{El-Bennich2006}. 
Here, we focus on the decays $B^\pm\to (K^\pm\pi^\mp)\pi^\pm,\ B^0\to(K^0\pi^+)\pi^-$ and $\bar B^0\to(\bar K^0\pi^-)\pi^+$. 
In the factorization approach the amplitudes can be expressed as the product 
of effective QCD coefficients~\cite{bene03} and  the  two matrix elements  of the vector currents 
$\langle K\pi \vert \bar{q}\gamma^\nu s\vert 0\rangle$  and 
$\langle\pi\vert\bar q\gamma_\nu b\vert B\rangle$, with $q=u$ or $d$. 
 One conspicuous consequence is that the $K\pi$ pair is restricted to be in either an $S$ or a $P$ state; no $D$ or higher waves being allowed. 
This fact is indeed supported by the Belle experiment of Ref.~\cite{Garmash2007}. 
In their analysis on $B^0\to K_S^0\pi^+\pi^-$ decays, one clearly observes a vector $K^*(892)^+$ and a scalar $K^*_0(1430)$ resonances, 
the signal for the tensor $K_2^*(1430)$ being small.

%
The matrix element of the vector current
$\langle K\pi \vert \bar{q}\gamma^\nu s\vert 0 \rangle$  involves two functions of the $K \pi$ effective mass squared, the strange scalar and vector form factors. 
We will perform a construction based on general properties of analyticity, QCD asymptotic counting rules~\cite{brodskylepage} and using accurate experimental data on $K \pi$ scattering,  both in the elastic and inelastic region. 
Such data have been obtained in the relatively high statistics 
LASS experiment on $K^-$ proton interaction~\cite{aston84,aston87,aston88}. 
A specific feature of $K \pi$ scattering at medium energies is that inelasticity is
dominated by two-body or quasi two-body channels. 
More specifically, it was shown by LASS that inelasticity in the $S$-wave is dominated 
by the $K\eta'$ state and in the $P$-wave by the $K^*\pi$ and $K\rho$ states. 
Combining dispersion relations with unitarity relations then leads to a set of coupled integral equations for the form factors.
These equations constitute a generalization of Watson's theorem of final state interactions, well known for a single channel case~\cite{Watson52}.

An analogous system of equations was studied and solved for the first
time by Donoghue, Gasser and Leutwyler~\cite{dgl} in the case of the pion 
scalar form factor. 
More recently, this method was applied to the $K \pi$ scalar form factor in Ref.~\cite{jop}. 
For our purpose, we have redone the calculation of Ref.~\cite{jop} and extended this framework to the case of the vector form factor. 
Both of these form factors, needed in our work on $B\to(K\pi)\ \pi$ decays, are treated exactly in the same way. 
This provides an unified treatment of different $B$ decays through coupled final-state channels and also of  $\tau\to K \pi\nu_\tau$ decays~\cite{jopffa,MoussallampiKSandPwave} using form factors constrained by accurate results on $K \pi$ scattering. 
It could allow, for instance, to give predictions for $B\to (K \eta')_S\ \pi,\ ( K^*\pi)_P\ \pi$ and $(K \rho)_P\ \pi$ decays as was done in Ref.~\cite{fkll} for $B\to (K\bar K)_S\ K$ decays connected to $B\to(\pi^+\pi^-)_S\ K$ decays.

 The introduction of form factors, constrained by  theory and other experiments than $B$ decays, is an alternative to the use of the isobar model.
 This latter approximation violates unitarity and the information about resonances, present in the final states, can be distorted by other nearby resonances due to interference.
 In our approach, we use the complex pole definition of a resonance which allows us to obtain its branching ratio and its decay constant. An alternative way, examined here too, is to integrate over the effective mass range where the resonance  dominates.
Recently charmless three-body decays of $B$ mesons have been extensively studied by Cheng, Chua and Soni in the factorization scheme~\cite{Cheng0704.1049}.
Their calculation proceeds via quasi two-body decays involving resonant states and nonresonant contributions.
 Breit-Wigner expressions are used to describe the appropriate resonance effects in the scalar and vector matrix elements.
Their $K^*(892)\pi,\ \rho(770)^0K$ and $K_0^*(1430)\pi$ branching ratios are too small compared to the data by a factor varying between 2 and 5.
 This is also the case in our QCD factorization approach.
 To improve agreement with experiment we introduce phenomenological corrections to the QCD penguin amplitudes.
 The latter could represent (in part) the contribution (not studied here) of the weak annihilation and hard-spectator contributions together with their phenomenological components~\cite{bene03}.
 They could also partially come from long distance charming penguin 
 amplitudes~\cite{BauerFPCP06,Ciuchini:1997hb} which, themselves,
could arise from intermediate $D_s^{(*)}D^{(*)}$ states, reminding that $B\to D_s^{(*)}D^{(*)}$ branching fractions are large.
Finally, they can come from unknown effects of new physics which could appear in the $b \to s$ quark loop transitions (see for instance Ref.~\cite{fleischer08}).

 Our paper is organized as follows. In Sec.~\ref{Decay_amplitudes} we derive, in
 the QCDF framework, the $B\to(K\pi)_{S,P}\ \pi^\pm$ decay amplitudes for
 $B^\pm$, $B^0$ and $\bar B^0$.
The charged $B$ ($b\to s\bar dd$ transition for $B^-$) decay amplitudes have only penguin diagram contributions, while the neutral ($b\to s\bar uu$ transition for $\bar B^0$) have an additional tree diagram.
In the weak amplitudes, we include penguin-correction terms represented by four complex parameters.
The amplitudes are expressed in terms of the product of the effective QCD coefficients by the $B$ to $\pi$ transition form factor and the  $K \pi$ strange form factors.
Section~\ref{ai} tabulates the values of the process and scale dependent effective QCD coefficients~\cite{bene03} we use in our amplitudes.
To the leading-order contribution in $\alpha_s$, we add the next-to-leading order short-distance vertex and penguin corrections. 
Their calculation is outlined in Appendix~\ref{appendixvpc} where we give also their values.

In Sec.~\ref{formfactors} we specify the model we use for the $B$ to $\pi$ transition form factors.
The unitary equations satisfied by the scalar and vector strange form factors are also presented.
We discuss briefly the low energy constraints and the two-channel description ($K \pi,\ K\eta'$) for the $S$-wave $K\pi$ scattering.
In the case of the $P$-wave, the necessary three channel ($K\pi,\  K^*\pi,\ K\rho$) description is then described.
We give the results for the strange scalar and vector $ K\pi$ form factors.
Using the complex pole definition of a resonance, a simple and unambiguous separation between the background and the resonant contributions of the $S$-wave and $P$-wave amplitudes, is given. 
This allows us to determine in a unambiguous way the $B\to K^*_0(1430)\pi$ and $B\to K^*(892))\pi$ branching fractions. 
A detailed derivation of the $S$-wave $T$-matrix elements, necessary to calculate the  strange scalar form factor, is presented in the Appendix~\ref{appendixSwavT}.
We calculate the values of the decay constants of $K^*(1430)$ and $K_0^*(892)$ from the knowledge of the vector and scalar $K\pi$ form factors in Appendix~\ref{appendixfdecaypole}.
In Appendix~\ref{appendix2a3body}, we show how the two  body amplitudes $B\to K^*_0(1430)\pi$ and $B\to K^*(892))\pi$ are related to the three-body ones.
Effective decay constants for  $K^*(1430)$ and $K_0^*(892)$ are also calculated.

In Sec.~\ref{results}
 we describe our fitting procedure on the additional penguin parameters.
We furthermore compare the results of our fit to the experimental $m_{K \pi}$ mass and helicity-angle distributions.
We present our fitted values for the branching ratios and $CP$ asymmetries for the $B\to K^*(892)\pi$ and $B\to K_0^*(1430)\pi$ decays. 
Note that, in the case of the scalar meson production, our branching ratios are predictions.
 Discussion of the results and comparison with experimental analyzes are also given.
A summary and some conclusions are presented in Sec.~\ref{summary}.

\section{Decay amplitudes and physical observables}
\label{Decay_amplitudes}

The amplitudes for the non-leptonic decays of the $B$ meson 
are given as matrix elements of the effective weak Hamiltonian
\begin{equation}
H_{eff}=\frac{G_F}{\sqrt2}\sum_{p=u,c}\lambda_p\, \Big[ C_1 O_1^p + C_2 O_2^p
+\sum_{i=3}^{10} C_i O_i + C_{7\gamma} O_{7\gamma} + C_{8g} O_{8g} \Big] +h.c., 
\end{equation}
where in the case of strangeness $S= \pm 1$ final states 
\begin{equation}
\lambda_u= V_{ub} V^*_{us},\ 
\lambda_c= V_{cb} V^*_{cs},
\end{equation}
the $V_{pp'}$ being Cabibbo-Kobayashi-Maskawa quark-mixing matrix elements.
For the Fermi coupling constant $G_F$ we take the value 1.16637\ $10^{-5}$ GeV$^{-2}$. 
In this work we use $\lambda_u=3.55\times 10^{-4}-i\ 7.49\times 10^{-4}$ and $\lambda_c=4.05\times 10^{-2}+i\ 6.5\times 10^{-7}$.
The $C_i(\mu)$ are the Wilson coefficients of the respective four-quark operators  $O_i(\mu)$ at a given renormalization scale $\mu$.
The explicit expression of the operators $O_i$ may be found e.g. in
Ref.~\cite{Beneke:2001ev}. 
Studies of $B$ decays into 
two-body~\cite{Beneke:2001ev} and 
quasi-two-body~\cite{bene03,Cheng:2005nb} 
final states have been performed in the QCDF framework. 
These studies show that 
naive 
factorization is a useful first order approximation which receives corrections
proportional to the strong coupling constant $\alpha_s(m_b)$, $\alpha_s(\sqrt{\Lambda_{QCD} m_b})$ 
and in inverse powers of $m_b$~\cite{Beneke2006}. Here, we
perform a heuristic extension of these results to a class of three body
decays $B\to (K \pi) \pi$.

\subsection{Charged {\boldmath $B$} decay amplitudes}
We focus on the process $B^-\to  ( K^- \pi^+)\ \pi^-$. To illustrate our
approach, let us write the matrix elements of the penguin operators $O_3$ and
$O_4$ at leading order factorization\footnote{In this derivation we have assumed that the $( K^- \pi^+)$ pair originates from a quark-antiquark state.}
\begin{equation}
\braque{ \pi^- ( K^- \pi^+) \vert C_3 O_3 +C_4 O_4 \vert B^-}=
a_4 \braque{\pi^- \vert \bar{d}\gamma^\nu (1-\gamma^5) b\vert B^-}
\braque{ K^- \pi^+\vert \bar{s}\gamma_\nu (1-\gamma_5) d\vert 0 }      
\end{equation}
with
\begin{equation}
a_4= C_4(\mu) +\frac{1}{N_c}\  C_3(\mu),
\end{equation}
where $N_c=3$ is the number of colors.
In this approximation, the dependence of the amplitude 
as a function of the two Dalitz-plot 
variables $m_{K\pi}$, $m_{\pi\pi}$
is completely determined in terms of $K\to\pi$ and $B\to\pi$ 
form factors. We will probe this prediction by employing a careful
determination of the $K\pi$ form factors described in detail in 
Sect.~\ref{formfactors} 
and in Appendix~\ref{appendixSwavT}. 
Our main assumption will be that
the corrections to naive factorization can be absorbed into 
effective-mass independent modifications of the parameters $a_i$. 
We will borrow parts of these corrections from quasi-two-body 
calculations and also append a phenomenological part.   

The $B$ to $\pi$ transition matrix element is written as
\begin{multline}
\label{BtopiFF}
\langle \pi^-(p_{\pi^-})\vert \bar d\gamma^\nu(1-\gamma^5)b\vert B^-(p_{B^-})\rangle\\
=\left[
(p_{B^-}+p_{\pi^-})^\nu-\dfrac{M_{B}^2-m_{\pi}^2}{q^2}q^\nu
\right] f_1^{B^-\pi^-}(q^2)
+\dfrac{M_{B}^2-m_{\pi}^2}{q^2}q^\nu f_0^{B^-\pi^-}(q^2) ,
\end{multline}
where $f_{0,1}^{B^-\pi^-}(q^2)$ are the scalar and vector $B^-$ to $\pi^-$ 
form factors. The four-momentum transfer is
\begin{equation}
q=p_{B^-}-p_{\pi^-}=p_{K^-}+p_{\pi^+} ,
\label{q}
\end{equation}
with $p_{\pi^-},\ p_{\pi^+},\ p_{K^-}$ and $p_{B^-}$ being the four momenta 
of the negative and positive pions, of the $K^-$ and of the $B^-$ mesons, 
respectively. In an analogous way, 
the matrix element for the transition from vacuum to the $K^-\pi^+$ state 
is given in terms of the scalar $f_0^{K^-\pi^+}(q^2)$ and 
vector $f_1^{K^-\pi^+}(q^2)$ form factors by
\begin{multline}
\label{Ktopiff}
\langle K^-(p_{K^-})\pi^+(p_{\pi^+})\vert\bar s\gamma_\nu(1-\gamma_5) d\vert 0\rangle \\
=
\left[
(p_{K^-}-p_{\pi^+})_\nu-\dfrac{m_{K}^2-m_{\pi}^2}{q^2}q_\nu
\right]
f_1^{K^-\pi^+}(q^2)
+\dfrac{m_{K}^2-m_{\pi}^2}{q^2}q_\nu f_0^{K^-\pi^+}(q^2).
\end{multline}
Above, $M_B,\ m_K$ and $m_\pi$ are the masses of the charged $B$ mesons, 
kaons and pions, respectively.
From Eqs.~(\ref{BtopiFF}) and (\ref{Ktopiff}) one obtains
\begin{multline}
\label{productFF}
\left\langle\pi^-\vert\bar d\gamma^\nu\left(1-\gamma^5\right)b\vert B^-\right\rangle
\langle K^-\pi^+\vert \bar s\gamma_\nu(1-\gamma_5) d\vert 0\rangle \\
=f_0^{B^-\pi^-}(q^2)f_0^{K^-\pi^+}(q^2)
(M_B^2-m_\pi^2)\ \dfrac{m_K^2-m_\pi^2}{q^2} \\
+f_1^{B^-\pi^-}(q^2)f_1^{K^-\pi^+}(q^2)
\left[
m_{K^-\pi^-}^2-m_{\pi^+\pi^-}^2-(M_B^2-m_\pi^2)\ \dfrac{m_K^2-m_\pi^2}{q^2}
\right],
\end{multline}
where  $m_{K^-\pi^-}$ and $m_{\pi^+\pi^-}$ are the $K^-\pi^-$ and $\pi^+\pi^-$ effective masses.
Note that $q^2$ is the square of the $K^-\pi^+$ effective mass.
In the $K^-\pi^+$ center of mass system
\begin{equation}
\label{costheta}
m_{K^-\pi^-}^2-m_{\pi^+\pi^-}^2-(M_B^2-m_\pi^2)\ \dfrac{m_K^2-m_\pi^2}{q^2}
=4\mathbf{p}_{\pi^+}\cdot \mathbf{p}_{\pi^-}=4\vert \mathbf{p}_{\pi^+}\vert\ \vert \mathbf{p}_{\pi^-}\vert\cos\theta ,
\end{equation}
where $\theta$ is the angle between the $\pi^+$ and $\pi^-$ three-momenta.
The first term of the right hand side of Eq.~(\ref{productFF}) corresponds 
to the $K^-\pi^+$ $S$-wave contribution while the second term to that of 
the $P$-wave since it depends linearly on $\cos\theta$. One observes that 
there are no contributions from $l\ge 2$ partial waves.

Finally, we introduce the matrix element of the complete weak 
effective Lagrangian.
We write the $K \pi$ $S$-wave contribution in the following form
\begin{multline}
\label{K-pi+Sampli}
\mathcal{M}_S^-\equiv \langle \pi^-\ (K^-\pi^+)_S\vert H_{eff}\vert B^-\rangle
=\dfrac{G_F}{\sqrt{2}} (M_B^2-m_{\pi}^2)\dfrac{m_{K}^2-m_{\pi}^2}{q^2}
f_0^{B^-\pi^-}(q^2) \ f_0^{K^-\pi^+}(q^2)\\
\times
\bigg\{
\lambda_u\left(a_4^{u}(S)-\dfrac{a_{10}^{u}(S)}{2}+c_4^{u}\right)
+\lambda_c\left(a_4^c(S)-\dfrac{a_{10}^c(S)}{2}+c_4^{c}\right)\\
 -\ \dfrac{2q^2}{(m_b-m_d)(m_s-m_d)} \left[
\lambda_u\left(a_6^{u}(S)-\dfrac{a_8^{u}(S)}{2}+c_6^{u}\right)
+\lambda_c\left(a_6^c(S)-\dfrac{a_8^c(S)}{2}+c_6^{c}\right)
\right]\bigg\}.
\end{multline}
Here the kinematical $q^2$ dependence associated with the terms $a_4^p-a_{10}^p/2+c_4^p$ ($p=u,\ c$) is different from that in front of  the $a_6^p-a_8^p/2+c_6^p$ ones.
In the latter case, due to a Fierz transformation, the matrix elements of scalar rather than  vector current are involved.
This $q^2$ dependence has an important consequence for the behavior of the effective mass distributions (see Sec.~\ref{SubSec:Bpmdecays}).
For the $P$-wave one has,
\begin{multline}
\label{K-pi+Pampli}
\mathcal{M}_P^-\ \mathbf{p}_{\pi^-}\cdot\mathbf{p}_{\pi^+}\equiv \
\langle\pi^-\ (K^-\pi^+)_P\vert H_{eff}\vert B^-\rangle
=2\sqrt{2}G_F 
 \ \mathbf{p}_{\pi^-}\cdot\mathbf{p}_{\pi^+}  
\ f_1^{B^-\pi^-}(q^2)\  f_1^{K^-\pi^+}(q^2) \\
\times
\bigg\{
 \lambda_u\left(a_4^{u}(P)-\dfrac{a_{10}^{u}(P)}{2}+c_4^{u}\right)+ 
 \lambda_c\left(a_4^c(P)-\dfrac{a_{10}^c(P)}{2}+c_4^{c}\right)\\
+2\dfrac{m_{K^*}}{m_b}\dfrac{f^\perp_V(\mu)}{f_V}\left[ 
 \lambda_u\left(a_6^u(P) -\dfrac{a_8^u(P)}{2} +c_6^u\right)
+\lambda_c\left(a_6^c(P) -\dfrac{a_8^c(P)}{2} +c_6^c\right) \right]
\bigg\}  .
\end{multline}
In Eqs.~(\ref{K-pi+Sampli}) and (\ref{K-pi+Pampli}) $a_i^p(S)$ and 
$a_i^p(P)$ are the leading order factorization coefficients to which $O(\alpha_s)$ vertex and penguin corrections are added
in the quasi-two-body approximation of  pseudoscalar-scalar, $PS$, or pseudoscalar-vector, $PV$, final states (see Sect.~\ref{ai}).
These coefficients will be evaluated at the scale $\mu=m_b$.
 The term proportional to $f_V^\perp(\mu)/f_V$ has been inferred from a similar term which
arises at order $\alpha_s$ in the $B\to PV$ 
amplitudes~\cite{bene03}.
In the following, we take the $K^*(892)$ decay constant $f_V=218$ MeV and $f^\perp_V (m_b) =175$ MeV with $m_{K^*}=893.8$ MeV.
 We have further assumed that the corresponding $P$-wave form factor (which is associated with the antisymmetric tensor current) is simply proportional to the one
associated with the vector current.

 It was observed in Ref.~\cite{bene03} that the calculated $O(\alpha_s)$ and $1/m_b$ corrections are insufficient to explain the experimental branching rate for the $B\to  K^* \pi$ decay. 
We have therefore allowed for four additional complex terms $c_4^u$,  $c_4^c$, $c_6^u$ and $c_6^c$ which one could partly interpret as non-perturbative contributions in the penguin diagrams.
The other part of these coefficients could represent hard spectator interaction in the pertubative regime and annihilation terms.
Annihilation diagrams, due to divergences inherent to the form of the twist-three amplitudes employed in QCDF~\cite{bene03,Beneke:2001ev}, are commonly parametrized by complex amplitudes very similar to these QCD penguin corrections.
This complex parametrization is also the case for the end-point divergence of the hard-spectator amplitudes.
The above parameters will be determined by performing detailed 
fits to the Dalitz plot in the region $m_{K\pi} < 1.8$ GeV,  
$m_{\pi\pi} > 1.5$ GeV where one believes that the QCDF formalism could apply.

The complete amplitude for the $B^-\to (K^-\pi^+)\pi^-$ decay is
\begin{equation}
\label{Mmoinstotal }
\mathcal{M^-} = \mathcal{M}_S^-+\mathcal{M}_P^-\ \mathbf{p}_{\pi^-}\cdot\mathbf{p}_{\pi^+} .
\end{equation}
Charge conjugation of $\mathcal{M^-}$ gives the  $B^+\to (K^+\pi^-)\pi^+$ decay amplitude
\begin{equation}
\label{Mplustotal}
\mathcal{M^+} = \mathcal{M^-}(\lambda_u\to\lambda_u^*,\ \lambda_c\to\lambda_c^*,\ B^-\to B^+,\ K^-\to K^+,\ \pi^\pm\to\pi^\mp).
\end{equation}

Let us specify here the value of the meson masses used in this work: $M_B=5279.2$~MeV, $m_K=495.66$~MeV (averaged $K^\pm$ and $K^0$ masses) and $m_\pi=139.57$~MeV.
For the quark masses, evaluated at the scale $\mu=m_b$, we take $m_b=4.2$~GeV, $m_s=84$ MeV and $m_u=m_d=3.4$~ MeV.
\subsection{ Neutral {\boldmath $B$} decay amplitudes}
In the $\bar B^0\to(\bar K^0\pi^-)_{S,P} \ \pi^+$ decay we have the quark transition $b\to s\bar uu$.
The derivation is similar to that just described above for the $B^-\to(K^-\pi^+)_{S,P}\ \pi^-$ case.
However, there is a tree diagram $a_1$ contribution.
One obtains for the $S$-wave amplitude
\begin{multline}
\label{K0Spi-Sampli}
\mathcal{\bar M}_S^0\equiv\langle\pi^+ \ (\bar K^0\pi^-)_S\vert H_{eff}\vert \bar B^0\rangle
=\dfrac{G_F}{\sqrt{2}}(M_{\bar B^0}^2-m_{\pi}^2)
\dfrac{m_{\bar K^0}^2-m_{\pi}^2}{q^2}
f_0^{\bar B^0\pi^+}(q^2)\  f_0^{\bar K^0\pi^-}(q^2)\\
\times 
\bigg\{
\lambda_u(a_1+a_4^u(S)+a_{10}^u(S)+c_4^u)
+\lambda_c(   a_4^c(S)+a_{10}^c(S)+c_4^{c})\\
 -\ \dfrac{2q^2}{(m_b-m_u)(m_s-m_u)} \left[
\lambda_u(    a_6^u(S)+a_8^u(S)+c_6^u)
+\lambda_c(   a_6^c(S)+a_8^c(S)+c_6^c)
\right]\bigg\} .
\end{multline}
For the $P$-wave amplitude one has
\begin{multline}
\label{K0Spi-Pampli}
\mathcal{\bar M}^0_P\ \mathbf{p}_{\pi^-}\cdot\mathbf{p}_{\pi^+}\equiv 
\langle\pi^+\ (\bar K^0\pi^-)_P\vert H_{eff}\vert \bar B^0\rangle
=2\sqrt{2}G_F 
\mathbf{p}_{\pi^-}\cdot\mathbf{p}_{\pi^+} 
\ f_1^{\bar B^0\pi^+}(q^2)\ f_1^{\bar K^0\pi^-}(q^2) \\
\times 
\bigg\{
 \lambda_u\left(a_1+ a_4^u(P)+a_{10}^u(P)+c_4^u\right)+ 
 \lambda_c\left(     a_4^c(P)+a_{10}^c(P)+c_4^c\right)\\
+2\dfrac{m_{K^*}}{m_b}\dfrac{f^\perp_V}{f_V}\left[ 
 \lambda_u\left(     a_6^u(P) +a_8^u(P)+c_6^u\right)
+\lambda_c\left(     a_6^c(P) +a_8^c(P)+c_6^c\right) \right]
\bigg\} .
\end{multline}
The full amplitude for the $\bar B^0\to(\bar K^0\pi^-)\pi^+$ decays is 
\begin{equation}
\label{M-0}
\mathcal{\bar M}^0=\mathcal{\bar M}_S^0+\mathcal{\bar M}_P^0\ \mathbf{p}_{\pi^-}\cdot\mathbf{p}_{\pi^+}. 
\end{equation}
The charge conjugation of the $\mathcal{\bar M}^0$ amplitude gives the $B^0\to (K^0\pi^+)\pi^-$ decay amplitude as 
\begin{equation}
\label{M0}
\mathcal{M}^0=\mathcal{\bar M}^0(\lambda_u\to\lambda_u^*,\ \lambda_c\to\lambda_c^*,\ \bar B^0\to B^0,\ \bar K^0\to K^0,\ \pi^\mp\to\pi^\pm).
\end{equation}
Isospin symmetry leads to
\begin{equation}
\label{Fisosym}
f_{0,1}^{\bar B^0\pi^+}(q^2)=f_{0,1}^{ B^0\pi^-}(q^2)=
f_{0,1}^{B^-\pi^-}(q^2)=f_{0,1}^{B^+\pi^+}(q^2)
\end{equation}
and to
\begin{equation}
\label{FisosymK}
f_{0,1}^{\bar K^0\pi^-}(q^2)=f_{0,1}^{ K^0\pi^+}(q^2)=
f_{0,1}^{K^-\pi^+}(q^2)=f_{0,1}^{K^+\pi^-}(q^2).
\end{equation}
In Eqs.~(\ref{K0Spi-Sampli}) and (\ref{K0Spi-Pampli}) we have introduced 
the same phenomenological parameters as those for the charged 
$B$-decay amplitudes, Eqs.~(\ref{K-pi+Sampli}) and (\ref{K-pi+Pampli}). 

\subsection{Physical observables}
The density distribution of the Dalitz plot for the $B^-\to K^-\pi^+\pi^-$ decay\footnote {Note that here and in the following, for simplicity, we suppress the parentheses between the two first mesons of the three-meson final state.} can be expressed in terms of the $K^-\pi^+$ and $\pi^+\pi^-$ effective masses, the latter being related to $\cos\theta$ as seen in Eq.~(\ref{costheta}).
The double differential $B^-\to K^-\pi^+\pi^-$ decay rate reads
\begin{equation}
\label{d2gamma}
\dfrac{d^2\Gamma^-}{d\cos\theta dm_{K^-\pi^+}}=\dfrac{m_{K^-\pi^+}\vert\mathbf{p}_{\pi^+}\vert\ \vert\mathbf{p}_{\pi^-}\vert}{8(2\pi)^3M_B^3}\vert\mathcal{M}^-\vert^2 ,
\end{equation}
where $\mathcal{M}^-$is given by Eq.~(\ref{Mmoinstotal }). The moduli of the $\pi^+$ and $\pi^-$ momenta are
\begin{equation}
\label{ppi+}
\vert\mathbf{p}_{\pi^+}\vert=\dfrac{1}{2m_{K^-\pi^+}} 
\sqrt{
\left[
m_{K^-\pi^+}^2-\left(m_{K}+m_{\pi}\right)^2
\right]
\left[m_{K^-\pi^+}^2-\left(m_{K}-m_{\pi}\right)^2\right]
} ,
\end{equation}
\begin{equation}
\label{ppi-}
\vert\mathbf{p}_{\pi^-}\vert=\dfrac{1}{2m_{K^-\pi^+}} 
\sqrt{
\left[
M_B^2-\left(m_{K^-\pi^+}+m_{\pi}\right)^2
\right]
\left[M_B^2-\left(m_{K^-\pi^+}-m_{\pi}\right)^2\right]
} .
\end{equation}
In the experimental analyzes, the helicity angle $\theta_H$ is usually defined as $\theta_H=\pi-\theta$.
Integrating the double differential distribution of Eq.~(\ref{d2gamma}) over $\cos\theta$ gives for the differential effective mass branching fraction
\begin{equation}
\label{dB-}
\dfrac{d\mathcal{B}^-}{dm_{K^-\pi^+}}=\dfrac{1}{\Gamma_{B^-}}
\dfrac{m_{K^-\pi^+}\vert\mathbf{p}_{\pi^+}\vert\ \vert\mathbf{p}_{\pi^-}\vert}{4(2\pi)^3M_B^3}
\left(
\vert\mathcal{M}_S^-\vert^2+\dfrac{1}{3}\vert\mathbf{p}_{\pi^+}\vert^2\ \vert\mathbf{p}_{\pi^-}\vert^2\vert\mathcal{M}_P^-\vert^2
\right) ,
\end{equation}
where $\Gamma_{B^-}$ denotes the total width of $B^-$.
This is a sum of the $S$- and $P$-wave contributions
\begin{equation}
\label{dB-bis}
\dfrac{d\mathcal{B}^-}{dm_{K^-\pi^+}}=\dfrac{d\mathcal{B}_S^-}{dm_{K^-\pi^+}}+\dfrac{d\mathcal{B}_P^-}{dm_{K^-\pi^+}} .
\end{equation}
The $CP$ violating asymmetry for the charged $B$ decays is defined as
\begin{equation}
\label{calA-CP}
{A}_{CP}=\dfrac{\mathcal{B}^--\mathcal{B}^+}{\mathcal{B}^-+\mathcal{B}^+}
\end{equation}

The integration of the double differential decay rate over $m_{K^-\pi^+}$ within the range ($m_{\mathrm{min}},\ m_{\mathrm{max}}$) gives the angular distribution
\begin{equation}
\label{dB-cos}
\dfrac{d\mathcal{B}^-}{d\cos\theta}=A+B\cos\theta+C\cos^2\theta ,
\end{equation}
where
\begin{equation}
\label{eqA}
A=\int_{m_{\mathrm{min}}}^{m_{\mathrm{max}}}
dm_{K^-\pi^+}\ \dfrac{m_{K^-\pi^+}\vert\mathbf{p}_{\pi^+}\vert\ \vert\mathbf{p}_{\pi^-}\vert}
{8(2\pi)^3M_B^3}\ \vert\mathcal{M}_S^-\vert^2 ,
\end{equation}
\begin{equation}
\label{eqB}
B=2\int_{m_{\mathrm{min}}}^{m_{\mathrm{max}}}
dm_{K^-\pi^+}\ \dfrac{m_{K^-\pi^+}\vert\mathbf{p}_{\pi^+}\vert^2\ \vert\mathbf{p}_{\pi^-}\vert^2}{8(2\pi)^3M_B^3}\ 
\mathrm{Re}(\mathcal{M}_S^-\mathcal{M}_P^{-*}) ,
\end{equation}
\begin{equation}
\label{eqC}
C=\int_{m_{\mathrm{min}}}^{m_{\mathrm{max}}}
dm_{K^-\pi^+}\ \dfrac{m_{K^-\pi^+}\vert\mathbf{p}_{\pi^+}\vert^3\ \vert\mathbf{p}_{\pi^-}\vert^3}{8(2\pi)^3M_B^3}\ 
\vert\mathcal{M}_P^-\vert^2\ .
\end{equation}
Similarly one can derive the above observables given in Eqs.~(\ref{d2gamma}), (\ref{dB-}) 
and (\ref{dB-cos}) for the $B^+\to K^+\pi^-\pi^+$, $\bar B^0\to\bar K^0\pi^-\pi^+$ and $B^0\to K^0\pi^+\pi^-$ reactions.

\subsection{{\boldmath $B\to K^*_0(1430)\pi$} and 
{\boldmath $B\to K^*(892))\pi$} amplitudes}
\label{bk*piamplitude}

The scalar $K_0^*(1430)$ and the vector $K^*(892)$ resonances are 
quite visible in the experimental effective $K\pi$ mass distributions 
of the $B\to K\pi\pi$ decays~\cite{AubertPRD72,Abe2005,Garmash:2005rv,AubertPRD73,Abe0509047,Garmash2007,AubertLP2007,Aubert:2008bj,Aubert:2007bs}.
Branching fractions for the  $B\to K^*_0(1430)\pi$ and  $B\to K^*(892)\pi$ 
reactions have been extracted by the experimental groups 
within the framework of the isobar model. 
Here, we will make use of the complex pole definition of a resonance in 
scattering theory. This method allows one to perform in a simple way a 
separation between the resonance contribution 
(defined to correspond to the pole part) and the background
contribution in the amplitude.

We obtain the $B\to K^*_0(1430)\pi$ amplitudes by replacing in Eqs.~(\ref{K-pi+Sampli}) and  (\ref{K0Spi-Sampli})  $f_0^{K\pi}(q^2)$ by $f_0^{pole} (q^2)$ defined in Eqs.~(\ref{f0pole}), (\ref{t0}) and (\ref{f0t0}).
Similarly  the $B\to K^*(892)\pi$ amplitudes are given by replacing in Eqs.~(\ref{K-pi+Pampli}) and  (\ref{K0Spi-Pampli})  $f_1^{K\pi}(q^2)$ by $f_1^{pole} (q^2)$  defined in Eqs.~(\ref{fpluspole}) and (\ref{t1fplus}).
Integration of the differential branching 
fraction~(\ref{dB-}) over the effective mass $m_{K\pi}$ of the pole 
parts of the scalar or vector  form factors will 
determine the $B^-\to\bar K_0^{*0}(1430)\pi^-$ and  
$B^-\to \bar K^{*0}(892) \pi^-$  branching fractions, respectively.
In this approach, we can also determine the decay constants of the vector
and the scalar resonances: this is exposed in Appendix~\ref{appendixfdecaypole}.

\begin{table}[htdp]
\caption{
Leading  and next-to-leading order effective QCD amplitudes  $a_i(m_b)$ and $a_i^p(m_b)$ for the  $K_0^*(1430)\pi$ and  $K^*(892)\pi$ final states [see Eqs.~(\ref{ailo}) and (\ref{eq:44})].
\label{aipmb} 
}
\begin{ruledtabular}
\begin{tabular}{ccccc}
&  \multicolumn{2}{c}{$B\to K_0^*(1430)\pi$} &  \multicolumn{2}{c}{$B\to K^*(892)\pi$} \\
& $a_i(m_b)$ & $a_i^p(m_b)$ &  $a_i(m_b)$ & $a_i^p(m_b)$ \\
\hline
$a_1$     & $1.018  $ & $1.029+i\ 0.063  $   & 1.018   & $1.045+i\ 0.014$\\
$a_4^{u}$ & $-0.031  $ & $-0.061-i\ 0.023 $  & $-0.031$ & $-0.030-i\ 0.015$ \\
$a_4^c$   & $-0.031  $ & $-0.069+i\ 0.057 $  & $-0.031$ & $-0.035-i\ 0.005$ \\
$a_6^{u}$ & $-0.039  $ & $-0.042-i\ 0.014 $  & 0         & $-0.006-i\ 0.002$ \\
$a_6^c$   & $-0.039  $ & $\ -0.045-i\ 0.004~~$  & 0         & $0.002+i\ 0.009$ \\
$a_8^{u}$ & $0.00044$ & $0.0005-i\ 0.0001$  & 0         & $-0.0001+i\ 0.0$ \\
$a_8^c$   & $0.00044$ & $0.0005-i\ 0.0   $  & 0         & $-0.0+i\ 0.0001$ \\
$a_{10}^u$& $-0.0015$ & $-0.002+i\ 0.003  $  & $-0.0015$ & $0.0001+i\ 0.0006$\\
$a_{10}^c$& $-0.0015$ & $-0.002+i\ 0.004  $  & $-0.0015$ & $0.0001+i\ 0.0007$
\end{tabular}
\end{ruledtabular}
\end{table}

\section{Effective QCD amplitudes}
\label{ai}

The short-distance physics of the weak decay amplitudes $b\to s \bar dd$ or
 $b\to s\bar uu$ is codified in the effective QCD amplitudes $a_i^{u,c}(\mu)$,
  each of which corresponds to a particular decay topology. 
The available QCDF calculations (see for instance Refs.~\cite{Beneke:2001ev,bene03,Cheng:2005nb}) concern two-body final states. 
They apply to
our case of a three-body final state $(K\pi)\pi$ when the effective mass of the $K\pi$ sub-system coincides with the mass of a vector or a scalar resonance.
We will make the approximation to use these calculated corrections also away from the resonance masses.
In this work, we take into account only vertex and penguin corrections.
At the leading order, these amplitudes are universal (i.e. do not depend on quark flavor) 
 and given by
  
\begin{equation}
\label{ailo}
a_i(\mu)=\left [C_i(\mu) + \dfrac{C_{i\pm 1}(\mu)}{N_c} \right ]\ N_i (M),
\end {equation}
where the plus (minus) sign corresponds to odd (even) values of the index $i$.
One finds $N_i(M)=0$ if the emitted meson $M$ (either the $K^*(892)$ or the $K^*_0(1430)$, which do not include the spectator quark) is a vector one and if
$i=6,8$, otherwise $N_i(M)=1$.
The next-to-leading order vertex and penguin corrections  are calculated  following Ref.~\cite{Cheng:2005nb} for $ K^*_0(1430) \pi$ final states and Ref.~\cite{bene03} for  $K^*(892) \pi$ ones.
We write
\begin{equation}
  a_i^p(\mu)  = a_i(\mu)
   +   \dfrac{C_{i\pm 1}}{N_c} \dfrac{C_F\ \alpha_s}{4\pi}\, V_i(M) + P_i^p(M)  ,
  \label{eq:44}
\end{equation}
with $C_F=(N_c^2-1)/2N_c$. 
 The relevant Wilson $C_i(\mu)$ coefficients were calculated in next-to-leading  order logarithmic approximation and
we take their values from Table~1 of Ref.~\cite{Beneke:2001ev} at the natural characteristic scale $\mu=m_b$.
The second and third terms of Eq.~(\ref{eq:44}) refer to the vertex  and penguin corrections and
for completeness we describe in Appendix~\ref{appendixvpc} the main steps of their calculations.
In  Table~\ref{aipmb} we list the values of the coefficients  $a_i^p(m_b)$ with $i=1,4,6,8,10$, used in the weak decay amplitudes discussed in the present paper.
These have been obtained adding to the $a_i(m_b)$ values (given in this table) those of the vertex and penguin corrections listed in  Table~\ref{nlovp}.
The next-to-leading order corrections are relatively small except  in the case of the $B\to K^*_0(1430) \pi$ decays for $a_4^u$ and $a_4^c$.
These arise mainly from the penguin terms.

\section{Scalar and vector {\boldmath $B\pi$} and {\boldmath $K \pi$} form 
factors \label{formfactors}}
\subsection{{\boldmath $B\pi$}  form factors \label{transitionff}}

The scalar and vector $B\pi$ transition form factors $f_0^{B\pi}(q^2)$ and $f_1^{B\pi}(q^2)$ were introduced in Eq.~(\ref{BtopiFF}) and are one of the ingredients in QCDF. These form factors contain parts of the non-perturbative physics which stems from the hadronization of quark currents. 
In the applications considered here, the variable 
$q^2$ is timelike and remains 
small compared to the physical threshold $(m_B+m_\pi)^2$. The form 
factors in this region are real and we expect their variation to be slow.
%
Several studies dedicated to these transition form factors have been carried out; on the phenomenological side, we refer to the various approaches to heavy-to-light transition amplitudes.  These include light-cone sum rules~\cite{Khodjamirian:2006st}, light-front~\cite{Lu:2007sg}, simple non-relativistic~\cite{Albertus:2005ud} or relativistic quark models~\cite{Ebert:2006nz,Ivanov:2000aj,Faessler:2002ut,Melikhov:2001zv}. 
Ab initio approaches such as lattice-regularized QCD have also been used to determine $B \pi$ form factors, however the lattice results are for large momenta $q^2> 10$~GeV$^2$ and need to be extrapolated to low momenta~\cite{Burford:1995fc,Bowler:1999xn} by means of pole dominance models. 

More recently, a comprehensive set of $B$-meson heavy- to light-transition form factors, calculated with truncated transition amplitudes based on Dyson-Schwinger equations in QCD, was reported in Ref.~\cite{Ivanov:2007cw}. The methods of Refs.~\cite{Khodjamirian:2006st,Lu:2007sg} yield form factors for a small domain of time-like momentum transfer $q^2$, while those in Refs.~\cite{Ebert:2006nz,Ivanov:2000aj,Faessler:2002ut} apply to the entire range of physical momenta. However, this is only possible by employing functional extrapolations of the transition form factors $f_{0,1}^{B\pi}(q^2=0)$ to the time-like $q^2$-range. On the other hand, the form factors obtained from double dispersion relations of spectral densities in the relativistic quark model~\cite{Melikhov:2001zv} or using Dyson-Schwinger equations in QCD~\cite{Ivanov:2007cw} are calculated for all physical $q^2$ values. 
A typical value found in~\cite{Melikhov:2001zv} is $f_1^{B\pi}(q^2=0)\simeq 0.2$
which is in agreement with the Dyson-Schwinger result $f_1^{B\pi}(q^2=0)=0.24$ 
as well as with lattice data extrapolations~\cite{Bowler:1999xn}. 

In the above studies, the $q^2$ dependence  of the $B\pi$ transition 
form factors, as $q^2$ varies  from threshold up to about $(1.8\ {\rm GeV})^2$, is found to be small. 
This, of course, is in contrast with the case of the $K\pi$ scalar 
or vector form factors which we discuss below. 
In practice, we take the following constant values:
$f_0^{B\pi}(q^2)=f_0^{B\pi}(m^2_{K_0^*(1430)})=0.266$ and 
$f_1^{B\pi}(q^2)=f_1^{B\pi}(m^2_{K^*(892)})  =0.250$.

\subsection{{\boldmath $K \pi$} form factors \label{svff}}
Another important ingredient of the QCD-factorized $B$-decay amplitudes are the $K \pi$ scalar and vector form factors. 
These also appear in semileptonic decays like
$\tau\to K\pi\nu_\tau$ or $K\to \pi l \nu_l$. 
Use of analyticity and unitarity allows one to relate them to the 
$S$ and the $P$ wave $K \pi$ scattering amplitudes. 
Indeed, rather accurate information on $K \pi$ scattering is available, in particular from the high statistics experiment of Estabrooks~\textit{et al.}~\cite{estabrooks} and from the LASS Collaboration~\cite{aston88}. 
This approach served to determine the scalar form factors of the pion and  kaon in Ref.~\cite{dgl}. 
Jamin, Oller and Pich applied it recently to evaluate the 
$K\pi$ scalar form factor~\cite{jop}. 
We perform also its calculation following, 
essentially, their work. 
We then develop an extension of this construction to the vector case, so
both form factors are handled in a similar way. 
The main approximation is to reduce the sum over the inelastic many-body
channels in the unitarity equations to a finite number of two-body channels.
This is supported by the experiments performed by the LASS Collaboration
in the center of mass energy range $m_{K \pi} < 2.5$ GeV~\cite{aston84,aston87,aston88,aston88b}.
They find that the inelasticity in the $S$-wave
is dominated by one channel, $K \eta'$, and in the $P$-wave by two channels,
$K^*\pi$ and $K \rho$. 
Then, from fits to the experimental data, one constructs
a $2\times 2$ scattering $T$-matrix for the $S$ wave and a $3\times 3$ one for
the $P$ wave. The form factors satisfy a set of $n$ coupled, homogeneous
singular integral equations with a kernel linear in the $T$-matrix ($n=2$ for 
the $S$ wave and $n=3$ for the $P$ wave). The mathematical properties of such
equations are derived in Muskhelishvili's book~\cite{muskhelishvili}.
In particular the number
of independent solutions $N$ is given by the index of the integral
operator which can be expressed in terms of the sum of the $S$-matrix 
eigenphases
$\delta_j(t)$ of the $S$-matrix,
\begin{equation}
\lbl{index}
\sum_{j=1}^n [ \delta_j(\infty) - \delta_j(0) ] = N \pi\ .
\end{equation}
$N$ is also the number of independent conditions that one must impose on the
form factors in order to determine them from the integral equations. 
We will use conditions at $t=0$ or near $t=0$ derived from chiral
symmetry. Asymptotic conditions will also be used.

\begin{figure}[ht]
\centering
\includegraphics[width=8.1cm]{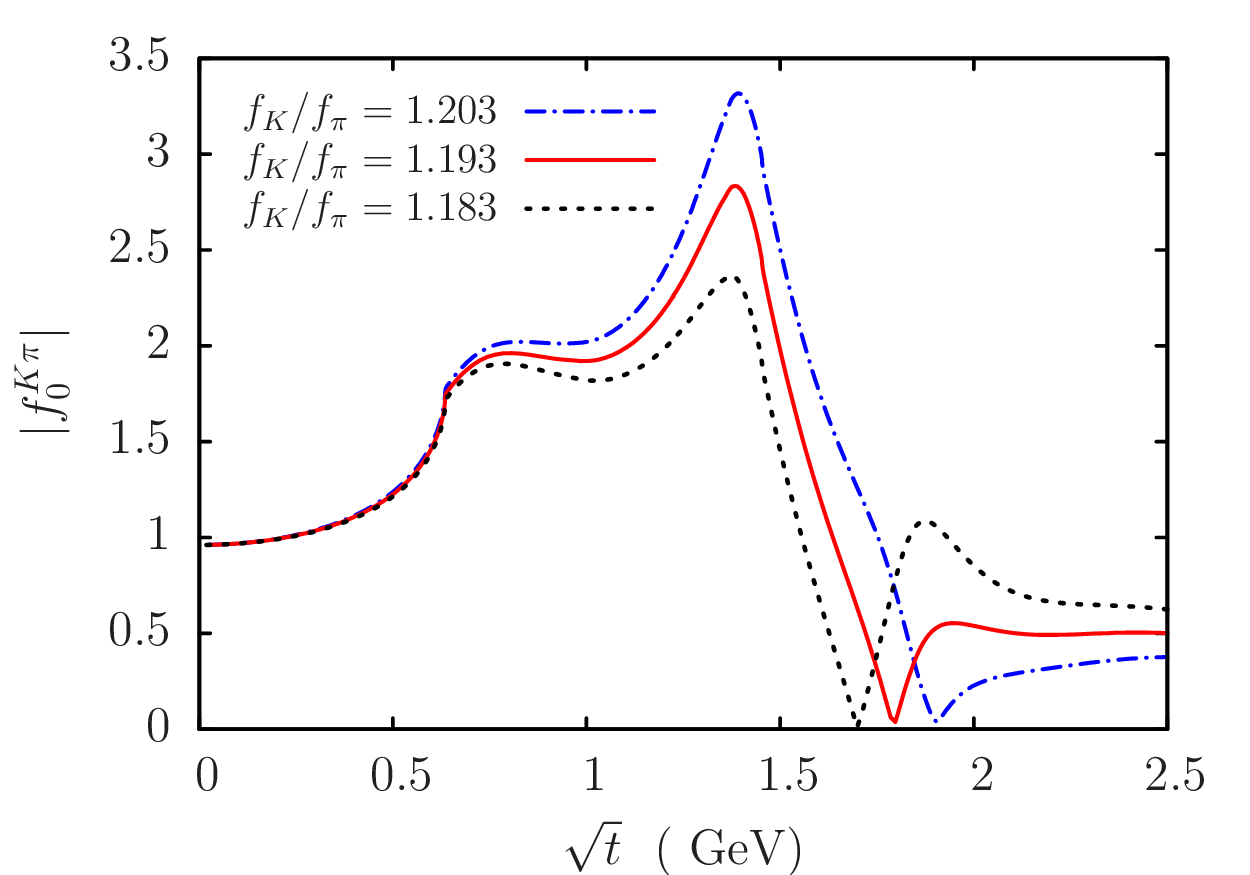}
\includegraphics[width=8.cm]{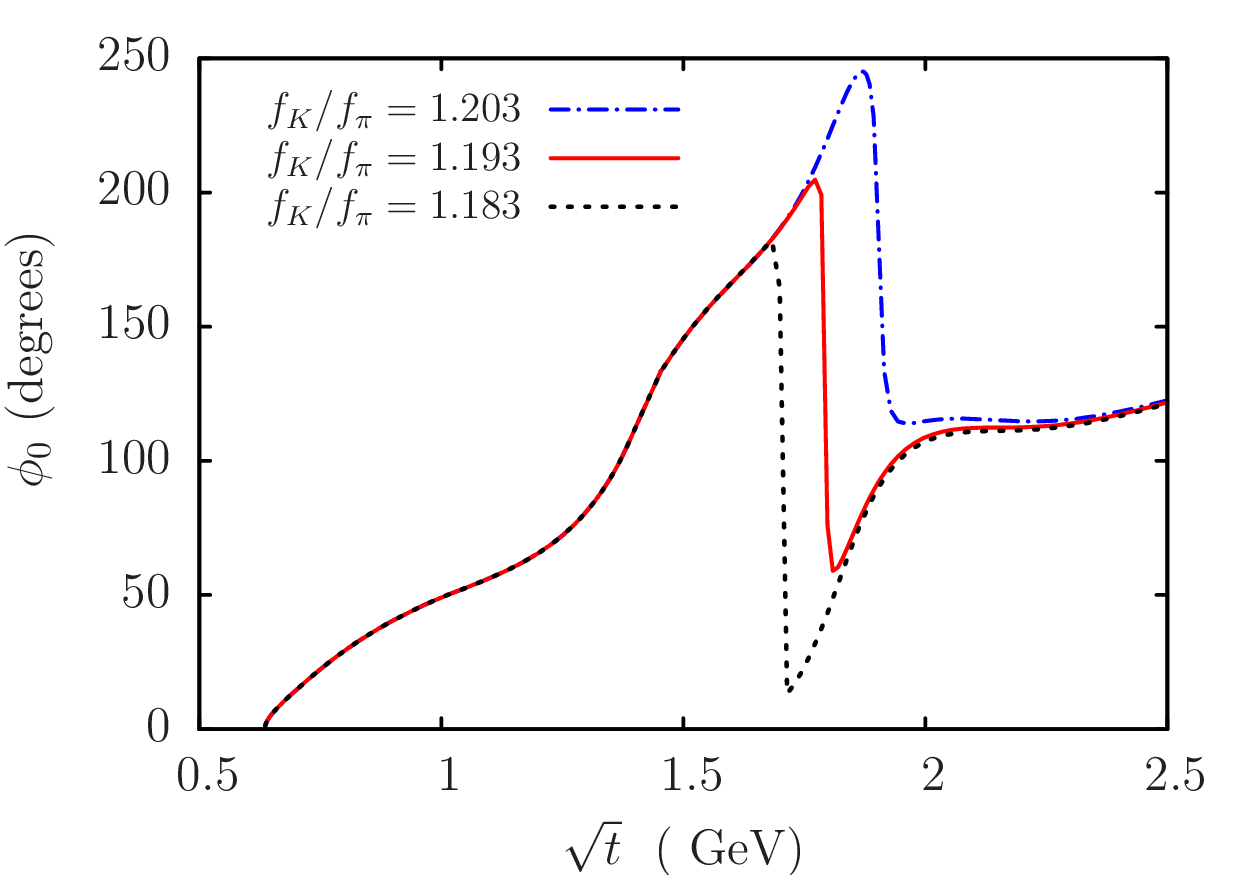}\
\caption{
\label{scalarff} 
Modulus (left) and phase (right) of the strange scalar
form factor $f_0^{K^-\pi^+}(t)$ obtained by solving a two-channel Muskhelishvili-Omn\`es equation system. 
Variation with the input $F_1(\Delta_{K\pi})= f_K/f_\pi - 3.1\times 10^{-3}$ at the Cheng-Dashen point $\Delta_{K\pi}=m_K^2-m_\pi^2$ is illustrated.
In the present work we use the form factor corresponding to $f_K/f_\pi= 1.193$.
}
\end{figure}

\subsubsection{\boldmath $Scalar~ form~ factor$ \label{scalarformfactor}}

Following the experimental analyzes, we assume the dominance of a single
inelastic channel $K\eta'$~\footnote[1]{The authors of Ref.~\cite{jop} have
studied the influence of an additional $K \eta$ inelastic channel and found
that it is rather small.}. 
The $K\eta'$ matrix element of the vector current can be written in terms of two
form factors $f_+^{K^+\eta'}(t)$ and $f_-^{K^+\eta'}(t)$ as
\begin{equation}
\braque{K^+\vert  \bar{u}\gamma^\mu s \vert\eta'}
=f_+^{K^+\eta'}(t) (p_K+p_{\eta'})^\mu+ f_-^{K^+\eta'}(t) 
(p_K-p_{\eta'})^\mu ,
\end{equation}
where $t=(p_K-p_{\eta'})^2=q^2$.
The coupled channel integral equations will involve the two components
\begin{eqnarray}
&& F_1(t)\equiv \sqrt2 f_0^{K^+\pi^0}(t)= f_0^{K^-\pi^+}(t) ,
\nonumber\\
&& F_2(t)= \sqrt{\frac{2}{3}} \left[ {\mkd-\metapd\over\mkd-\mpid} 
f_+^{K^+\eta'}(t)  +  {t\over \mkd-\mpid} f_-^{K^+\eta'}(t)\right].
\end{eqnarray}
We have used here isospin symmetry  to express the form factor $f_0^{K^-\pi^+}(t)$,
 introduced in Eq.~(\ref{Ktopiff}), in terms of  $f_0^{K^+\pi^0}(t)$.

The form factors $F_1(t)$ and $F_2(t)$ are analytic functions in the complex 
$t$ plane
with a cut along the real axis $ (m_\pi+m_K)^2 \le t \le \infty$.
Asymptotic counting rules in QCD~\cite{brodskylepage} imply that the
dispersion relations satisfied by the functions $F_i(t)$ converge without
subtractions 
\begin{equation}
\lbl{dispfi}
F_i(t)= {1\over\pi}\int_{(m_\pi+m_K)^2}^\infty {\im\,F_i(t')\, dt'
\over t'-t}\quad,\quad i=1,2\ .
\end{equation}
Unitary equations of the two coupled channels $K \pi\ (i=1)$ and $K \eta'\ (i=2)$ 
allow one to express the imaginary parts as follows (see~\cite{dgl,jop})
\begin{eqnarray}
\lbl{unitscal}
&&\im F_1(t)=\theta(t-t_1) {2 q_{K\pi}(t)\over \sqrt{t}}\, T_{11}^*(t)\, F_1(t)
+\theta(t-t_2) {2 q_{K\eta'}(t)\over \sqrt{t}}\, T_{12}^*(t)\, F_2(t),
\nonumber\\
&&\im F_2(t)=\theta(t-t_1) {2 q_{K\pi}(t)\over \sqrt{t}}\, T_{12}^*(t)\, F_1(t)
+\theta(t-t_2){2 q_{K\eta'}(t)\over \sqrt{t}}\, T_{22}^*(t)\, F_2(t),
\end{eqnarray}
where
\begin{equation}
t_1= (m_K+m_\pi)^2,\quad 
t_2= (m_K+\metap)^2
\end{equation}
and $q_{K\pi}(t)$,  $q_{K\eta'}(t)$  are the center of mass momenta. 
The set of integral equations obtained by inserting Eqs.~\rf{unitscal}
into~\rf{dispfi} are often called Muskhelishvili-Omn\`es 
equations. 
The detailed determination of the matrix elements $T_{11}$,
$T_{12}$, $T_{22}$ is given in Appendix~\ref{appendixSwavT}. 
We use experimental data 
and theoretical constraints at low energy, in particular we employ
a systematic combination of the chiral and $1/N_c$ expansions~\cite{mou95,kaiserleutwyler}. 
At medium and high energies, we employ a $K$-matrix parametrization
as in Ref.~\cite{jop} which guarantees unitarity. This provides
a smooth interpolation with index $N=2$ in the
asymptotic region, where no experimental data exist [see Eq.~\rf{index}].
 One must therefore provide two initial conditions for the form factors. 
As in Ref.~\cite{jop}, we use the values of $F_1(t)$ at $t=0$
and at the Cheng-Dashen point $t=\mkd-\mpid$ which are precisely constrained 
by chiral symmetry 

\begin{equation}
\label{scalconds}
F_1(0)=0.961,\quad F_1(\mkd-\mpid)= {f_K\over f_\pi} -3.1 \cdot 10^{-3},
\end{equation}
where $f_K$ and $f_{\pi}$ are the kaon and pion decay constants, respectively. The value at $t=0$ is derived from chiral perturbation theory at order $p^4$~\cite{gl85,gl85ff} and includes an estimate for the $p^6$ corrections made in Ref.~\cite{leutwyler-roos}. 
The value at $t=\mkd-\mpid$ was obtained in Ref.~\cite{gl85ff} at order $p^4$. In that case, the $p^6$ corrections are expected to be very small, of order $10^{-3}$. 
For the ratio $f_K/f_\pi$, the latest Review of Particle Physics~\cite{pdg08}
(see the note by J. Rosner and S. Stone, p.~821)  gives
\begin{equation}
\frac{f_K}{f_\pi}=1.193\pm 0.002\pm 0.006\pm 0.001
\end{equation} 
and we will use the central value in Eq.~(\ref{scalconds}).
We solve the set of integral equations for $F_1$ and $F_2$ using the numerical method of Ref.~\cite{L6}. 
The results for the modulus and for the phase of $F_1$ are displayed in Fig.~\ref{scalarff}. 
Our numerical results are in a fair agreement with those of Refs.~\cite{jop} and \cite{jopffa}. 
The phase of the form factor satisfies, as it should, Watson's theorem in the energy region below the $K\eta'$ threshold. 
Above this point it displays a sharp drop. 
Correspondingly, the modulus of $F_1$ displays a dip.
These structures reflect interference effects between the independent solutions which are linearly combined. 
Due to the linearity of the equations, the dependence on the initial
conditions can be studied by varying $F_1(m_K^2-m_\pi^2)$ and keeping
$F_1(0)$ fixed. The figure illustrates the variation of the scalar form factor
$f_0^{K \pi}$ as a function of $f_K/f_\pi$.

\subsubsection{{\boldmath $K^*_0(1430)~ pole~ part~ of~ the~ scalar~ form~ factor$}
\label{polescal}}

The separation between resonant and background contributions has some
arbitrariness, in particular if the background contribution is 
significant, depending on how each contribution is parametrized.
In the case of $B\to K\pi\pi$ amplitude, 
we propose a simple and unambiguous way to define this separation
based on the well known property that a resonance can be associated 
with a pole of the scattering matrix in the complex energy plane on the second
Riemann sheet (e.g.~\cite{taylorq}). 
This pole also appears in form factors and current correlation functions.
We first study the scalar $K_0^*(1430)$ resonance
and then the vector $K^*(892)$ resonance.

Let us analyze the scalar form factor $f_0^{K\pi}(t)$.  
We want to define the extrapolation
to the second Riemann sheet in the $t$ variable.  We recall that we have
assumed, in constructing the $T$ matrix, that scattering was elastic
up to the $\eta' K$ threshold. We can write the
discontinuity of the form factor upon crossing the cut as follows,
\begin{equation}
\lbl{discf0}
f_0^{K\pi}(t+i\epsilon)- f_0^{K\pi}(t-i\epsilon)= 
-2\sigma_{K \pi}(t+i\epsilon) 
T_{11}^S(t+i\epsilon) f_0^{K\pi}(t-i\epsilon),
\end{equation} 
for $t$ real and in the range
$(m_\pi+m_K)^2 \le t \le (m_{\eta'}+m_K)^2 $.
Here,
$\sigma_{K \pi}(t)={\sqrt{((m_K+m_\pi)^2-t)(t-(m_K-m_\pi)^2)}/ t }$
and  $T_{11}^S$ is  the  $S$-wave $T$-matrix element of the $K\pi \to K\pi$ process.
$T_{11}^S(t)$ satisfies a discontinuity equation similar
to Eq.~\rf{discf0},
\begin{equation}\lbl{discT11}
T_{11}^S(t+i\epsilon)- T_{11}^S(t-i\epsilon)= -2\sigma_{K \pi}(t+i\epsilon) 
T_{11}^S(t+i\epsilon) T_{11}^S(t-i\epsilon) .
\end{equation}
Eqs.~\rf{discf0} and \rf{discT11} 
allow us to find the extension of $f_0^{K\pi}$ on the second Riemann sheet
\begin{equation}
\lbl{f0II}
f_0^{II}(t)= {f_0^{K\pi}(t)\over 1-2\sigma_{K \pi}(t) T_{11}^S(t) } ,
\end{equation}
which, by definition, must satisfy 
\begin{equation}
f_0^{II}(t-i\epsilon)=f_0^{K\pi}(t+i\epsilon)
\end{equation}
along the cut.
Eq.~\rf{f0II} shows that $f_0^{II}(t)$ displays a pole whenever the denominator
function 
\begin{equation}
\lbl{zerofonc}
D(t)=1-2\sigma_{K \pi}(t) T_{11}^S(t) 
\end{equation} 
displays a zero. In a similar way, one can define the extension to the second
sheet of the $T$-matrix, $(T_{11}^S)^{II}$, which because of Eq.~\rf{discT11}
has exactly the same denominator function $D(t)$. A point $t_0$
such that $D(t_0)=0$ corresponds to a pole of $(T_{11}^S)^{II}$ 
and thus can be associated with a resonance~\cite{taylorq}.
From this, it is simple to isolate the pole part of the form factor
\begin{equation}
\label{f0pole}
f_0^{pole}(t)= {f_0^{K\pi}(t_0)\over\alpha\,(t-t_0) }, 
\end{equation}
where  $\alpha= d D(t)/dt$ at $t=t_0$. 
In the numerator,  $f_0^{K\pi}(t_0)$ can be computed using its dispersive representation. 

\begin{figure} [ht]
\begin{center}
\includegraphics[width=10cm]{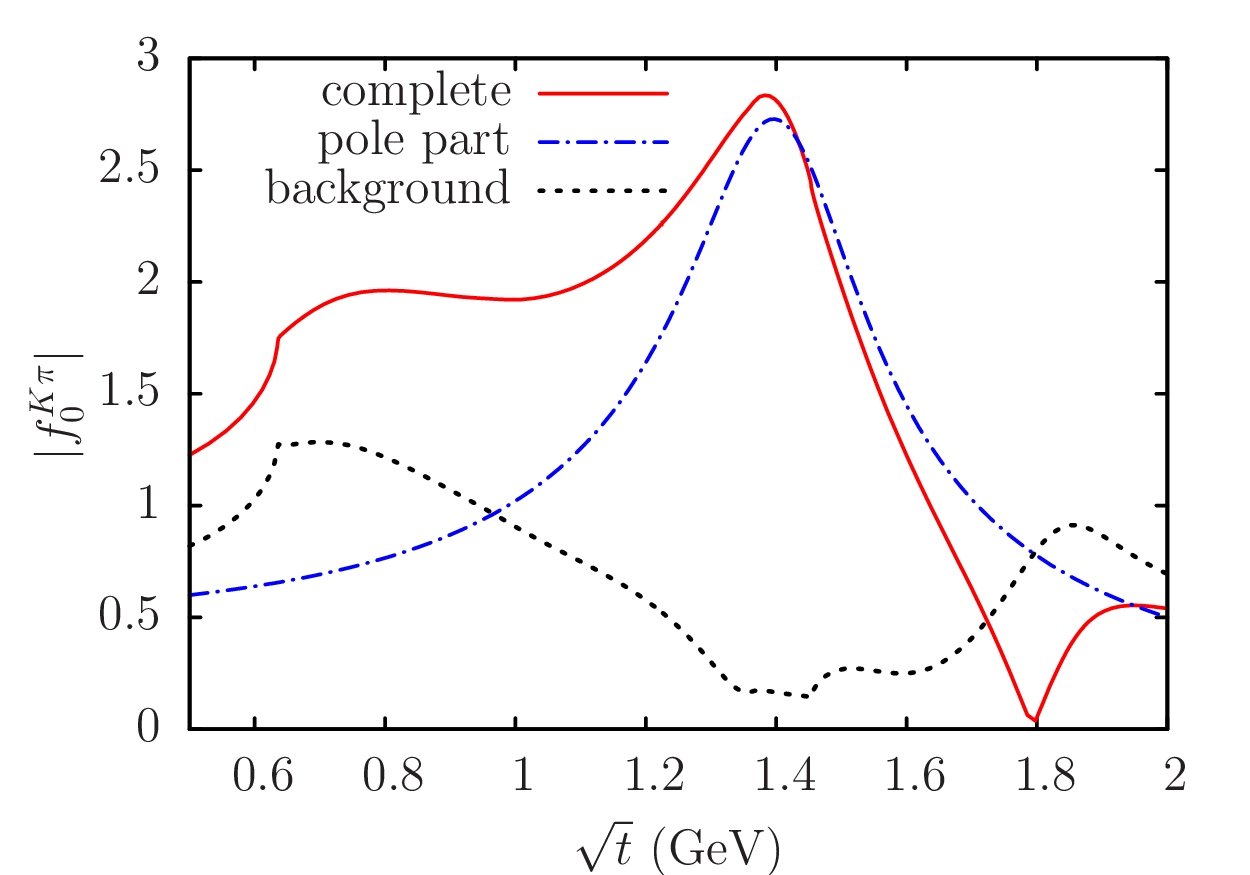}
\caption{
Modulus of the scalar form factor $f_0^{K^-\pi^+}(t)$ compared
with its pole part, as defined in the text.
\label{ffpole}
}
\end{center}
\end{figure}

In practice, in order to do so, 
we must be able to define $T_{11}^S(t)$ for complex values of $t$. In our work
we have defined $T_{11}^S(t)$ on the real axis in the range $1.25 \le \sqrt{t}
\le 2.5$ GeV from a $2\times2$ $K$ matrix 
fit to the experimental data. 
By construction, the elements of the $K$-matrix have no branch cut on the
positive real axis. The meromorphic function parametrization which we used
should be valid in some complex domain of the $t$ variable.  
It seems reasonable to assume that it remains reliable in the region of
the $K^*_0(1430)$ resonance pole since this pole lies rather close to 
the real axis. 
Numerically, we find the following result for the location of the pole
\begin{equation}
\label{t0}
t_0= (1.9487  -i\ 0.3825)\ {\rm GeV}^2,
\quad \sqrt{t_0}= (1.4026 -i\ 0.1364) \ {\rm GeV}.
\end{equation}
These results compare reasonably 
well with the values of the mass $M_R=( 1.414\pm 0.006)$~GeV and of the half-width $\Gamma_R/2= (0.145\pm 0.011)$ GeV of the 
$K^*(1430)$ given by the PDG~\cite{pdg08}.
For the other quantities needed for $f_0^{pole}(t)$ we obtain
\begin{equation}
\label{f0t0}
f_0^{K\pi}(t_0)= -0.3242 -i\,1.4679,\quad \alpha= (0.8381+i\,1.1713)\ \mbox{\rm{GeV}}^{-2}.
\end{equation} 
We remind that, in the dispersive construction of $f_0^{K\pi}(t)$ 
from Ref.~\cite{jop}, 
followed here, the result depends on the value of $ f_K/ f_\pi$ which
controls one of the initial conditions [see Eq.~(\ref{scalconds})]. 
The number given above for $f_0(t_0)$ corresponds to $ f_K/ f_\pi=1.193$ which we used as
central value. Varying $ f_K/ f_\pi$ we would obtain:
\begin{eqnarray}
\lbl{fkfpivar}
&& f_0^{K\pi}(t_0)=-0.4901-i\ 1.6652  \ \ \mbox{\rm{for}}\  { f_K\over  f_\pi}=1.203,
\nonumber \\
&& f_0^{K\pi}(t_0)=\ -0.1586-i\ 1.2710  \ \ \mbox{\rm{for}}\ { f_K\over  f_\pi}=1.183.
\end{eqnarray}
Figure~\ref{ffpole}  
shows the modulus of $f_0^{pole}(t)$ compared with
the modulus of $f^{K^-\pi^+}_0(t)$ and that of the ``background'' which we may define
just as the difference 
\begin{equation}
\label{f0back}
f_0^{back}(t)= f^{K\pi}_0(t)-f_0^{pole}(t).
\end{equation}
We can define the isolated  $K^{*}_0(1430)$ resonance contribution to the 
$B \to  K \pi^+ \pi^-$ decay amplitudes
by replacing the scalar form factor 
$f_0^{K\pi}(t)$
by $f_0^{pole}(t)$. 
For instance, for the $B^- \to  K^- \pi^+ \pi^-$ case, this substitution can be done in Eq.~(\ref{K-pi+Sampli}).

\subsubsection{\boldmath $Vector~ form~ factor$} \label{vff}
 We perform a construction using the same  method as in the case of
the scalar form factor. 
Here the main points involved in this construction are given below, a more detailed discussion can be found in Ref.~\cite{MoussallampiKSandPwave}. 
For  $K \pi$ scattering in $P$-wave at
medium energies ($m_{ K \pi}\le 2.4$ GeV), we assume
the dominance of two inelastic channels, $K^*\pi$ and $K\rho$. We treat vector
mesons as stable particles in our unitary equations. For the three coupled
channels $K \pi\ (i=1), K^*\pi\ (i=2)$ and $K \rho\ (i=3)$ three form factors 
$H_i(t)$ enter the calculation. 
The first one was defined in Eq.~(\ref{Ktopiff}):
\begin{equation} 
\lbl{H1}
 H_1(t)\equiv \sqrt2  f_1^{K^+\pi^0}(t)= f_1^{K^-\pi^+}(t).
\end{equation}
The other two are defined by the following matrix elements:

\begin{equation}
\lbl{H2}
\braque{K^{*+}(p_V,\lambda) \vert \bar{u}\gamma_\mu s\vert \pi^0(p_\pi)}=
\epsilon_{\mu\nu\alpha\beta}\, \varepsilon^{*\nu}(\lambda) p_V^{\alpha} 
p_{\pi}^{\beta}\,H_2(t), 
\end{equation}
\begin{equation}
\lbl{H3}
\braque{\rho^0(p_V,\lambda) \vert \bar{u}\gamma_\mu s\vert K^-(p_K)}=
-\epsilon_{\mu\nu\alpha\beta} \varepsilon^{*\nu}(\lambda) p_V^{\alpha} 
p_{K}^{\beta} H_3(t). 
\end{equation}
In the above equations $\varepsilon^{\nu}$ is the polarization four-vector of  
the $K^*$ or the $\rho$ meson, $p$ denote four-momenta of mesons and 
$\epsilon_{\mu\nu\alpha\beta}$ is the completely antisymmetric tensor. 
The components $H_2$ and $H_3$ have dimension mass$^{-2}$
while $H_1$ is dimensionless. As for the $S-$wave case each form factor
satisfies an unsubtracted dispersion relation. 
We have now to determine six independent matrix elements of the $3\times3$
$P$-wave $T$-matrix, $T^P$, from fits to the experimental data. 
The most complete data exist for the elastic channel $ K \pi\to K \pi $~\cite{aston88}. 
Some information is also available on the inelastic amplitudes  
$ K \pi\to  K^* \pi$ and $ K \pi\to  K \rho$ in the regions of the resonances $K^*(1410)$ and $K^*(1680)$~\cite{aston87,aston88b}. 
As in the case of the $S$-wave, the $K$-matrix method
is used to enforce three-channel unitarity.
The unitarity equations obeyed by the three form factors $H_i(t)$  can be 
written as
\begin{equation}
\lbl{Puniteq}
\im
H_i(t)
= \sum_{j=1}^3\, \left( \tau^{-1} \left(T^P\right)^*\, Q^2\, \Sigma\, \tau\right)_{ij} 
H_j(t) ,
\end{equation}
where $\tau$, $Q$ and $\Sigma$ are diagonal matrices with $\tau= \mathrm {diag}[1,\,\sqrt{t},\,\sqrt{t}]$,  $\Sigma= {2Q/\sqrt{t}} $
and
\begin{equation}
Q=\mathrm{diag}[ 
\theta(t-t_1)q_{K\pi}(t),\, 
\theta(t-t_3)q_{K^*\pi}(t),\,
\theta(t-t_4)q_{K\rho}(t)]
\end{equation}
with $t_3=(m_{K^*}+ m_\pi)^2$ and $t_4= (m_K +m_\rho)^2$.
The normalization of the $T^P$-matrix here is such that its relation with the
$S$-matrix is
\begin{equation}
S^P = 1 +2i \sqrt{\Sigma}\, Q\, T^P\, Q\,  \sqrt{\Sigma} .
\end{equation}
In the asymptotic region, $m_{K\pi } \ge 2.4 $ GeV, we impose a smooth 
interpolation for the scattering amplitudes with the index $N=4$ [see 
Eq.~\rf{index}]. 
In order to enforce exact unitarity in the interpolation region we 
write the $S$-matrix in exponential form, $S^P= \exp(2ih)$, where
$h$ is real and symmetric matrix and we interpolate the non-diagonal elements of $h$ to zero and the diagonal ones to multiples of $\pi$.
Illustration of this interpolation of the eigenphases of the $S^P$-matrix is given in the Fig. 4 of Ref.~\cite{MoussallampiKSandPwave}. 
\begin{figure}[hbt]
\centering
\includegraphics[width=8.1cm]{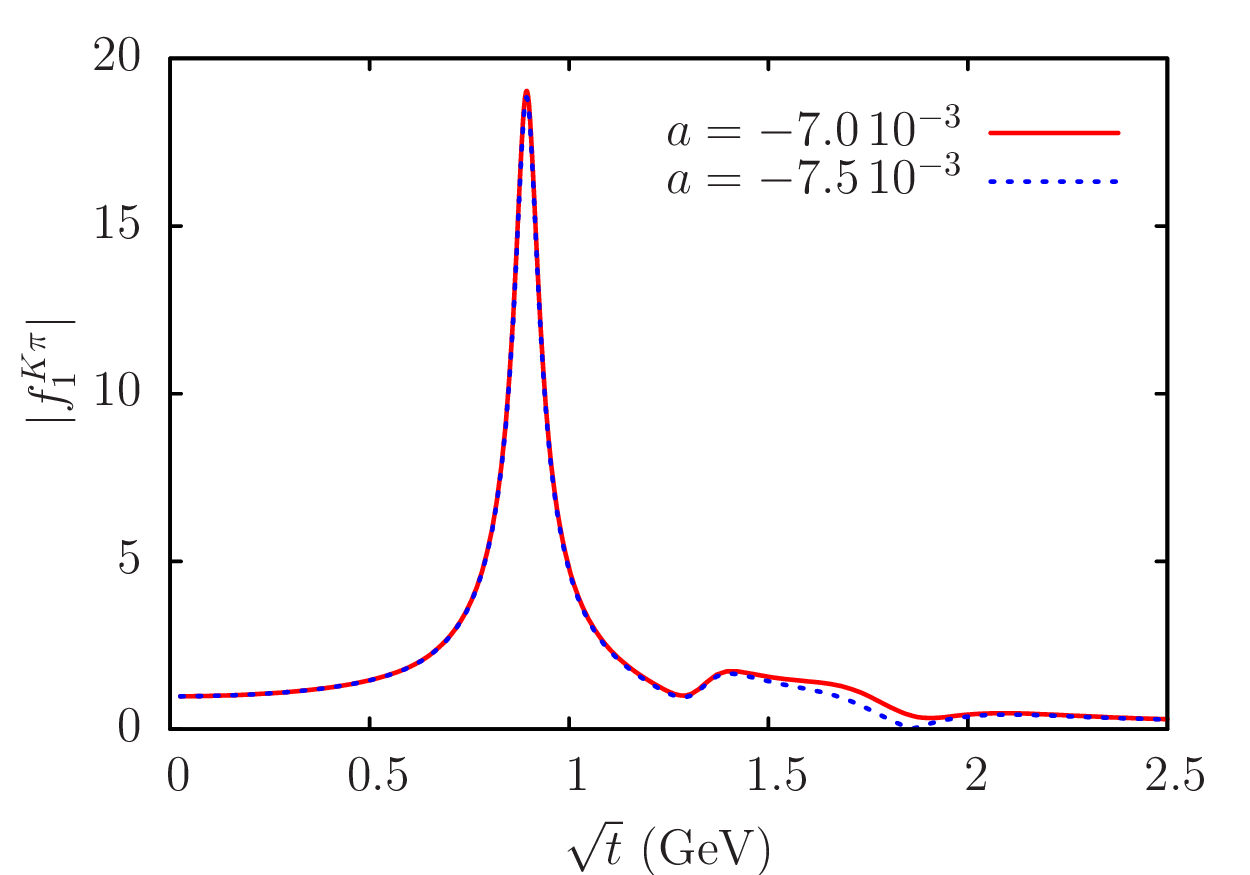}
\includegraphics[width=8.1cm]{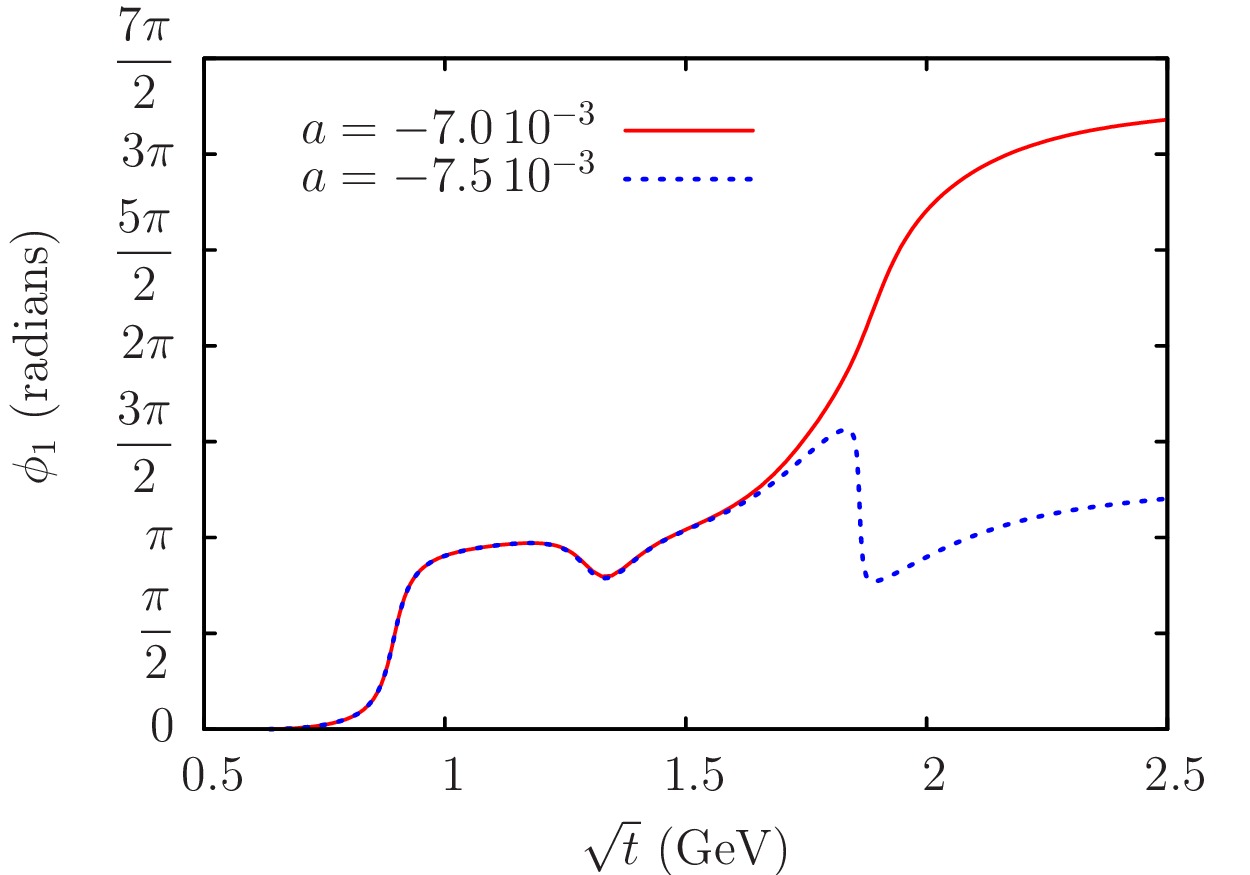}\
\caption{Modulus (left panel) and phase (right panel) of the strange vector
form factor $f_1^{K^-\pi^+}(t)$ obtained by solving a three-channel  Muskhelishvili-Omn\`es equations system. 
Dependence on the symmetry breaking parameter $a$ is illustrated [see
Eq.~\rf{H0a}].
 \label{vectorff}}
\end{figure} 

For $N=4$, in order to solve the system of integral equations we must impose 
four conditions
on the form factors. We use the three values of $H_i$ at $t=0$ and, as
the  fourth constraint, we implement an asymptotic condition for $t \to -\infty$
on the form factor $H_1$. According to Refs.~\cite{brodskylepage,duncanmueller} 
and ignoring flavor symmetry breaking, one should have,
\begin{equation}
\lbl{asyqcd}
\left.H_1(-q^2)\right\vert_{q^2\to\infty} 
\sim{16\pi\sqrt2\, \alpha_s(q^2)  f_\pi^2
\over q^2 }.
\end{equation}
We do not attempt to reproduce the logarithmic running of $\alpha_s$ 
and actually implement Eq.~\rf{asyqcd} with the constant value $\alpha_s=0.2$.
At $t=0$,  one has $H_1(0)=F_1(0)$ and we will use the value given in 
Eq.~(\ref{scalconds}). The values of $H_2(0)$ and 
$H_3(0)$ are not known as precisely.  In the $SU(3)$ chiral limit, flavor
symmetry leads to the following relation between the charged current matrix
element and the electromagnetic one, $j^\mu_{EM}$
\begin{equation}
\braque{ K^{*\,+} \vert \bar{u}\,\gamma^\mu\, s\vert \pi^0}=
{3\sqrt{2}\over2} \braque{ \rho^+\vert j^\mu_{EM} \vert \pi^+},
\end{equation}
which allows one to relate $H_2(0)$ and $H_3(0)$ to the radiative decay width
of the $\rho^+$, yielding
\begin{equation}
H_2(0)=-H_3(0)= (1.54\pm0.08)\ {\rm GeV^{-1}}.
\end{equation}
The relative sign is determined by  vector meson dominance arguments.
We have studied the influence of flavor symmetry breaking to first order
in the quark masses.
There are three independent symmetry breaking parameters and two of them
can be determined from experiment (see~\cite{MoussallampiKSandPwave}). As a consequence, one can express $H_2$ and $H_3$ in terms of the third, unknown parameter $a$, as follows:
\begin{equation}
\lbl{H0a}
H_2(0)= (1.41 \pm 0.09 -65.4\, a)\ {\rm GeV^{-1}},\quad
H_3(0)= (-1.34 \pm 0.07 -65.4\, a)\ {\rm GeV^{-1}}.
\end{equation}
The magnitude of $a$ is expected to be a few times $10^{-3}$. 
We have estimated  $a$ from the sum $R$ of the decay rates of the $\tau^-$ into 
$K^- \pi^0$ and $ \bar K^0 \pi^-$
$R(\tau\to K \pi \nu_\tau) =(13.5\pm 0.5)\times 10^{-3}$~\cite{pdg08}      
which gives $a= (-7.0^{+0.7}_{-2.0})\times 10^{-3}$. 
Results for
the vector form factor are displayed in Fig.~\ref{vectorff}. Its phase shows a
sharp transition as a function of the parameter $a$
close to $a=-7 \times 10^{-3}$. It goes from a regime where its value is 
$3\pi$ at infinity to one where it is $\pi$ displaying a sharp drop.
 The $K^*(1680)$
resonance appears to be suppressed but the properties of the form factor
in this energy region depend significantly on the $S$-matrix  
interpolation parameters
in the region $m_{K\pi} \ge 2.4$ GeV. In our application to $B$ decays
we will use the form factor in the region $\sqrt{t} \lapprox 1.8$ GeV,
where the sensitivity to the asymptotic interpolation is small. 

\subsubsection{\boldmath $K^*(892)~ part~ of~ the~ vector~ form~ factor$}
\label{polevec}

As before, the starting point is the discontinuity equation satisfied by
$ f_1^{K\pi}(t)$ across the elastic unitarity cut, which reads
\begin{equation}
 f_1^{K\pi}(t+i\epsilon)-  f_1^{K\pi}(t-i\epsilon)= 
-2\sigma_{K \pi}(t+i\epsilon) q^2_{K \pi}(t+i\epsilon)
T_{11}^P(t+i\epsilon)  f_1^{K\pi}(t-i\epsilon), 
\end{equation}
and from this we deduce the expression of $ f_1^{K\pi}(t)$ on the second
Riemann sheet
\begin{equation}
\lbl{fplusII}
 f_1^{II}(t)= { f_1^{K\pi}(t)\over 1-2 \sigma_{K \pi}(t) q^2_{K \pi}(t)
T_{11}^P(t)}.
\end{equation}
The position of a resonance in the complex plane, $t_1^{pole}$,   
corresponds to a zero of the denominator function in Eq.~\rf{fplusII}
and this allows one to isolate a pole in $ f_1^{II}(t)$,
\begin{equation}
\label{fpluspole}
 f_1^{pole}(t)= { f_1^{K\pi}(t_1^{pole})\over\beta\,(t-t_1^{pole})}.
\end{equation}
In the case of the $K^*(892)$ we obtain, based on our fit, the following
values for the pole parameters
\begin{eqnarray}
&& t_1^{pole} = (0.7982-i\ 0.0504)\ {\rm GeV}^2,\qquad \sqrt{t_1^{pole}}= (0.8939-i\  0.0282)\ {\rm GeV},
\nonumber\\
&& \beta= (-1.8874+i\ 9.5726)\ {\rm GeV}^2, \qquad  f_1^{K\pi}(t_1^{pole})= 0.8244-i\ 9.0784.
\label{t1fplus}
\end{eqnarray}
\begin{figure} [ht]
\begin{center}
\includegraphics[width=10cm]{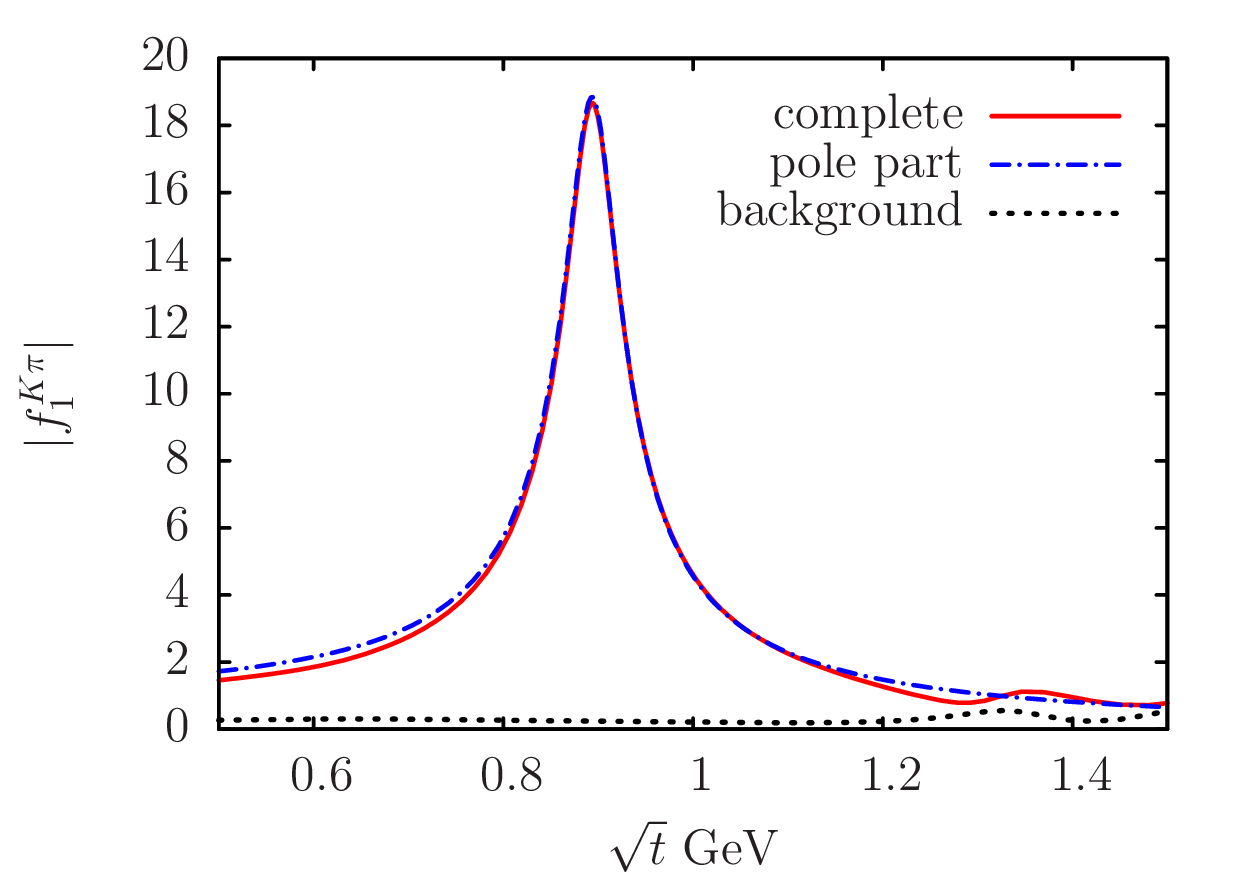}
\caption{
Modulus of the vector form factor $ f_1^{K^-\pi^+}(t)$ compared
with the $K^*(892)$ complex pole component 
\label{ffPpole}
}
\end{center}
\end{figure}
The $K^*(892)$ resonance being very narrow, we expect the pole component
to strongly dominate the form factor below one GeV. This is illustrated in 
Fig.~\ref{ffPpole}
which shows that, indeed, the background component is
very small in that case.
As before we can define the $B$ decay amplitude $B\to K^*(892)\pi$ from
the three-body amplitude $B\to K\pi\pi$ by substituting $ f_1^{K\pi}(t)$
by $f^{pole}_1(t)$ in the relevant formulas, for example in Eqs.~(\ref{K-pi+Pampli}) and (\ref{K0Spi-Pampli}). 

\section{Fit, results and discussion \label{results}}

\subsection{Fitting procedure}
\label{SubSect:FitProc}

In this paper we use a fitting procedure similar to that described in
Ref.~\cite{El-Bennich2006}. 
We perform a $\chi^2$ fit on experimental data from Belle~\cite{Garmash2007,Garmash:2005rv,Abe0509047,Abe2005} and BaBar~\cite{AubertPRD73,AubertLP2007,Aubert:2008bj,Aubert:2007bs} collaborations.
 We use six  $m_{K\pi}$ and five $\cos \theta_H$   distributions where the background is subtracted. 
These data are extracted from the figures of the first seven papers just cited.
We also exploit the four branching fractions for the \Kvec$\pi$, 
the three $CP$ asymmetries for \Kvec$\pi$ and the three $CP$ asymmetries for \Kscal$\pi$ calculated by experimentalists in their data 
analyzes~\cite{Garmash:2005rv,Garmash2007,Aubert:2008bj,Aubert:2007bs}.
Branching fractions are necessary to determine the absolute size of  decay amplitudes.
However, the branching fractions for the ${B} \to K^*_0(1430)\pi$ are not well determined due to the large width of the $K^*_0(1430)$ resonance.
Therefore we use here only the well measured branching fractions for the ${B} \to K^*(892)\pi$, the $K^*(892)$ being a narrow resonance. 
The phases of the decay amplitudes can be constrained by the phase difference, $\Delta \Phi_0$, between the decay amplitudes of $B^0 \to K^{*+}(892) \pi^-$ and $\bar{B}^0 \to K^{*-}(892) \pi^+$. Here we use the preliminary result of Ref.~\cite{AubertLP2007}.
The total $\chi^2$ reads
\begin{eqnarray}
\chi^2_{tot} = \chi^2_{m_{K\pi}} + \chi^2_{\cos \theta_H} + w\left (\chi^2_{BR} + \chi^2_{A_{CP}}+\chi^2_{\Delta \Phi_0}\right),
\end{eqnarray} 
where the coefficient $w$ is introduced in order to increase the weight of  the  branching fractions, $CP$ asymmetries and the phase difference, 
 which form a significantly smaller  
data set than the \mKpi and $\cos \theta_H$ distributions.
We have verified that varying $w$ between 5 and 20 leads to very similar fits. In this  analysis, to perform our best fit, we choose $w=10$.

The $\chi^2$ for a given distribution with $n$  bins is defined by
\begin{eqnarray}
\chi^2  = \sum_{i=1}^{n} \left[ \dfrac{Y_{exp}(x_i)-Y_{th}(x_i)}{\Delta Y_{exp}(x_i)} \right]^2, 
\end{eqnarray} 
where $Y_{exp}(x_i)$ and $\Delta Y_{exp}(x_i)$   are  the number of experimental events and associated error in each bin $x_i$.
Here $x$ denotes either \mKpi  or $\cos \theta_H$. 
Integration of the differential distributions  $d\mathcal {B}(x)/dx$ 
(see Eqs.~(\ref{dB-}) and (\ref{dB-cos})) over the bin width [$x_{i-1}, x_i$] yields the  
theoretical number of events $Y_{th}(x_i)$,
\begin{eqnarray}
Y_{th}(x_i) = N \int_{x_{i-1}}^{x_{i}} \dfrac{d\mathcal {B}(x)}{dx} dx.
\end{eqnarray} 
Our theoretical distributions  are normalized to the number of experimental 
events in the analyzed range from $x_0$ to $x_n$ with
\begin{equation}
N = \dfrac{\sum\limits_{i=1}^{n} Y_{exp}(x_i)}{\int\limits_{x_{0}}^{x_{n}}\dfrac{d\mathcal {B}(x)}{dx}dx}.
\end{equation} 
Altogether in our fit we include 308 bins in the $m_{K \pi}$ and $\cos \theta_H$ distributions, four branching fractions, six $CP$ asymmetries and one phase difference.

In our minimization, the branching fractions and asymmetries are calculated in limited \mKpi regions. 
For the \Kvec we choose a \mKpi range from 0.82 to 0.97 GeV and for the \Kscal
one from 1.0 to 1.76~GeV. These ranges have been used in 
Refs.~\cite{Abe2005,Abe0509047,Garmash2007} to obtain the helicity angle distributions in the regions where the resonances dominate. 
The experimental  branching fractions, which we use in our fit, are 
calculated in the above \mKpi regions from the models presented in the experimental analyzes (see in particular Refs.~\cite{Abe2005,Garmash2007,Aubert:2008bj}). 
In our analysis we have excluded two  experimental points in the \mKpi distributions and three in the $ \cos \theta_H$   ones. 
As will be seen below these points lie significantly far from the general trend of the data.

\begin{table}[h]
\caption{Phenomenological parameters of the decay amplitudes (see e.g.
 Eqs.~(\ref{K-pi+Sampli}), (\ref{K-pi+Pampli}) (\ref{K0Spi-Sampli})  and (\ref{K0Spi-Pampli})).\label{cparam}}
\begin{ruledtabular}
\begin{tabular}{ccc}
& Real part & Imaginary part \\
$c_4^{u}$ & $-0.402\pm 0.244$ & $-3.641\pm 0.054$ \\
$c_4^c$ & $+0.015\pm 0.003$ & $+0.033\pm 0.004$ \\
$c_6^{u}$ & $-0.051\pm 0.153$ & $-0.161\pm 0.184$ \\
$c_6^{c}$ & $+0.075\pm 0.009$ & $-0.033\pm 0.007$
\end{tabular}
\end{ruledtabular}
\end{table}

\begin{figure}[h!]
\includegraphics*[width=7.5cm]{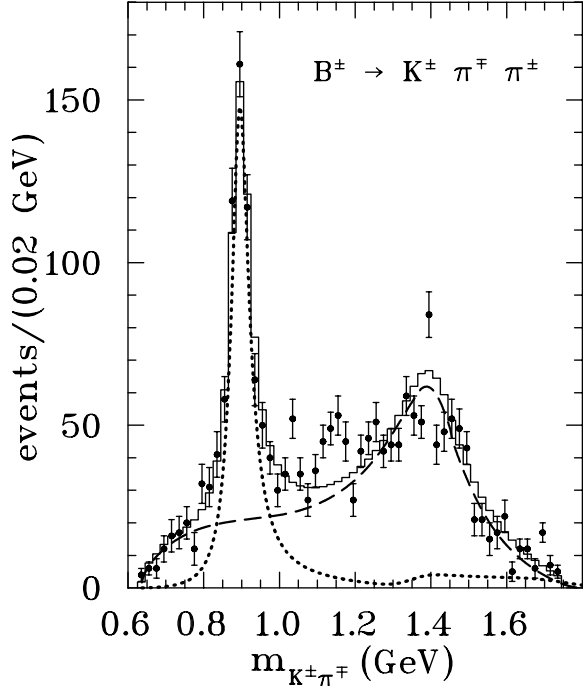}
\caption{
The  $K^\pm \pi^\mp$ effective mass distributions in the  $B^\pm \to K^\pm \pi^\mp \pi^\pm$ decays from the fit to the experimental data as described in Sec.~\ref{SubSect:FitProc}.
 Data points are from  Ref.~\cite{Garmash:2005rv}. The
dashed line represents the $S$-wave contribution of our model, the dotted line that of the $P$-wave and  the  histogram corresponds to the coherent sum of the $S$- and $P$-wave contributions.
\label{fig:MKpiBpmBelle}
}
\end{figure}

\subsection{Results}
\label{SubSec:results}

\begin{figure}[h!]
\includegraphics*[width=7.5cm]{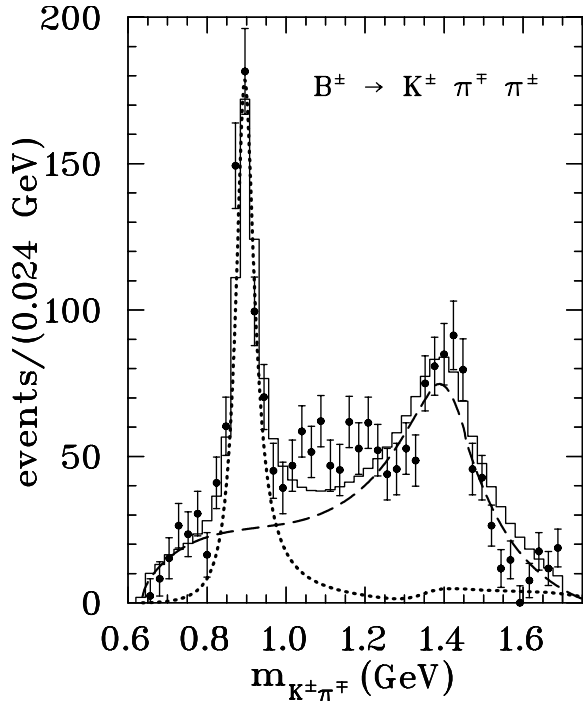}
\caption{
As in Fig.~\ref{fig:MKpiBpmBelle}  but for the data points from  Ref.~\cite{Aubert:2008bj}. \label{fig:MKpiBpmBaBar}
}
\end{figure}

In Table~\ref{cparam} we give the values of  the phenomenological  $ c_{4,6}^{u,c}$ parameters and their errors obtained from our best fit. 
These parameters enter our amplitudes defined in Sec.~\ref{Decay_amplitudes}.
The large values of the $c_{4,6}^u$ coefficients should not be directly compared to those of  $c_{4,6}^c$, since in the amplitudes the latter are multiplied by $\lambda_c$ and the former by $\lambda_u$ with $\vert \lambda_c \vert \simeq 50\vert \lambda_u\vert$. 
The results of our fits, with $\chi^2$= 541 for 308 experimental $K\pi$ effective mass and helicity angle distribution points, and   $\chi^2$=9.3  for ten  experimental branching ratios and asymmetries, are presented in Tables~\ref{tab:branch} to \ref{tab:branch2} and in figures~\ref{fig:MKpiBpmBelle}  to \ref{fig:CosB0Babar}.
For  the phase difference $\Delta \phi_0$ our fit gives $-199 ^\circ\pm 6^\circ$ to be compared with ($-164\pm24\pm12\pm15)^\circ$ found in the experimental analysis of Ref.~\cite{AubertLP2007}.
In the calculation of distributions we take into account all the $\pi\pi$ effective mass cuts around the $D$, $J/\Psi$ and $\Psi(2S)$ meson masses.
As described in
 Refs.~\cite{Abe2005,Garmash:2005rv,AubertPRD73,Abe0509047,Garmash2007,AubertLP2007,Aubert:2008bj}, these cuts are introduced in the experimental analyzes in order to eliminate the decay
 contributions from these resonances.
Our results, shown as histograms, take into account the above cuts.
In all figures presented in this section, the dashed and dotted lines describe the $S$-  and $P$-wave contributions, respectively. In these latter cases the cuts are not taken into account. 
Note that, following the experimental procedure, for the $m_{K \pi}$ plots the
requirement is made that $m_{\pi^+\pi^-}$ is greater than 1.5 GeV for the
Belle data and 2.0 GeV for the Babar results. 
Thus, the contributions unrelated to $K \pi$ rescattering and arising, for example, from $B \to \rho(770) K$ and $B \to f_0(980) K$ decays, are removed from the data samples.
In our model such contributions are omitted.

\subsubsection{\boldmath $The~  K\pi~ effective~ mass~ and~ helicity~ angle~
 distributions~ in~ 
{B^\pm \to K^\pm \pi^\mp \pi^\pm}~ decays$}
\label{SubSec:Bpmdecays}

The  $K^\pm \pi^\mp$ effective mass  distribution for the $B^\pm \to K^\pm \pi^\mp \pi^\pm$  decays, obtained from our fit, 
is compared to the experimental distributions of Belle~\cite{Garmash:2005rv}  and BaBar Collaborations~\cite{Aubert:2008bj} in Figs.~\ref{fig:MKpiBpmBelle} and  \ref{fig:MKpiBpmBaBar}, respectively. The mass distributions are averaged over charge conjugate states.
In both cases, our model describes quite well the \mKpi distributions in
the \Kvec and \Kscal  regions.
It also depicts quite well the sizable enhancement  below 1 GeV related to the  $K_0^*(800)$  state, often called $\kappa$~\cite{pdg08,Descotes06}.
In our amplitude, its contribution is present in the relatively large background found in the modulus of the scalar form factor at low $m_{K\pi}$ (see Eq.~(\ref{f0back}) and Fig.~\ref{ffpole}).
If one  approximates  the quasi-two body $B \to (K\pi)_S\ \pi$ by the two-body  $B \to K^*_0(1430)\ \pi$ amplitudes, then $q^2$ is fixed at the resonance mass  $m^2_{K^*_0(1430)}$ in the third lines of Eqs.~(\ref{K-pi+Sampli}) and (\ref{K0Spi-Sampli}). 
With this replacement, one cannot reproduce the low $m_{K\pi} $ distributions below about 1 GeV, the $q^2$-term contribution becoming much too large.
This justifies the form of our three-body approach, within the QCD factorization framework, to these decays.
The origin of this  $q^2$ term is given below Eq.~(\ref{K-pi+Sampli}).

In the isobar model, used in experimental analyzes, the above $q^2$ dependence is approximated by one fitted constant parameter.
For the description of the wide $K^*_0(1430)$ resonance this may not be a very good approximation since the $q^2$ varies by a factor of eight from the $K\pi$ threshold to the $m_{K\pi}$ limit of about 1.8 GeV, close to the sum of the $m_{K^*_0(1430)}$ mass and its width.
 
\begin{figure}[h!]
\includegraphics*[width=7.5cm]{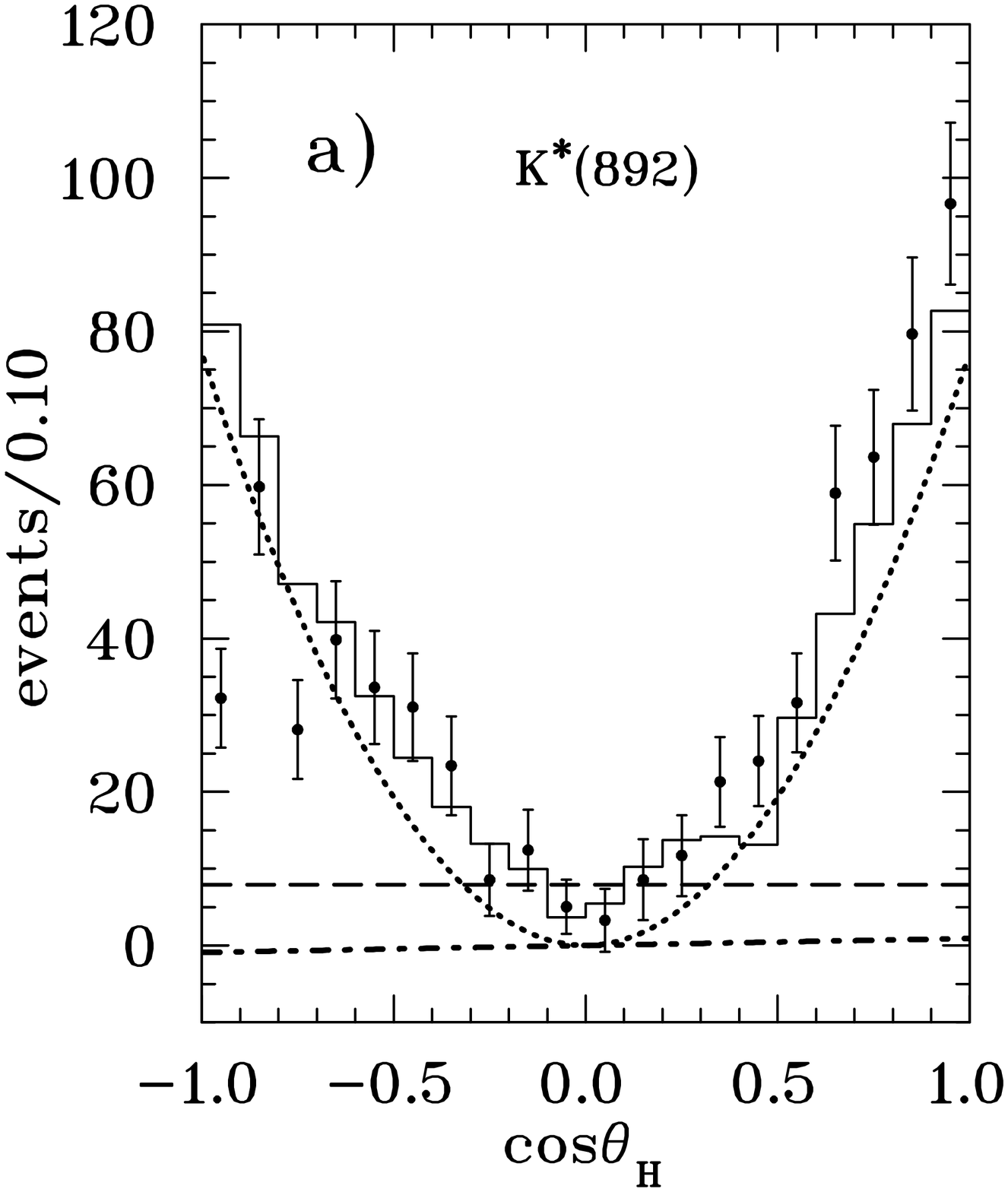}
\hspace{0.4cm}
\includegraphics*[width=7.5cm]{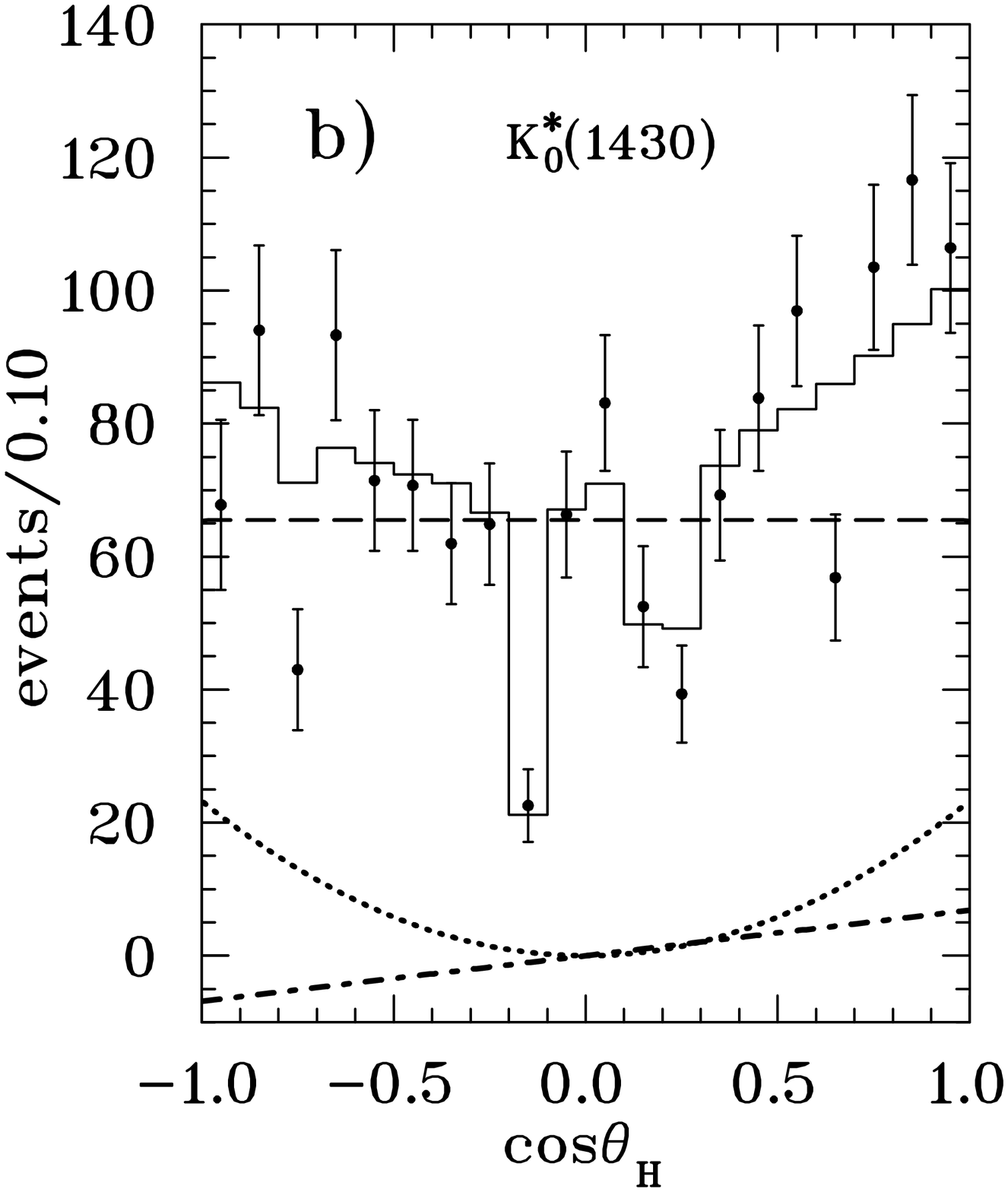}
\caption{
Helicity angle distributions for  $B^\pm \to K^\pm \pi^\mp \pi^\pm$ decays  calculated from the averaged double differential distribution  integrated over $m_{K^\pm \pi^\mp}$ mass from 0.82 to 0.97 GeV in the \Kvec case   a) and  from 1.0 to 1.76 GeV in the \Kscal one b).  
 Data  points are from  Ref.~\cite{Abe2005}. 
Dashed lines represent the $S$-wave contribution of our model, dotted lines that of the $P$-wave, the dot-dashed that of the interference term.
The histograms correspond to the sum of these three contributions.
\label{fig:CosBpmBelle}
}
\end{figure}

The results of the fit to  the  $ \cos \theta_H $ distributions of the Belle Collaboration~\cite{Abe2005} around the \Kvec and \Kscal resonances are shown in 
Figs.~\ref{fig:CosBpmBelle}a  and~\ref{fig:CosBpmBelle}b.
 Here we show the contributions of the $S$ and $P$ waves as their interference given by Eq.~(\ref{eqB}).
 In the $\chi^2$ fit corresponding to Fig.~\ref{fig:CosBpmBelle}a, we have excluded one bin at $\cos \theta_H = -0.95$. 
 This bin is not related to any cut and its  $\chi^2$ value 
is almost twice as large as   the value of   the total $\chi^2$  for this distribution.
In Fig.~\ref{fig:CosBpmBelle}a, the $P$-wave contribution dominates,
 those of the   $S$ wave  and of   the  interference term 
 being rather  small.
On the contrary, and as expected, it can be seen in Fig.~\ref{fig:CosBpmBelle}b that the $S$-wave contribution is much larger than that of the $P$ wave.
Contribution of the interference term leads  to a visible left-right asymmetry.
The minima in the histograms of  Figs.~\ref{fig:CosBpmBelle}b and  \ref{fig:CosB0Belle}b at $\cos {\theta}_H \simeq -0.15$ and $+0.2$ correspond to the cuts related to background events of $J/\Psi$ and $\Psi(2S)$, respectively.
There is also a cut in  Figs.~\ref{fig:CosBpmBelle}a and~\ref{fig:CosBpmBelle}b at  $\cos {\theta}_H \simeq -0.75$ corresponding to the $D$ meson.
\begin{figure}[h!]
\includegraphics*[width=7.5cm]{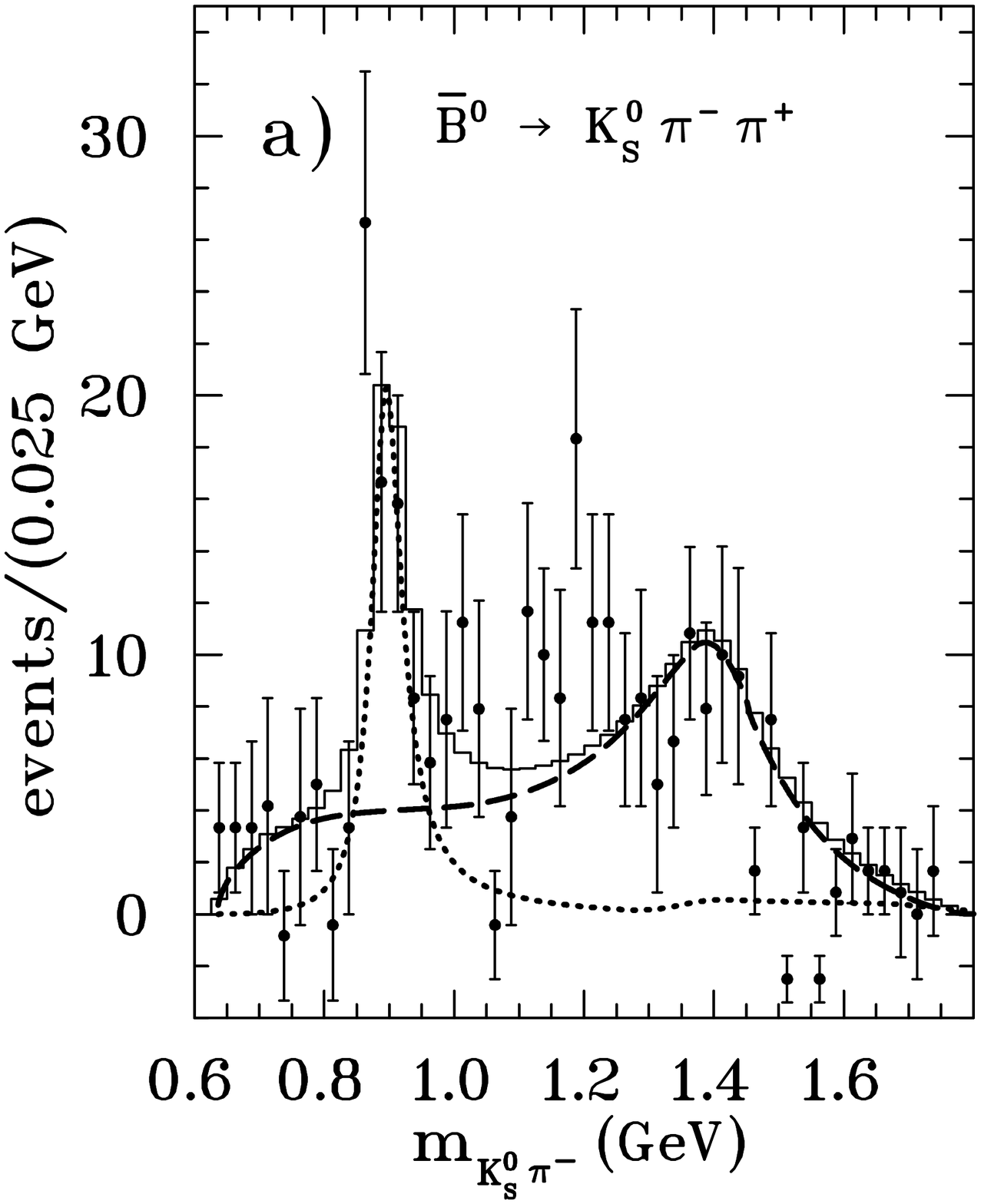}
\hspace{0.4cm}
\includegraphics*[width=7.5cm]{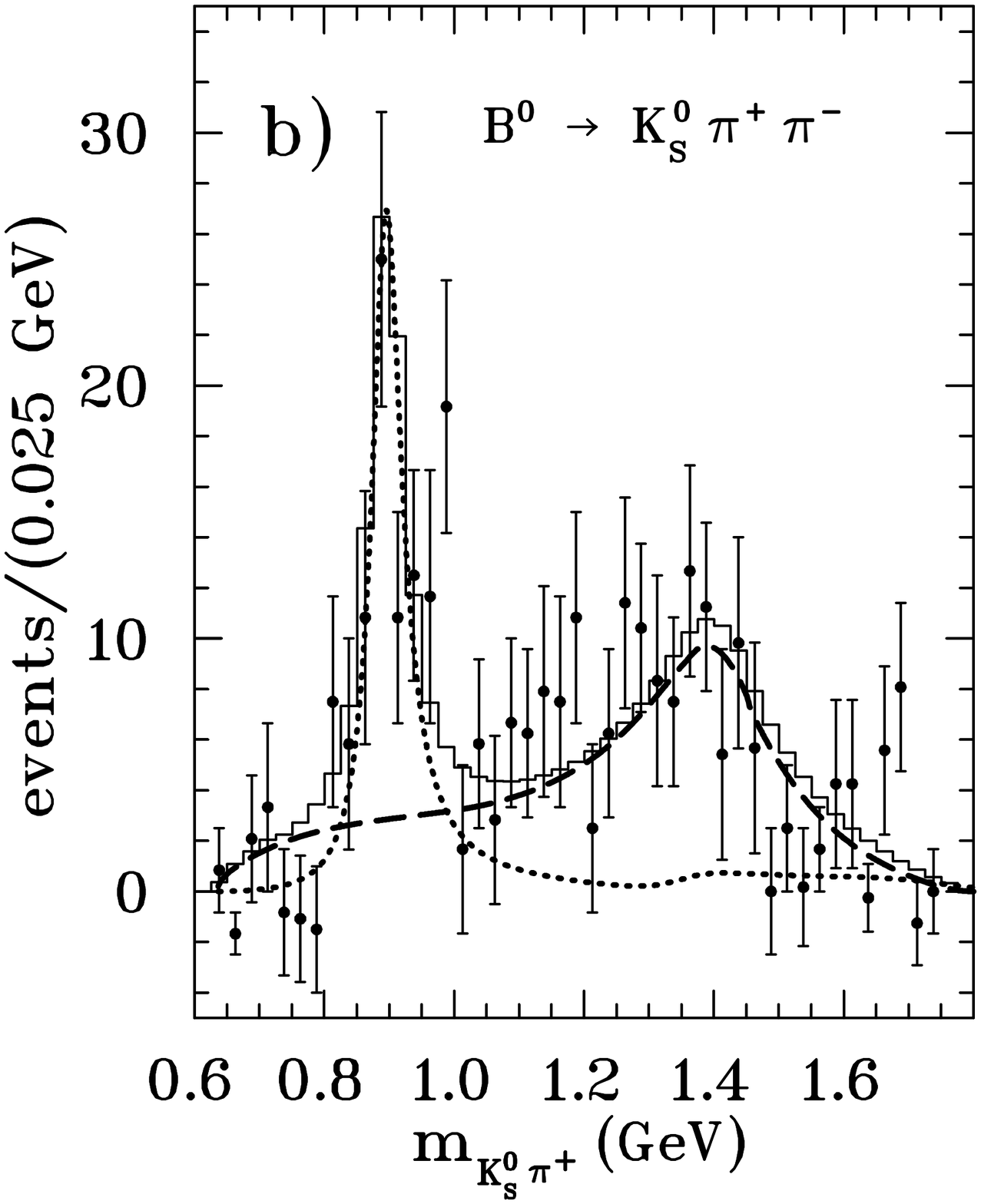}
\caption{
As in Fig.~\ref{fig:MKpiBpmBelle} but for $\bar B^0 \to K^0_S \pi^- \pi^+$ decays a), for $B^0 \to K^0_S \pi^+ \pi^-$ ones  b) and for the data of Ref.~\cite{Garmash2007}.
\label{fig:MKpiB0Belle}
}
\end{figure}

\begin{figure}[h!]
\includegraphics*[width=7.5cm]{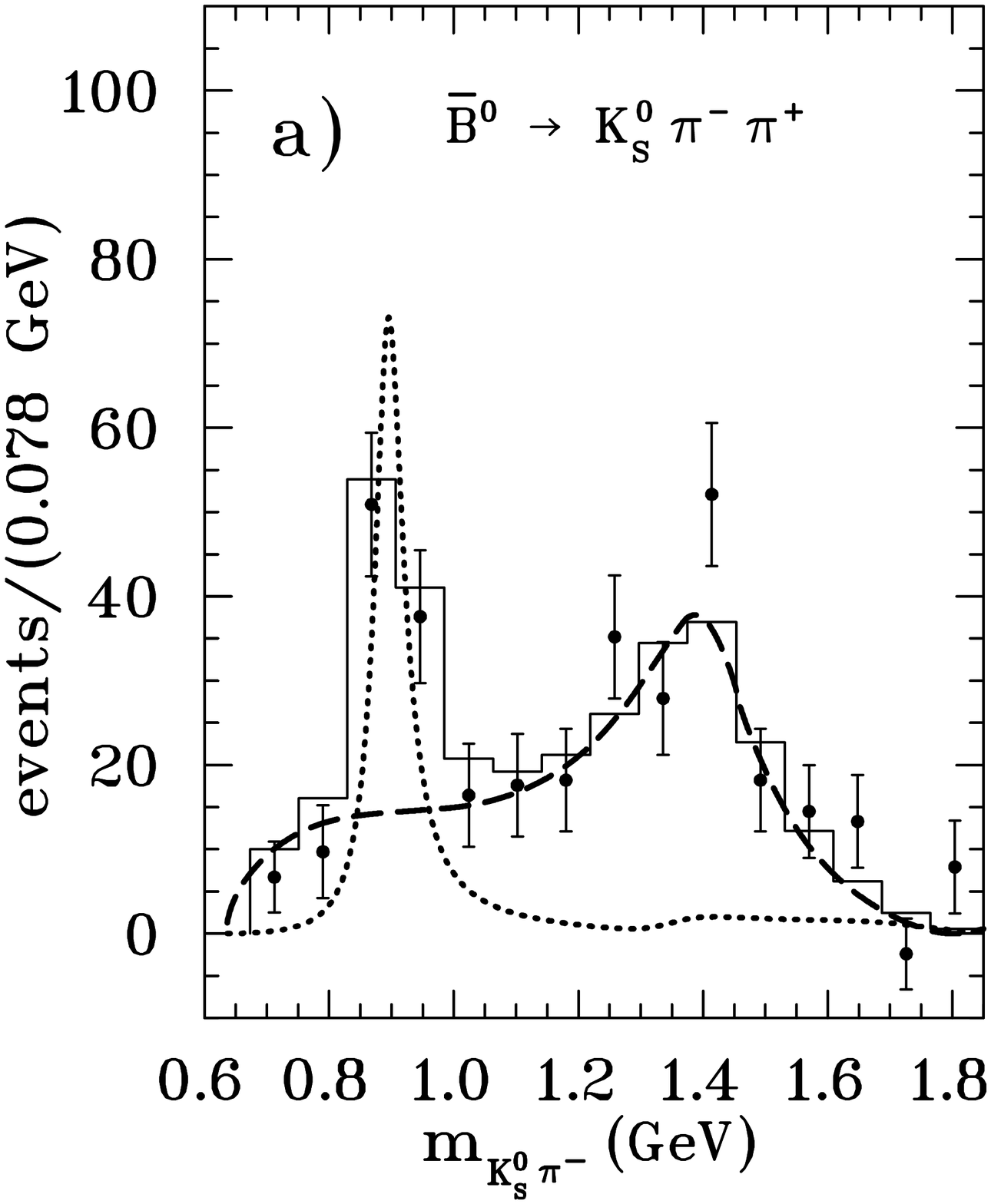}
\hspace{0.4cm}
\includegraphics*[width=7.5cm]{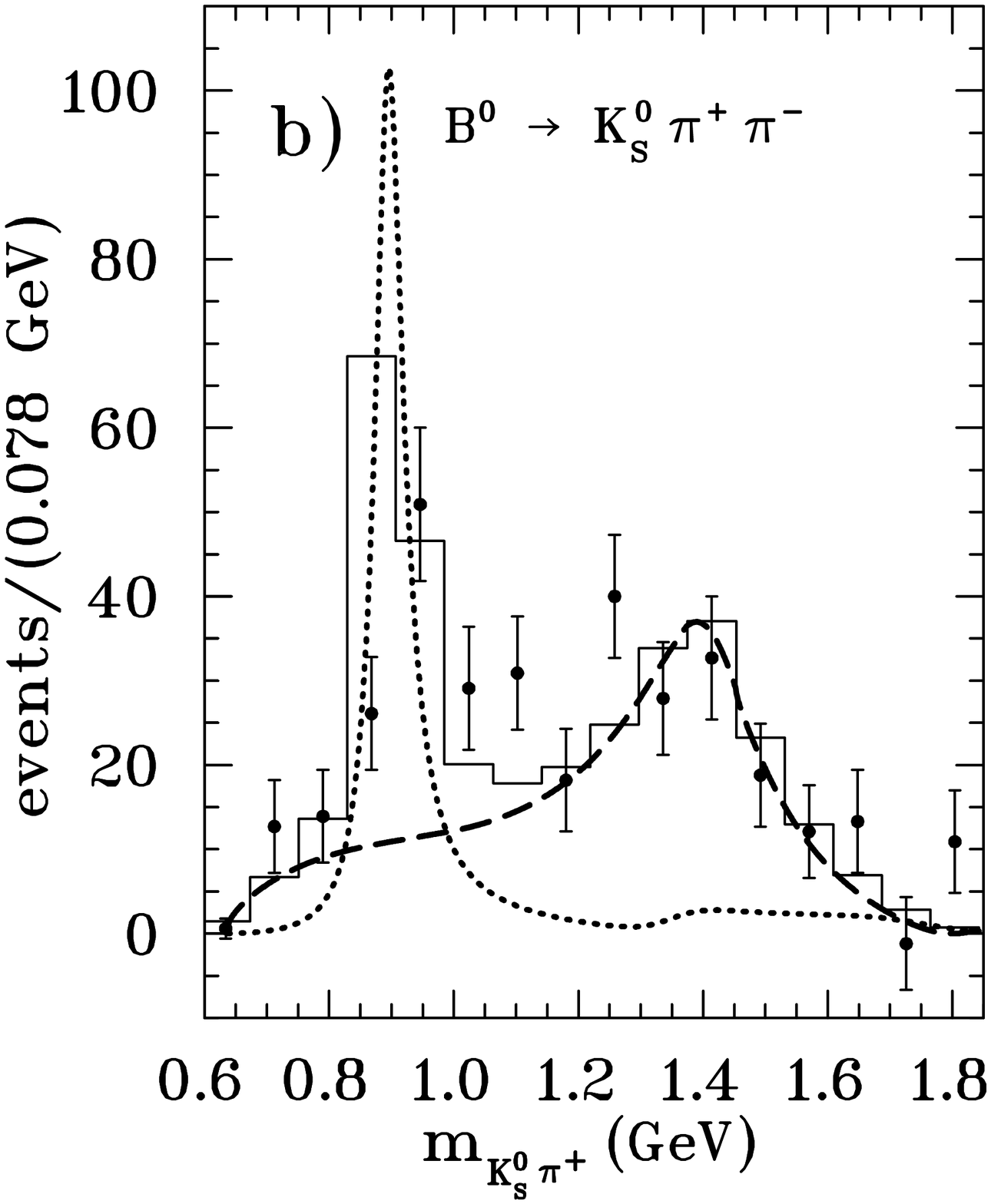}
\caption{
As in Fig.~\ref{fig:MKpiBpmBelle}  but for $\bar B^0 \to K^0_S \pi^- \pi^+$ decays in a) and for $B^0 \to K^0_S \pi^+ \pi^-$ ones in b), and for the data of Ref.~\cite{AubertLP2007}.
\label{fig:MKpiB0BaBar}
}
\end{figure}

\begin{figure}[h!]
\includegraphics*[width=7.7cm,height=9.25cm]{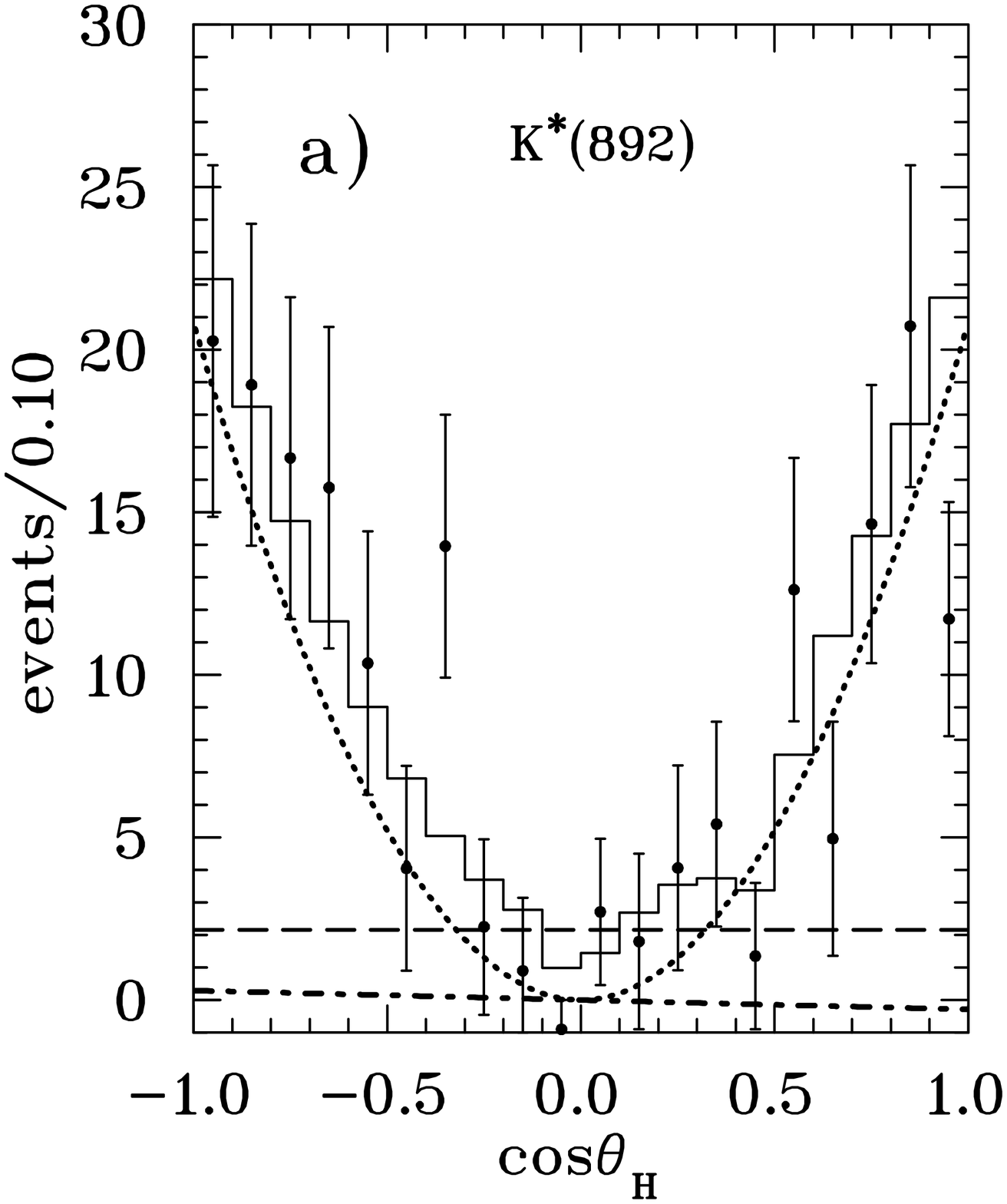}
\hspace{0.4cm}
\includegraphics*[width=7.7cm,height=9.25cm]{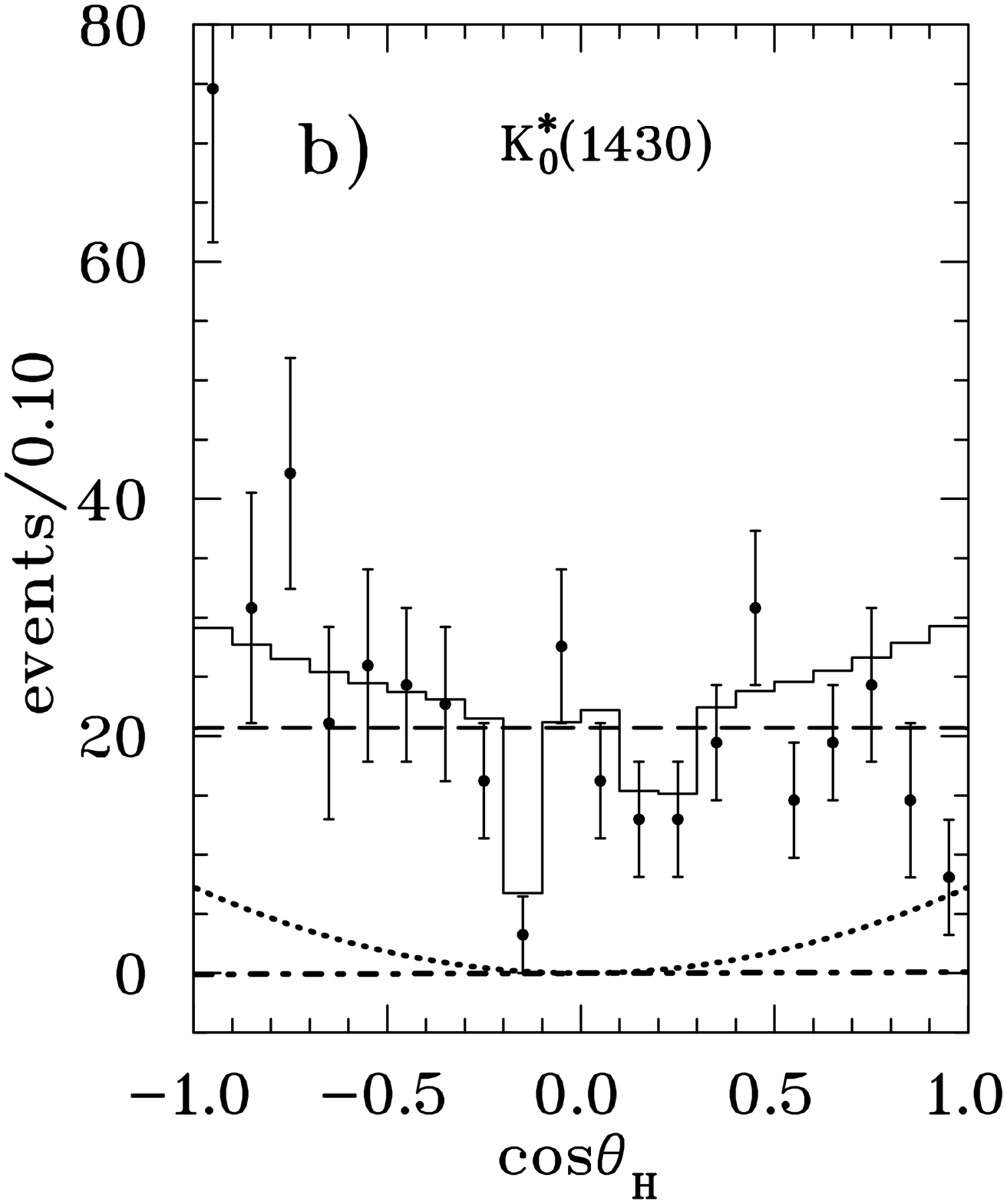}
\caption{As in Fig.~\ref{fig:CosBpmBelle} but for $\bar B^0 \to K^0_S \pi^- \pi^+$ and $B^0 \to K^0_S \pi^+ \pi^-$  averaged distributions, and for the Belle data of Ref.~\cite{Garmash2007} in a) and that of Ref.~\cite{Abe0509047} in b).\label{fig:CosB0Belle}}
\end{figure}

\begin{figure}[h!]
\includegraphics*[width=7.7cm,height=9.25cm]{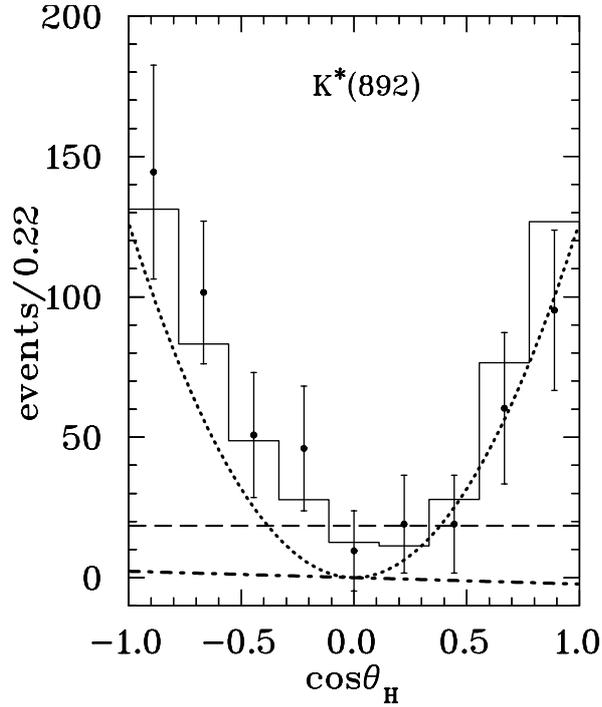}
\caption{As in Fig.~\ref{fig:CosBpmBelle} a) but for $\bar B^0 \to K^0_S \pi^- \pi^+$ and $B^0 \to K^0_S \pi^+ \pi^-$  averaged distributions, and for the BaBar data of Ref.~\cite{AubertPRD73}.\label{fig:CosB0Babar}}
\end{figure}

\subsubsection{\boldmath $The ~K\pi~ effective~ mass~ and~ helicity~ angle~
 distributions~ for~ B^0 \to K^0 \pi^+ \pi^-~   and~$ \\ 
{\boldmath $ \bar B^0 \to \bar K^0 \pi^- \pi^+ ~ decays$}}
\label{SubSec:B0decays}

The results of the fits to the \mKpi distributions for  the $\bar B^0 \to K^0_S \pi^- \pi^+$ and $B^0 \to K^0_S \pi^+ \pi^-$ decays of  the  Belle Collaboration~\cite{Garmash2007} are shown in Figs.~\ref{fig:MKpiB0Belle}a and~\ref{fig:MKpiB0Belle}b, respectively.
  In the $\chi^2$ fit to the $\bar B^0 \to K^0_S \pi^- \pi^+$ distribution, we have excluded the two bins at $m_{K\pi} = 1538$ and 1588~MeV.  
 They have quite small errors and  negative numbers of events. 
 In the  \mKpi range around the $K^*(892)$, the histogram of the model has
  less events for $\bar B^0$ case than for $ B^0$.  
  One then expects a negative $CP$ asymmetry,  
 which is  confirmed by the value given in Table~\ref{tab:CP}. 
 It is in agreement with the result of the experimental analysis.

Our $m_{K\pi}$ distributions for  the same neutral $B$ decays are compared to those of the BaBar Collaboration~\cite{AubertLP2007} in Fig.~\ref{fig:MKpiB0BaBar}. 
As previously, the  \Kvec and \Kscal are well described by our model.
 Here, the width of the bins is larger than that of Belle in 
Fig.~\ref{fig:MKpiB0Belle} which explains why the maximum of 
the $P$-wave contribution is above the experimental points close to the $K^*(892)$ position.
 
The  averaged $\bar B^0 \to K^0_S \pi^- \pi^+$ and $B^0 \to K^0_S \pi^+ \pi^-$ helicity angle
distributions for the \mKpi regions of  the \Kvec and \Kscal  are compared in Fig.~\ref{fig:CosB0Belle} to the Belle data~\cite{Abe0509047,Garmash2007}.  In the $\chi^2$ fit to the distribution shown in Fig.~\ref{fig:CosB0Belle}b we have excluded  two bins at $\cos \theta_H$ equal to $\pm 0.95$.
 These two data  lie  rather outside the general trend of the distribution.
As in the charged $B$ decays, the $P$ wave dominates the $K^*(892)$ region and the $S$ wave the $K^*_0(1430)$ one. In both cases the $S$-$P$ interference is rather small. 

Figure~\ref{fig:CosB0Babar} shows our helicity angle distribution fitted to the BaBar experimental data~\cite{AubertPRD73} which results from the integration of the double differential distribution over \mKpi  from 0.776 to 1.01 GeV.
The contributions of the $S$ and $P$ wave  and of their interference  are similar to those observed in 
Figs.~\ref{fig:CosBpmBelle}a and \ref{fig:CosB0Belle}a.

\subsubsection{\boldmath $Branching~ fractions~ and~ CP~ asymmetries$
 \label{subsub:CP}}

\begin{table}[t]
\caption{
Branching fractions for the $B\to K^*(892)\pi$ decays averaged over charge conjugate reactions  in units of $10^{-6}$ . 
In the second column, giving the experimental branching ratios, the 2/3 factor arises from isospin symmetry.
The values of the model calculated by the integration on $m_{K\pi}$ from 0.82 to 0.97 GeV are compared to the corresponding Belle and BaBar results given in the fourth column.
Model errors stem from the
phenomenological parameter uncertainties obtained through the minimization procedure.
The last column corresponds to the model without phenomenological parameters.
\label{tab:branch}
 }
\begin{ruledtabular}
\begin{tabular}{cccccc}
Decay mode & $\mathcal{B} ^{\mbox{exp}}$ & Ref. &  $\mathcal{B}^{\mbox{exp}}(0.82,0.97)$ & model & model [$c_i^p\equiv 0$] \\
\hline
$B^-\to [\bar K^{*0}(892)\to K^-\pi^+]\ \pi^-$ & $6.45\pm 0.71$ & \cite{Garmash:2005rv} & $5.35\pm 0.59$ & $5.73\pm 0.14$ &1.42 \\
                       & $7.20\pm 0.90$ & \cite{Aubert:2008bj} & $5.98\pm 0.75$ & & \\
$\bar B^0\to[\bar K^{*-}(892)\to\bar K^0\pi^-]\ \pi^+$ & $5.60\pm 0.93$ & \cite{Garmash2007} & $4.65\pm 0.77$ &$5.42\pm 0.16$ & 1.09 \\
                       & $\dfrac{2}{3}(11.7\pm 1.30)$ & \cite{Aubert:2007bs} & $6.47\pm 0.72$ & & 
\end{tabular}
\end{ruledtabular}
\end{table}

In Table~\ref{tab:branch}, our branching fractions for the $B\to K^*(892)\pi$ decays are compared to the corresponding experimental  Belle and BaBar values.
As already mentioned in Sec.~\ref{SubSect:FitProc} these are obtained from integration on $m_{K\pi}$ from 0.82 to 0.97 GeV of the double differential branching fractions of our model and those of the experimental analyzes.
The theoretical errors are calculated using covariance matrix elements and the corresponding derivatives of the branching fractions over all fitted parameters.
We did not include other uncertainties entering our amplitudes, so our  theoretical errors are underestimated.
Our model branching fractions for $B \to K^*(892) \pi$ decays agree   quite well  with  the corresponding experimental ones within their errors.
Using the pole part of our $P$-wave form factor (see Eqs.~(\ref{fpluspole}) and (\ref{t1fplus})) and integrating over the full range, we obtain for the charged  averaged  branching fraction $6.95\times 10^{-6}$.
This value compares quite well with those of Belle and BaBar given in Table~\ref{tab:branch}.
Let us stress however, that when the phenomenological parameters $c_i^p\ (i=4,6$ and $p=u,c$), are set to zero, the theoretical branching fractions are too small by a factor 4 or 5 (see the last column of Table~\ref{tab:branch}). 
This indicates that QCDF $P$-wave amplitudes are too small by a factor of about 2.

\begin{figure}[h!]
\includegraphics*[width=7.5cm]{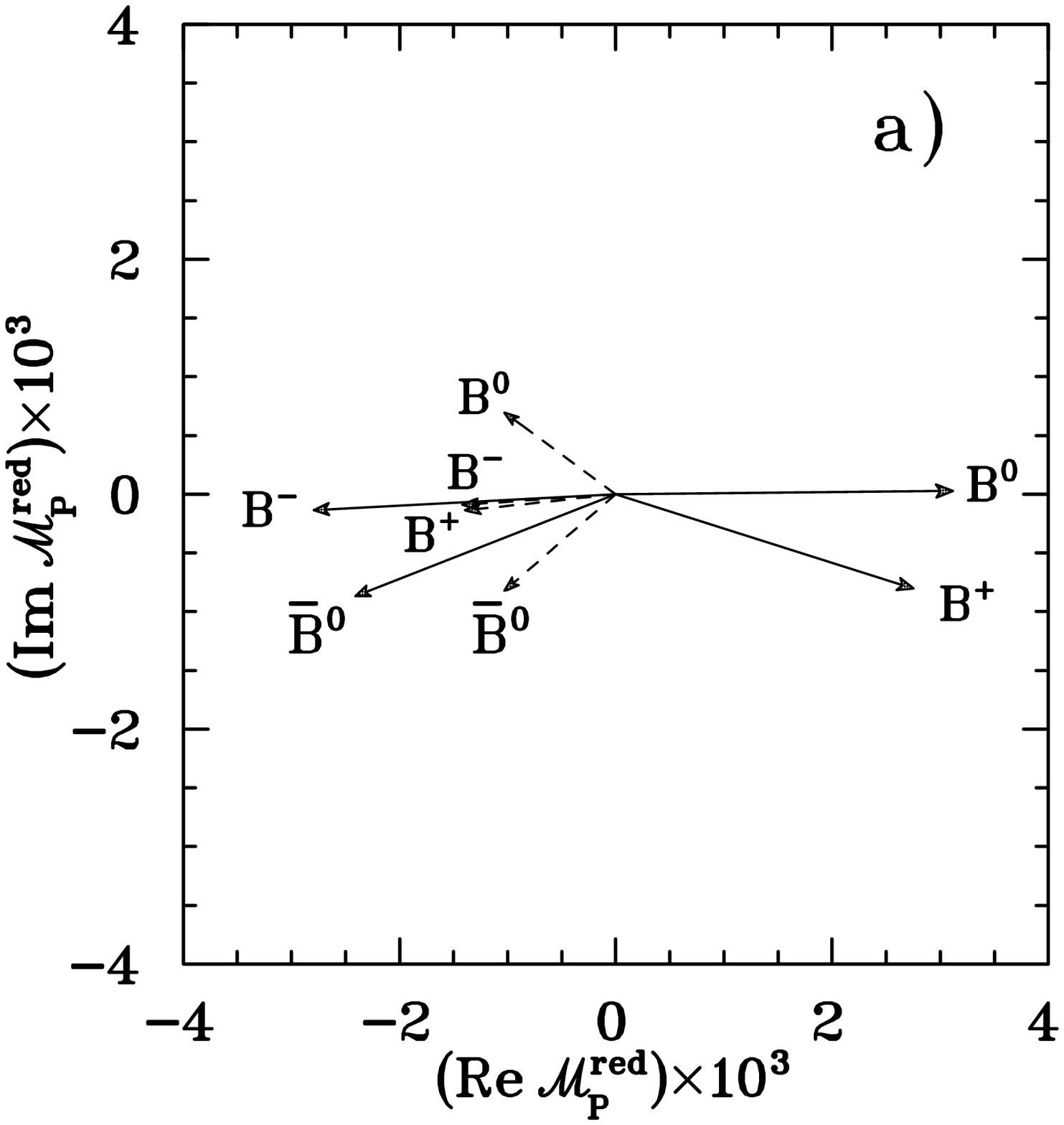}
\hspace{0.4cm}
\includegraphics*[width=7.5cm]{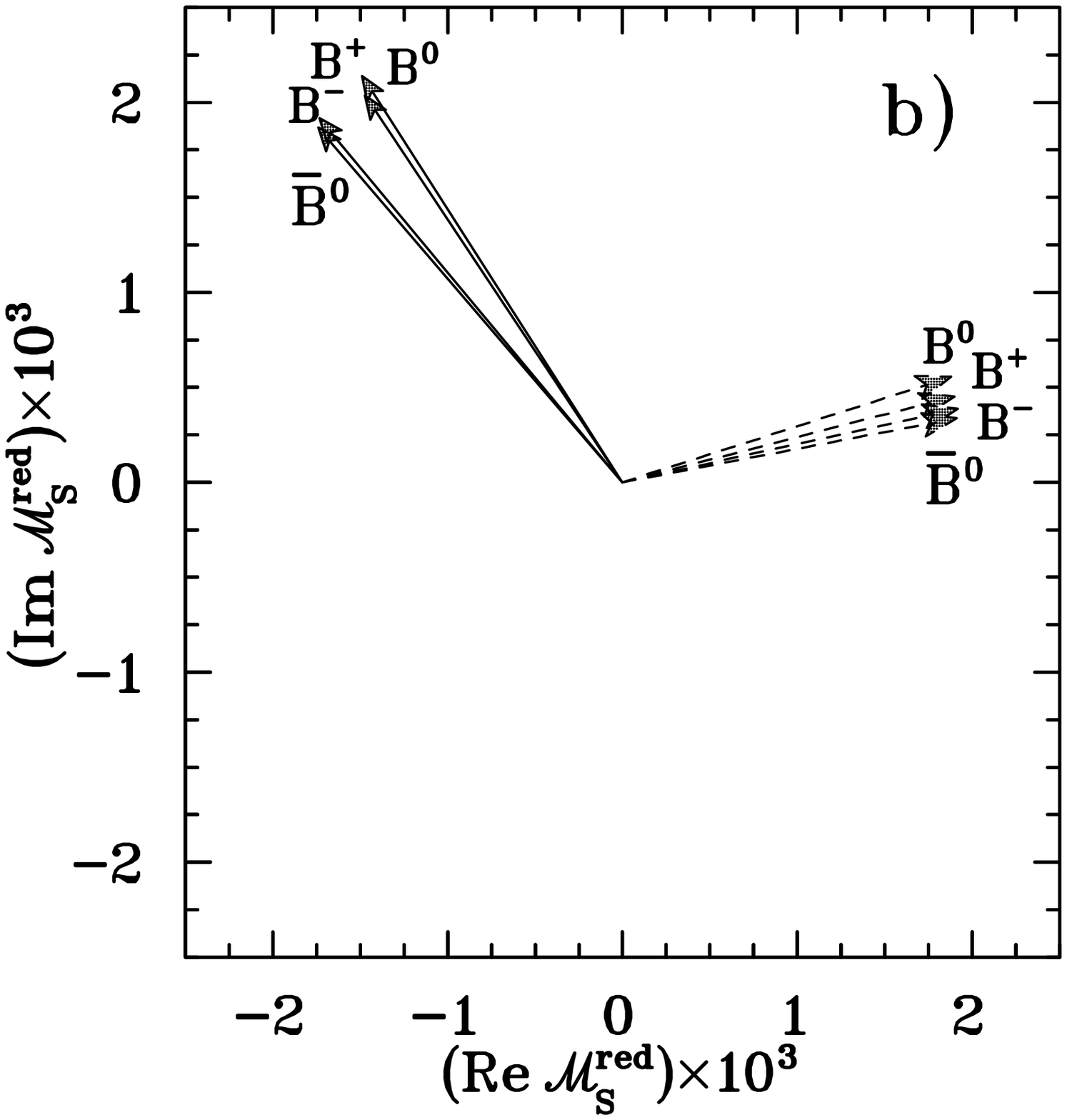}
\caption{
Complex plane representation of the parts of the amplitudes depending on the CKM matrix elements, on the effective QCD coefficients $a_i^p$ and on the fitted penguin parameters $c_i^p$.
a) $P$ wave, b)  $S$ wave at $m_{K\pi}$= 1.414 GeV ($K^*_0(1430)$ mass). 
Dashed lines: amplitudes with $c_i^p$ =0, solid lines: results of the fit. 
See text for the exact definition of these reduced amplitudes.
\label{ampPS}
}
\end{figure}

In Fig.~\ref{ampPS}a we present the reduced complex $P$-wave amplitudes, $\mathcal{M}_P^{red}$, which are  given by the expression between the curly brackets in Eqs.~(\ref{K-pi+Pampli}) and (\ref{K0Spi-Pampli}) for $B^-$ and $\bar B^0$ decays, respectively.
The reduced $S$-wave amplitudes, $\mathcal{M}_S^{red}$, for $B^-$ and $\bar B^0$ decays, shown in Fig.~\ref{ampPS}b, are defined similarly from Eqs.~(\ref{K-pi+Sampli}) and (\ref{K0Spi-Sampli}). 
The corresponding $B^+$ and $B^0$ amplitudes can then be obtained through the conjugation $\lambda_u\to\lambda_u^*,\ \lambda_c\to\lambda_c^*$.
The dashed arrows correspond to the reduced amplitudes before the fit ($c_i^p =0$) while the solid ones to the result of the fit.
In Fig.~\ref{ampPS}a,  the  $B^-$ and $B^+$ reduced amplitudes  without phenomenological parameters are degenerate due to the dominance of the almost real  $\lambda_c$ term over the  $\lambda_u$ one.
Moreover, the  fact that the $\bar B^0$ and $B^0$ reduced amplitudes for $c_i^p=0$ have almost opposite imaginary parts, comes from the presence of the tree term $\lambda_u\ a_1$ with a large real part of the $a_1$ coefficient close to 1 (see Table~\ref{aipmb}).

\begin{table}[h]
\caption{
Direct $CP$ asymmetries averaged over charge conjugate reactions. 
The values of the model, calculated over the $m_{K\pi}$ range from 0.82 to 0.97 GeV for the $K\pi$ $P$-wave and from 1.0 to 1.76 GeV for the $S$-wave, are compared to the Belle and BaBar results.
Concerning the errors of the model and the last column, see the caption in Table~\ref{tab:branch}.\label{tab:CP}
 }
\begin{ruledtabular}
\begin{tabular}{lcccc}
Decay mode & exp. (\%) & Ref. & model (\%) & model (\%) [$c_i^p\equiv 0$] \\
\hline
$B^-\to [\bar K^{*0}(892)\to K^-\pi^+]\ \pi^-$ & $-14.9\pm6.8$ & \cite{Garmash:2005rv} & $-2.5\pm 1.3$ & 1.4 \\
                          & $3.2\pm 5.4$ & \cite{Aubert:2008bj} & & \\
$B^-\to [\bar K^*_0(1430) \to  K^- \pi^+]\ \pi^-$ & $7.6\pm 4.6$ & \cite{Garmash:2005rv} &  &  \\
  $B^-\to (K^-\pi^+)_S\ \pi^-$  & $3.2\pm 4.6$ & \cite{Aubert:2008bj} & 5.4$\pm$1.0&0.2 \\
$\bar B^0\to  [\bar K^{*0}(892)\to \bar K^0\pi^-]\ \pi^+$ & $-14\pm 12$ & \cite{Aubert:2007bs} & $-19.6\pm 3.0$ & 6.1 \\
$\bar B^0\to (\bar K^0\pi^-)_S\ \pi^+$ & $17\pm 26$ & \cite{Aubert:2007bs} &
$-0.2\pm 1.3 $ & $-1.7$
\end{tabular}
\end{ruledtabular}
\end{table}

Our direct  $CP$ violating asymmetries are compared with the experimental  ones  in Table \ref{tab:CP}.
Their errors are calculated in the same way as for the branching fractions.
For $B^- \to \bar K^{*0}(892)\pi^-$  decays our asymmetries lie between those of Belle and BaBar. 
The results for $B^-\to (K^-\pi^+)_S\ \pi^-$  and for $\bar B^0\to (\bar K^0\pi^-)_S\ \pi^+$ decays agree with the experimental values of both collaborations.
For the $S$-wave, the variation with the range of integration is within the
experimental errors.
For instance, if one calculates the asymmetries over the $m_{K\pi}$ range from
threshold (0.64 GeV) to 1.76~ GeV, our result for the charged B decays varies
from (5.4$\pm$1.0)\% to (9.8$\pm$1.1)\% and for the neutral ones from (-0.2$\pm$1.3)\% to (2.7$\pm$1.3)\% .
There is no variation for the $P$ wave.

\subsection{Discussion\label{discussion}}

In the case of the $K^*_0(1430)$ resonance, it is difficult to 
extract the quasi-two body branching fraction from experimental 
data due to the presence of a significant background 
which can mainly be attributed  to the broad $K^*_0(800)$ resonance.
Contrary to the $P$-wave amplitude, entirely dominated by the 
$K^*(892)$ resonance below 1 GeV, the $S$ wave is more complex.
This is exemplified in Fig.~\ref{ffpole} by the comparison of the pole part of the scalar form factor to the complete form factor.
We remind that our $S$-wave amplitudes are proportional to this 
form factor as readily seen, for instance, in Eqs.~(\ref{K-pi+Sampli}) 
and~(\ref{K0Spi-Sampli}).
This $S$-wave complexity results in different parameterizations 
in Belle and BaBar analyzes.
The Belle group uses a Breit-Wigner amplitude to represent the $K\pi$ $S$-wave 
interaction. 
They have furthermore a large contribution from a nonresonant part.
The BaBar Collaboration has introduced a term proportional to the $K\pi$ $S$-wave
$T$-matrix, and used the LASS parametrization of the latter. 
It consists of an effective range nonresonant component plus a 
$K^*_0(1430)$ Breit-Wigner term. 
Since this parametrization is fitted
to experimental $K\pi$ scattering data (in the range $0.8 \leqslant m_{K\pi} \leqslant 1.53$ GeV), this method provides an improved 
treatment of the final state interaction as compared to the 
Belle parametrization (see e.g. Ref.~\cite{Dunwoodie06}). 
However, factorization implies that the $B$ decay 
amplitude should involve the scalar form factor rather than simply the $T$-matrix.
Note that the $T$-matrix and the associated form factor have the same
phase in the elastic region ($m_{K\pi}\lapprox 1.45$ GeV), 
but not the same modulus. 
Above the $K\pi$ elastic region both the phase and
the modulus are different. 

Our model, based on factorization, allows us to calculate, in an unambiguous way, 
the branching fractions making use of either the complete $K\pi$ 
$S$-wave contribution or of the  $K^*_0(1430)$ resonance only, 
described here as the pole position
of the scalar form factor on the second Riemann sheet 
(cf. Sec.~\ref{polescal}).
\begin{table}
\caption{
Branching fractions averaged over charge conjugate reactions $B\to (K\pi)_S\pi$  in units of $10^{-6}$.
The second column gives the experimental results.
The predictions of our model, calculated by the integration of the  $m_{K\pi}$  distribution
over $m_{K\pi}$ from threshold  (0.64 GeV) to 1.76 GeV, 
are compared to the corresponding Belle and BaBar results given in the fourth column. 
In the first two lines, the Belle branching fractions~\cite{Garmash:2005rv,Garmash2007}, calculated with a Breit-Wigner amplitude, are compared to our predictions obtained from the $K^*_0(1430)$ pole part of the scalar form factor (see Sec.~\ref{polescal}). 
In the last two lines we show the BaBar branching 
fractions~\cite{Aubert:2008bj,Aubert:2007bs} for $B\to (K\pi)_S\ \pi$ calculated, in  their parametrization, 
with the part of the decay amplitude 
proportional to the $K\pi$ $S$-wave $T$-matrix. 
This is compared to the results of our model, where  the $B\to (K\pi)_S\pi$ amplitude
corresponds to the part proportional to the 
scalar form factor (see Sec.~\ref{scalarformfactor}).
See caption of Table~\ref{tab:branch} for the factor of 2/3 in the first column, for the errors of the model and for the last column. 
\label{tab:branch2}
} 
\begin{ruledtabular}
\begin{tabular}{lccccc}
Decay mode & $\mathcal{B} ^{\mbox{exp}}$ & Ref. &  $\mathcal{B}^{\mbox{exp}}(0.64,1.76)$ & model & model [$c^p_i\equiv 0$] \\
\hline
$B^-\to [\bar K^{*0}_0(1430)\to K^-\pi^+]\ \pi^-$ & $32.0\pm 3.0$ & \cite{Garmash:2005rv} & $27.0\pm 2.5$ & $11.6\pm 0.6$ &6.1 \\
$\bar B^0\to[\bar K^{*-}_0(1430)\to\bar K^0\pi^-]\ \pi^+$ & $30.8 \pm 4.0$ & \cite{Garmash2007} & $26.0\pm 3.4$ &$11.1 \pm 0.5$ & 5.7 \\
$B^-\to (K^-\pi^+)_S\ \pi^-$ & $24.5\pm 5.0$ & \cite{Aubert:2008bj} & $22.5\pm 4.6$ & $16.5\pm 0.8$&7.5 \\
$\bar B^0\to (\bar K^0\pi^-)_S\ \pi^+$ & $\dfrac{2}{3}(28.2\pm 7.5)$ & \cite{Aubert:2007bs} &$17.3\pm 4.6$  &$15.8 \pm 0.7$ & 7.1
\end{tabular}
\end{ruledtabular}
\end{table}
Our predictions, using the $m_{K\pi}$ range from threshold (0.64 GeV) to 1.76 GeV, are shown in the fifth column of Table~\ref{tab:branch2}.
In the two first lines we use the pole-part contribution of the scalar form factor whereas in the last two lines the full scalar form factor contributes.
Our values have to be compared with those of the fourth column calculated by us using the experimental parameterizations with our range of integration.
The values of the experimental analyzes are given in the second column. They correspond to integration over the full $m_{K\pi}$ range.
In our model, if one integrates also over the full range using the pole part of
the scalar form factor (see Eqs.~~(\ref{f0pole}), (\ref{t0}) and (\ref{f0t0})),
the averaged charge branching ratio for the $B^{\pm}$ decays is $12.7\times10^{-6}$.

For both charged and neutral  $B \to K^*_0(1430)\pi$ decays, the predictions of our model are smaller than the corresponding Belle results~\cite{Garmash:2005rv,Garmash2007} by a factor of 2.3. 
It is worthwhile to mention that the Belle Collaboration has found two solutions in their Dalitz-plot analyzes.
For example, in the Table IV 
of the second paper of Ref.~\cite{Garmash:2005rv}, the value of the solution 2 is smaller by a factor of about 5 than that of the retained solution 1.
In the case of their solution 1, there is a strong negative interference between the resonant $K^*_0(1430)$ contribution and the nonresonant term.
In their  $B^0(\bar B^0)\to K^\pm\pi^\mp\pi^0$ analysis~\cite{Aubert:2007bs} the Babar Collaboration has found four degenerate solutions and the quoted errors of the final result (see their Tables~\ref{tab:CP} and \ref{tab:branch2}) include the spread  of these four solutions.

\begin{figure}[h!]
\includegraphics*[width=7.5cm]{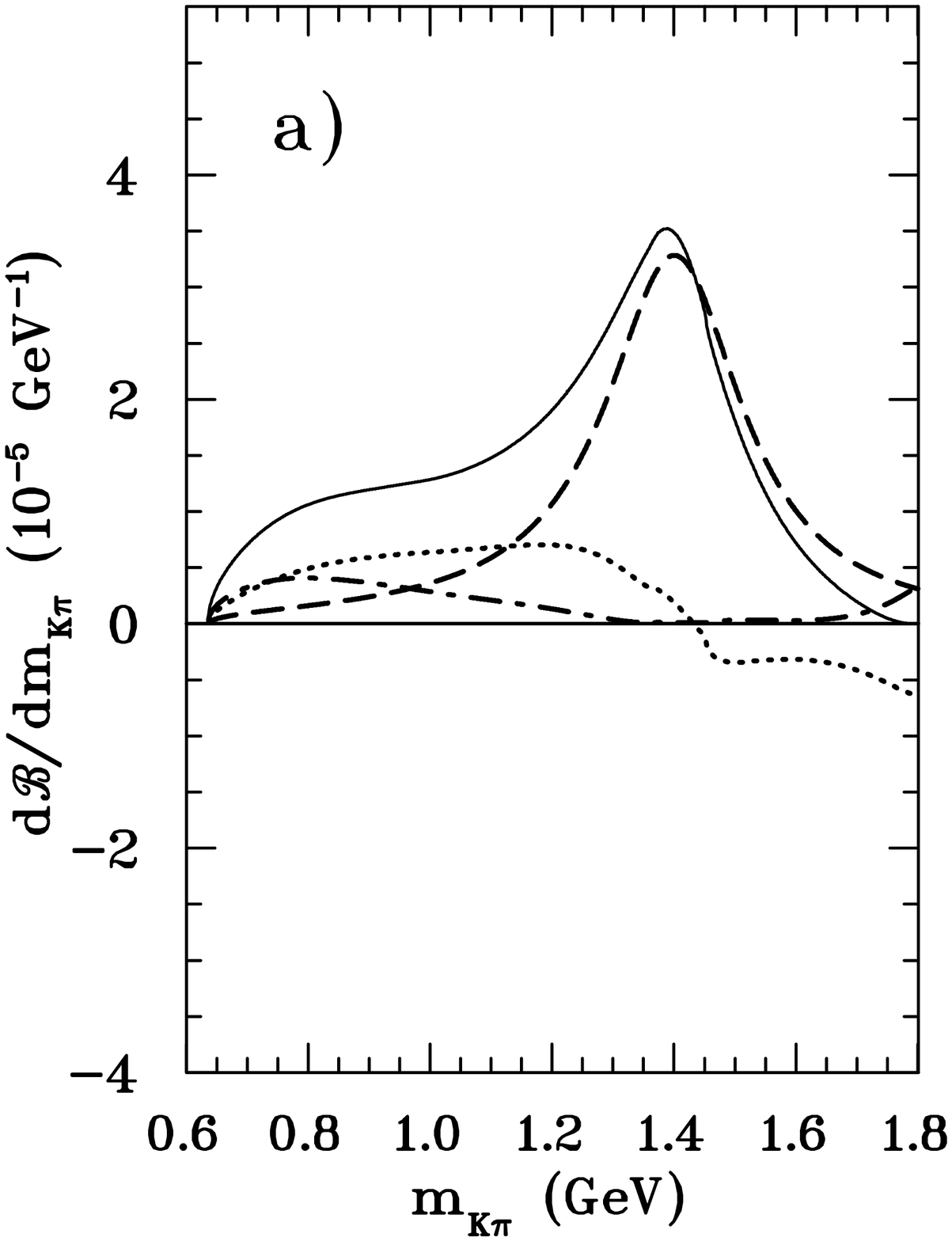}
\includegraphics*[width=7.5cm]{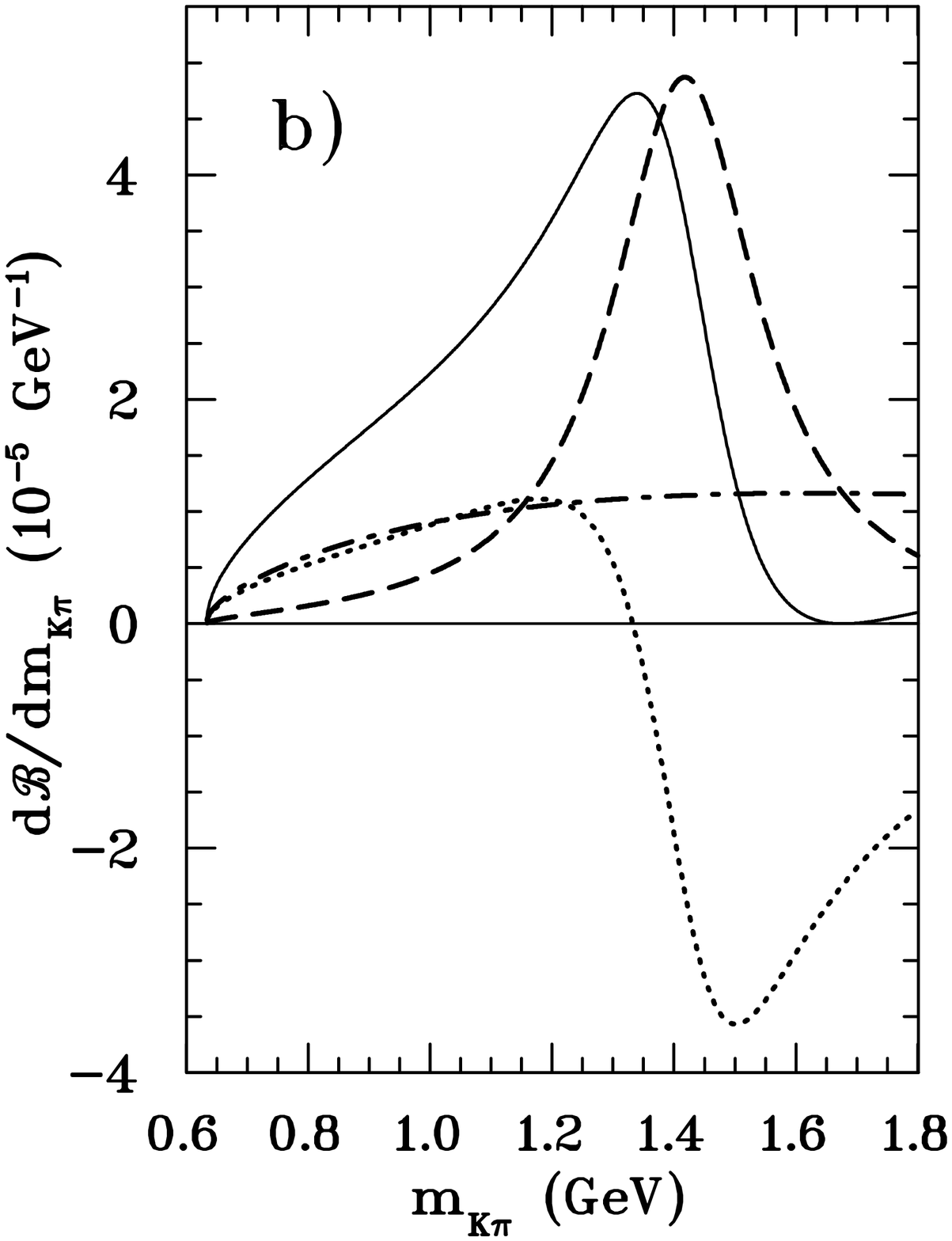}

\caption{
Comparison of the different components of the averaged $m_{K\pi}$ distributions  of the $B^\pm \to (K^\pm\pi^\mp)_S \pi^\pm$ decays: a) our model, b) BaBar's LASS parametrization~\cite{Aubert:2008bj}.
In this calculation our amplitude is proportional to the scalar $K\pi$ form factor but that of BaBar is the part proportional to the $S$-wave  $K\pi$ $T$-matrix.
The dashed lines correspond to the resonant $K^*_0(1430)$ contributions, the dotted-dashed lines to the background, dotted lines to the interference and the solid lines to their sum.
\label{Modulae}
}
\end{figure}

In Fig.~\ref{Modulae}, we compare the $m_{K\pi}$ distributions of the  averaged $B^\pm \to (K^\pm\pi^\mp)_S\  \pi^\pm$ decays corresponding to our model and to  the BaBar parametrization calculated with the central values of their parameters.
As mentioned above, our  $B\to (K\pi)_S\ \pi$ amplitude is proportional to the strange $K\pi$ scalar form factor (see Eqs.~(\ref{K-pi+Sampli}), (\ref{Mmoinstotal }) and (\ref{Mplustotal})) but in BaBar's case it is the part proportional to the $K\pi$ $S$-wave $T$-matrix.
In Fig.~\ref{Modulae}a, we show the resonant $K^*_0(1430)$ contribution (dashed line) of our model together with the background part (dotted line) and the interference term
(dashed-dotted line) between the resonant and background contributions. 
In Fig.~\ref{Modulae}b, the corresponding three components for the BaBar parametrization~\cite{Aubert:2008bj} are shown, their effective range term (dashed-dotted) line playing the role of the background.
In our case the interference term is \textit{positive}, its contribution being close to 19\%, that of the resonance about 70\% and that of the background part around 11 \%. 
This can be compared with the BaBar parametrization which, for the range $m_{K\pi}$ between 0.64 and 1.76 GeV, gives $-25$ \% for their \textit{negative} interference contribution, 78 \%  for  their resonance part  and 47 \% for the nonresonant effective range term.
These numbers are very close to the corresponding values, $-26$ \%, 81 \% and 45 \% given in Ref.~\cite{Aubert:2008bj} obtained integrating over the full range of $m_{K\pi}$.
Although these effective mass spectra are significantly different, our integrated value for the 
$B^-\to(K^-\pi^+)_S\ \pi^-$ branching fraction $(16 .5\pm 0.8)\times 10^{-6}$ is within one standard deviation with respect to the experimental BaBar result $(22.5\pm 4.6)\times 10^{-6}$. 
For the neutral $B$ decays the comparison is even better: we obtain  $(15.8\pm 0.7)\times 10^{-6}$ and the BaBar result, recalculated for the $m_{K\pi}$ range from 0.64 to 1.76 GeV, is  $(17.3\pm 4.6)\times 10^{-6}$ (see last line of Table~\ref{tab:branch2}).
We then suggest below [see Eq.~(\ref{MSparam})] a parameterization (based on our amplitude) proportional to the $K\pi$ scalar form factor and which could be used, instead of a  parameterization proportional to the  $K\pi$ $S$-wave $T$-matrix, in experimental analyzes of $B \to K \pi \pi$ decays.

Our theoretical QCDF predictions ($c_i^p=0$), shown in the last column of Table~\ref{tab:branch2}, give too low branching fractions for all $B$ decays into $(K\pi)_S\  \pi$ or $K^*_0(1430)\ \pi$  by a factor close to 2.
Figure~\ref{ampPS}b illustrates the influence of the phenomenological parameters on the theoretical  reduced $S$-wave amplitudes.
The modulus of the amplitudes increases by a factor of about $\sqrt{2}$ and there is also an important phase change.
The fact that,  at the $K\pi$ mass equal to the $K^*_0(1430)$ mass, the magnitudes and phases of all  reduced $S$-wave amplitudes (without and with $c_i^p$) are similar, comes from the dominance of the $q^2$ term proportional to $\lambda_c$ and from the smallness of the $a_8^{u,c}(S)$ coefficients (see, for instance, Eqs.~(\ref{K0Spi-Sampli}) and Table~\ref{aipmb}).

As was discussed in Sec.~\ref{scalarformfactor} and shown  Fig.~\ref{scalarff},  the $S$-wave form factor depends on the value of the $f_K/f_\pi$ ratio.
We found that the form factors corresponding to $f_K/f_\pi =1.193$ (form factor of the present work) and 1.183 lead to fits of comparable good quality,  with however a slightly better $\chi^2$ for the form factor calculated with $f_K/f_\pi$=1.183.
Use of the form factor with a ratio equal to 1.203 gives a poorer fit.

Effective mass and helicity distributions together with branching ratio data allow to determine mainly the moduli of the decay amplitudes. 
Their phases can be constrained by measurements of direct $CP$ asymmetries and by time dependent Dalitz-plot analyzes.
In our fit, besides asymmetries, we use the preliminary value, obtained by the BaBar group~\cite{AubertLP2007}, of the phase difference $\Delta\Phi_0$ between the $B^0$ and $\bar B^0$ decay amplitudes to $K^*(892)\pi$.
Had we not imposed this constraint, we would obtain four different solutions with equivalent $\chi^2$ and with almost unchanged moduli for $S$- and $P$-wave amplitudes but with different phases.
In a just published analysis of $B^0 \to K^0_S\pi^+\pi^-$ decays the Belle Collaboration~\cite{Belle1108} has proposed values for $\Delta\Phi_0$.
We have checked that we can, reasonably well, reproduce the value of their solution 2, viz. $(14.6^{+19.4}_{-20.3}\pm11\pm17.6)^\circ $ with a global fit of the same quality as the present one.
Comparing the results of the two fits, we found indeed that the $S$- and $P$-wave amplitudes had basically unchanged moduli but modified phases.
Note the large difference between the preliminary result of Ref.~\cite{AubertLP2007}  $, \Delta\Phi_0=(-164\pm24\pm12\pm15)^\circ$ and the above value of Belle.
Our $A_{CP}$ value  for $\bar B^0 \to (\bar K^0 \pi^-)_P\ \pi^+$ given in
Table~\ref{tab:CP},  ($-19.6\pm3.0$)\%, agrees well with that of the solution~2  of this Belle analysis, ($-20\pm11\pm5\pm5$)\%.

As just mentioned above and as found in their previous analysis~\cite{Garmash:2005rv,Garmash2007}, the Belle Collaboration, in this latest time-dependent analysis~\cite{Belle1108}, has retained 2 solutions which is consequence of the interplay between the quite broad $K^*_0(1430)$ resonance and their phenomenological nonresonant background. 
Their solution~1 has a large $K^*_0(1430)\pi$ fit fraction with a sizable negative resonant-nonresonant interference term while their solution~2 is characterized by a fit fraction smaller by a factor of 3.5 and a small interference term, as we found in our model. Using the $K_S^0\pi^+\pi^-$ charmless total branching fraction $(47.5\pm2.4\pm3.7)\times 10^{-6}$, as given in Table III of their previous publication~\cite{Garmash2007}, one obtains a branching fraction of 8.3 $\times 10^{-6}$ for their solution 2, value close to our result cited in Table~ \ref{tab:branch2}.

\section{Summary and Concluding Remarks \label{summary}}

With this analysis of $S$- and $P$-wave pion-kaon interactions in $B\to K \pi\pi $ decays, we have extended and completed previous studies on final-state interactions in these three-body decays~\cite{fkll,El-Bennich2006}. 
We have concentrated on the scalar $K^*_0(1430)$ and vector $K^*(892)$ resonances following the logic of these previous works that treated the $\pi\pi$ interactions in $S$- and $P$-waves as well as their interferences. 
In Sec.~\ref{Decay_amplitudes} the weak decay amplitudes were again derived in the QCD factorization approach~\cite{bene03,Beneke:2001ev}, which  express them as  a product of two currents multiplied by a sum of effective coefficients which includes non-factorizable contributions.
These coefficients, representing perturbative QCD leading order amplitudes and next-to-leading order vertex and penguin corrections, were studied and given in Sec.~\ref{ai}.
The contribution of these higher order terms is not sufficient to obtain a good description of data.
Therefore, we have introduced phenomenological parameters which can simulate on one hand long-distance charming penguin amplitudes~\cite{Ciuchini:1997hb} and on the other hand hard-spectator scattering and weak annihilation contributions~\cite{bene03}.
These phenomenological amplitudes could receive also, through $b$ to $s$ quark transitions involved here, some new-physics contributions. 

The different models for  the matrix element of the first current, expressing the $B$ to $\pi$ transition in terms of the scalar and vector transition form factors have been briefly reviewed  
in Sec.~\ref{transitionff}.
The  creation of a pion-kaon pair in an $S$- or $P$-wave from vacuum is mediated by the second current, 
and accordingly described by a $K\pi$ scalar and a  vector form factor. 
These control the dependence of the decay amplitude as a function of the
$K\pi$ invariant mass, because the $B\pi$ form factors are nearly constant
in the region considered. 
In Sec.~\ref{scalarformfactor} the scalar $K\pi$ form factor was calculated along similar lines as in Ref.~\cite{jop} and the extension to the case of a vector form factor was developed in Sec.~\ref{vff}.
 
We treat both the $S$- and $P$-wave on the same footing, namely relating 
the form factors using their analyticity and unitarity relations 
to pion-kaon scattering properties known from experiments. A simplified, but
realistic treatment of inelasticity is also implemented.
In the determination of these form factors  
we also use chiral symmetry and QCD constraints at low and high energies 
respectively. As a byproduct of the scalar form factor study we predict  
for the modulus of the $K^*_0(1430)$ decay constant a value of $32\pm5$ MeV. 
Our theoretical amplitudes go beyond the usual two-body approach 
applied to decays such as $B\to K^*\pi$ 
and correctly accounts for the  $K\pi$  
final-state interaction both on and away from resonance peaks. 
A nonresonant background can be isolated from the 
resonant one in our model as illustrated in Sec.~\ref{polescal}, and compared to those introduced by Belle as well as BaBar collaborations 
in their amplitude parametrization~\cite{Garmash:2005rv,Garmash2007,Aubert:2008bj,Aubert:2007bs}.
A comparison between our resonant, nonresonant and interference term splitting and that of BaBar was presented in Sec.~\ref{discussion}.

Furthermore, our model correctly reproduces the enhancement of the decay
distributions  in the low-mass region as observed in 
Figs.~\ref{fig:MKpiBpmBelle}, \ref{fig:MKpiBpmBaBar}, \ref{fig:MKpiB0Belle} 
and \ref{fig:MKpiB0BaBar}. 
This enhancement may be attributed to the broad $K^*_0(800)$ resonance
which is present in the $T$-matrix that we use.
As can be seen from Figs.~\ref{ffpole} and \ref{Modulae} the $K^*_0(800)$ 
is responsible, in our model, for the behavior of our $S$-wave amplitude 
for $m_{K\pi}$ from threshold to about 1.2 GeV. 

Our theoretical QCDF amplitude predicts branching fractions for the $B \to K^*(892)\ \pi$ 
and $B \to K^*_0(1439)\ \pi$ decays too small by factors of about 5 and 2, respectively.
The inclusion of four  complex phenomenological penguin parameters allows us to have a realistic model.
These parameters, common for $B^+, B^-, B^0$ and $\bar B^0$ decays, have been fitted to numerous experimental data, which includes 308 data for the $K\pi$ effective mass and helicity angle distributions, four $B \to K^*(892)\ \pi$ branching fractions, six direct $CP$ violating asymmetries and the phase difference between the $B^0$ and $\bar B^0$ decay amplitudes to  $K^*(892)\ \pi$.
Our model reproduces rather well these 319 data with total $\chi^2$ of 551.5
corresponding to a  $\chi^2$ per degree of freedom equal to 1.77.
This good reproduction of the data makes it possible to predict the $B \to (K\pi)_S\ \pi$ and $B \to K^*_0(1430)\ \pi$ branching fractions.
We can obtain, without ambiguities, the pole contribution of the  $K^*_0(1430)$ resonance.
This contribution is smaller than the experimental determination by Belle (see Table~\ref{tab:branch2}) and BaBar~\cite{Aubert:2007bs} by factors of  2.3 and 1.4, respectively.
The determination of these branching fractions, within the isobar model is problematic since the resonance $K^*_0(1430)$ is wide and the nonresonant part difficult to assess.
The non-uniqueness of the  parametrization of the nonresonant contribution leads to a large systematic uncertainty of the  $B \to K^*_0(1430)\pi$  branching fraction as seen in the Particle Data Tables~\cite{pdg08}.
In our approach, with a scalar form factor well constrained by theory and experiments other than $B$-decays studies, we can describe the data over the $m_{K\pi}$ range from threshold to 1.8 GeV.

Therefore, to diminish ambiguities in data analyzes, we propose to use the following $S$-wave amplitude parametrization for $B \to (K\pi)_S\pi$ decays:
\begin{equation}
\label{MSparam}
\mathcal{M}_S\left(m_{K\pi}\right) = f_0^{K\pi}\left(m_{K\pi}^2\right)\ \left ( \dfrac {c_0}{m_{K\pi}^2}+c_1 \right ),
\end{equation}
which follows from Eq.~(\ref{K-pi+Sampli}).
Here, $c_0$ and $c_1$ are constant complex  parameters to be determined through the Dalitz-plot analysis of a given $B$-meson decay.
Upon request, we can provide a numerical table for the scalar form factor $ f_0^{K\pi}\left(m_{K\pi}^2\right)$. 
To calculate the $K_0^*(1430)$ resonance contribution, one can replace, once the $c_0$ and $c_1$ parameters have been determined,  $ f_0^{K\pi}\left(m_{K\pi}^2\right)$ by its pole part $f_0^{pole}\left(m_{K\pi}^2\right)$ given in Eqs.~(\ref{f0pole}), (\ref{t0}) and (\ref{f0t0}).

The direct $CP$ violating asymmetries and the time dependent $CP$ asymmetries are related to the not very well determined angle $\gamma$ of the unitary triangle.
Our amplitudes are sensitive to $\gamma$ via their dependence on $\lambda_u$ (see Eqs.~(\ref{K-pi+Sampli}), (\ref{K-pi+Pampli}),
(\ref{K0Spi-Sampli}) and (\ref{K0Spi-Pampli})).
Precise measurements of the Dalitz plot distributions could allow to constrain $\gamma$ using our model.

\begin{acknowledgments}
 We would like to thank Maria R\'o\.{z}a\'nska and Olivier Leitner for very helpful discussions. 
  We acknowledge quite useful exchanges with Patricia Ball and Alexei Garmash. 
We thank T. E. Latham to have provided us plots concerning the results of the Babar Collaboration~\cite{Aubert:2008bj}.
B.~E. is grateful to C.~D.~Roberts and T.~S.~Lee for thoughtful comments. 
This work has been supported in part by the Polish Ministry of Science and
Higher Education (grant No N N202 248135), by the IN2P3-Polish Laboratories
Convention (project No 08-127), by the CNRS - Polish Academy of SciencesÄ
agreement (project No 19481) and by the Department of Energy, Office of Nuclear Physics, contract No DE-AC02-06CH11357.  
\end{acknowledgments}



\appendix
\section{Vertex and penguin corrections for $\mathbf{PS}$ and $\mathbf{PV}$ final
 states \label{appendixvpc}
 }
Here we show how we calculate the next-to-leading order vertex and penguin corrections entering the effective QCD amplitudes $a_i^p(\mu)$ of Eq.~(\ref{eq:44}). We compute their values for $PS$ and $PV$ final states, namely $K^*_0(1430)\pi$ and $K^*(892)\pi$. Let us first quote some general results from Ref.~\cite{bene03} for the vertex correction terms $V_i(M)$:
\begin{equation}
V_i(M) =  \int_0^1 \Phi_{M}(x)\, \left [ 12 \ln\dfrac{m_b}{\mu}-18+g(x)\right ]  dx
\label{v410}
\end{equation}
for $i=1,4,10$ and
\begin{equation}
V_i(M)  =  \int_0^1 \Phi_{m}(x)\, [ -6 + h(x) ]  \ dx \, 
\label{v68}
\end{equation}
if $i = 6,8$, where $\Phi_M$ and $\Phi_m$ are the leading-twist and twist-3 
distribution amplitudes, respectively, of the emitted meson. 
The integration is over the longitudinal meson-momentum fraction $x$. 
As in Sec.~\ref{ai}, $M$ stands for the emitted meson that does not include the spectator quark. 
The functions $g(x)$ and $h(x)$ are given in Eq.~(38) of Ref.~\cite{bene03}. 

Light cone distribution amplitudes (LCDA) for scalar mesons were derived making
 use of QCD sum rules~\cite{Cheng:2005nb}. We use these distributions to 
 calculate vertex corrections for the case where $M$  is a $K^*_0(1430)$.
  The leading twist Gegenbauer expansion for scalar mesons is given 
  by~\cite{Cheng:2005nb}
\begin{equation}
 \Phi_S(x) = 6x(1-x)\left [ \alpha_0^S+\sum_{n=1}^\infty \alpha_n^S(\mu)C_n^{3/2}(2x-1)\ \right ] ,
\label{scalarLCDA}
\end{equation}
while the scalar twist-3 amplitude is $\Phi_s(x) = 1$. Here, the 
$\alpha_n^S(\mu)$ are related to the scalar Gegenbauer moments $B_n(\mu)$ by 
virtue of 
\begin{equation}
   \alpha_n^S(\mu) = \mu_S B_n^S(\mu)\ , \quad\quad \mu_S = \dfrac{m_S}{m_2(\mu)-m_1(\mu)}  , 
\end{equation}
with the quark masses $m_1\neq m_2$, as in the case of the $K^*_0(1430)$, and where $m_S$ is the scalar meson mass. 
The normalization condition $\int_0^1 \Phi_S(x) dx=1$ yields $B_0^S=\mu_S^{-1}$ and thus $\alpha_0^S=1$ if the small even Gegenbauer moments are neglected. 
With this, the vertex corrections to order $\alpha_3^S$ for a $K^*_0(1430)\pi$ final state with $M=K^*_0$ are   given, for  $i=1,4,10$,  by the Eq.~(4.3) of  Ref.~\cite{Cheng:2005nb} and by $V_i(K^*_0)= -6$ for $i=6,8$.  

The leading twist LCDA for a vector meson $M=V$ reads
\begin{equation}
 \Phi_V(x) = 6x(1-x)\left [ \ 1+\sum_{n=1}^\infty  \alpha_n^V(\mu) C_n^{3/2}(2x-1)\ \right ],
\end{equation}
which is given by the same expansion in Gegebauer polynomials $ C_n^{3/2}(2x-1)$
as the one for pseudoscalar mesons $\Phi_P(x)$ but with different moments
 $\alpha_n^V(\mu)$. Thus, the vertex corrections Eq.~\eqref{v410} for a 
 $K^*(892)\pi$ final state where $M=K^*$,  are given by the Eqs.~(47) and (48) of 
 Ref.~\cite{Beneke:2001ev} for $i=1,4,10$.
Taking into account the twist-3 LCDA, $P_{n}(x)$ being the usual Legendre polynomials,
\begin{equation}
\Phi_v(x)  =  3 \sum_{n=0}^\infty \alpha_{n\perp}^V(\mu) P_{n+1}(2x-1),
\end{equation}
we obtain the $i=6,8$ corrections as
\begin{equation}
  V_i(K^*) =  \left ( 9-6i\pi\right) \alpha^{K^*}_{1\perp} +\left ( \dfrac{19}{6}-i\pi
   \right ) \alpha^{K^*}_{2\perp},
\end{equation}
where we have made use of the property $\int_0^1  \Phi_v(x) dx =0$.

At order $\alpha_s$, corrections from penguin contractions with the various operators $O_i(\mu)$ exist for $i=4,6$ for QCD penguins and $i=8,10$ for electroweak penguins but not for $i=1$. The expressions for these contributions can be found in integral form in Ref.~\cite{bene03} for the $B\to PV$ decay in Eqs.~(39) to (46). We apply them using the latest results on Gegenbauer moments for the $K^*(892)$~\cite{Ball:2007rt}.
The $PS$ penguin corrections have the same expressions as those for $PP$ final states~\cite{Beneke:2001ev} but one must employ the LCDA introduced in Eq.~\eqref{scalarLCDA}. Nonetheless, since even Gegenbauer moments are suppressed, we take into account corrections up to 
$\alpha_3^{K_0^*}(\mu)$ as for the vertex corrections.

\begin{table}
\caption{
\label{nlovp} 
Next-to-leading order vertex and penguin corrections entering $a_i^p(m_b)$ [see Eq.~(\ref{eq:44})] for  $ B \to K^*_0(1430)\pi$ and $B \to K^*(892) \pi$ decays.  
Note that there are no penguin corrections for $i=1$.
}
\begin{ruledtabular}
\begin{tabular}{ccccc}
&  \multicolumn{2}{c}{$B\to K_0^*(1430)\pi$} &  \multicolumn{2}{c}{$B\to K^*(892)\pi$} \\
             &       Vertex      &    Penguin           &      Vertex       & 
       Penguin      \\
\hline
$a_1$        & $0.011+i\ 0.063$  & $    0.0          $  & $ 0.028+i\ 0.014$      &  $ 0.0            $ \\
$a_4^{u}$    & $-0.001-i\ 0.005$ & $-0.029-i\ 0.019  $  & $-0.002-i\ 0.001$      &  $0.004-i\ 0.014  $ \\
$a_4^c$      & $-0.001-i\ 0.005$ & $-0.037+i\ 0.061  $  & $-0.002-i\ 0.001$      &  $-0.002-i\ 0.004 $ \\
$a_6^{u}$    & $-0.0004+i\ 0 $   & $-0.003 -i\ 0.014 $  & $ 0.001-i\ 0.001$      &  $-0.007-i\ 0.001 $ \\
$a_6^c$      & $-0.0004+i\ 0 $   & $-0.006 -i\ 0.004 $  & $ 0.001-i\ 0.001$      &  $0.001+i\ 0.011  $ \\
$a_8^{u}$    & $0.0 +i\ 0.0$     & $0.0  -i\ 0.0001$    & $-0.00001+i\ 0.00001 $ &  $-0.0 +i\ 0.0    $ \\
$a_8^c$      & $0.0+i\ 0.0 $     & $0.0  -i \ 0.0  $    & $-0.00001+i\ 0.00001 $ &  $-0.0+i\ 0.0001  $ \\
$a_{10}^{u}$ & $0.0006+i\ 0.0032$& $-0.0006-i\ 0.0001$  & $0.0014+i\ 0.0007$     &  $0.0002-i\ 0.0001$ \\
$a_{10}^c$   & $0.0006+i\ 0.0032$& $-0.0007+i\ 0.0003$  & $0.0014+i\ 0.0007$     &  $0.0002-i\ 0.0$
\end{tabular}
\end{ruledtabular}
\end{table}

Finally, the input parameters entering our computation of the $a_i^{u,c}(\mu)$  include the $u$-, $s$-, $c$- and $b$-quark masses, the strong coupling constant $\alpha_s$ and the Gegenbauer moments of the leading twist and twist-three light cone distribution amplitudes for the scalar $K^*_0(1430)$ and vector $K^*(892)$ mesons. 
We use the scale  $\mu=m_b$ with $\alpha_s(m_b)=0.223$. 
The corresponding values of the quark masses have been given below Eq.~(\ref{Mplustotal}).
In order to calculate the Gegenbauer moments associated with the LCDA of the scalar meson, we start from the values at $\mu=m_b/2$ from Table X of Ref.~\cite{Cheng:2005nb}.
For the vector meson we use the recently determined moments $\alpha_i^{K^*}$  and  $\alpha_{i\perp}^{K^*}$ at $\mu=1$~GeV~\cite{Ball:2007rt}. 
After evolution to the scale $\mu=m_b$, one obtains the following values: $\alpha_1(K^*_0)=5.26$, $\alpha_3(K^*_0)=-8.24$ for the scalar meson $K^*_0(1430)$ and $\alpha_1(K^*)=0.018$, $\alpha_2(K^*)=0.065$, $\alpha_{1\perp}(K^*)=0.026$, $\alpha_{2\perp}(K^*)=0.065$ for the vector meson $K^*(892)$. In Table~\ref{nlovp}, we give  our results for the next-to-leading order vertex and penguin corrections of Eq.~(\ref{eq:44}) and from which are calculated the $a_i^p(m_b)$ listed in Table~\ref{aipmb}.

\section{Determination of the {\boldmath $S$}-wave {\boldmath $T$}-matrix elements\label{appendixSwavT}}

Below, we describe the determination of the $S$-wave $T$-matrix elements
$T_{11}$, $T_{12}$ and $T_{22}$, the channel $K \pi$ being labeled as 1 and
$K \eta' $ as 2.

\subsection{Fit above the inelastic threshold}

\begin{figure}[ht]
\centering
\includegraphics*[width=7.5cm]{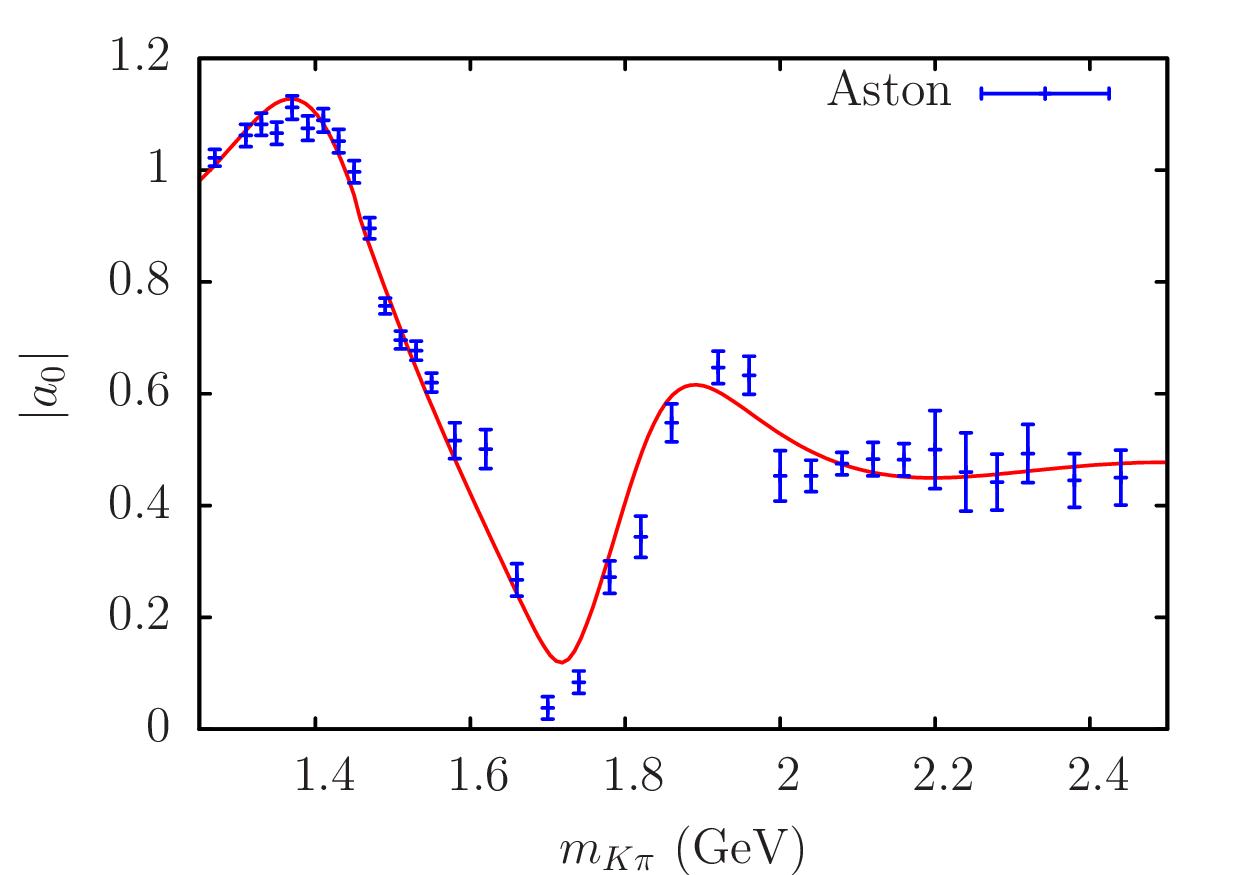}
\includegraphics*[width=7.5cm]{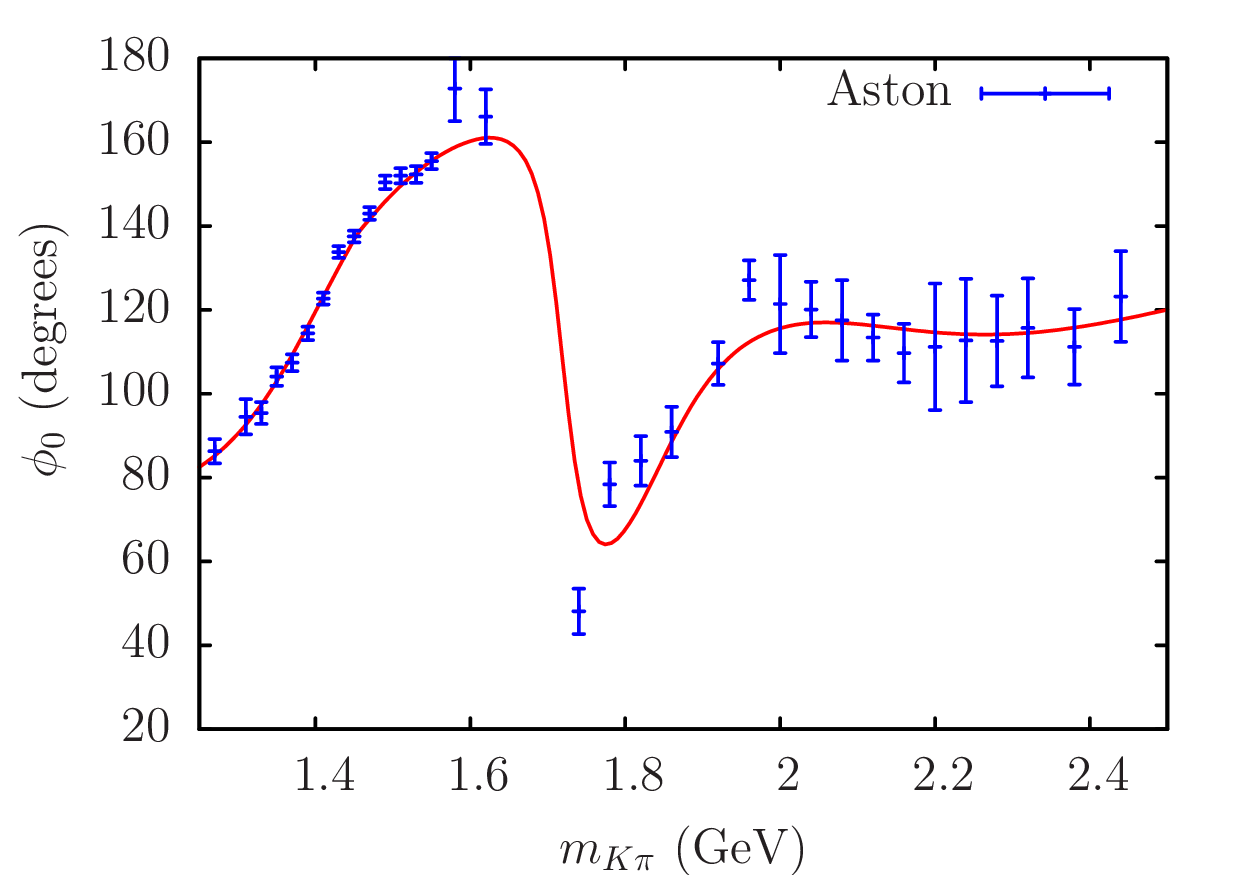}
\caption{Results of the $K$-matrix fit for the modulus, $\vert a_0 \vert$, and the phase, 
$\phi_0$, of the $K^+\pi^- \to K^+\pi^-$ amplitude.\label{a0phi0}}
\end{figure}

Precise experimental data on $K\pi$ scattering is 
available~\cite{estabrooks,aston88} and cover 
approximately the range $0.9\lapprox m_{K\pi} \lapprox~2.5$ GeV. 
Experiment shows that inelasticity effectively
sets in at the $\eta' K$ threshold and we make the assumption
that it is saturated by the $\eta' K$ channel.
Above the inelastic threshold, the three components of the $T$-matrix, 
$T_{11}$, $T_{12}$ and  $T_{22}$ are needed in the unitarity equations. 
A two-channel $S$-matrix which is unitary and satisfies time reversal
invariance can be parametrized, in terms
of three observable quantities: 
two phase-shifts ($\delta_{K\pi}$, $\delta_{K\eta'}$)
and one inelasticity parameter ($\eta_{K\pi}$),
\begin{equation}
\lbl{smatparam}
S=\left(
\begin{matrix} 
\eta_{K\pi} {\rm e}^{2i \delta_{K\pi}} & 
 \sqrt{1- (\eta_{K\pi})^2}{\rm e}^{i( \delta_{K\pi}+\delta_{K\eta'}) } \cr
 \sqrt{1- (\eta_{K\pi})^2}{\rm e}^{i( \delta_{K\pi}+\delta_{K\eta'}) } &
\eta_{K\pi} {\rm e}^{2i \delta_{K\eta'}} 
\end{matrix}
\right).
\end{equation}
The relation between the partial-wave  $T$ and $S$  matrices is
\begin{equation}
\lbl{stmat}
S_{mn}= \delta_{mn}+4i \left( {q_m(s) q_n(s)\over s} \right)^{1\over2}
T_{mn}, \ m,\ n= 1,\ 2 \ \mbox{\rm with}\ s=(p_K+p_\pi)^2\equiv m^2_{K\pi}, 
\end{equation}
and where $q_i(s)$  is the center-of-mass momentum for channel $i$. 
A simple way to enforce unitarity is to use a $K$-matrix type 
representation of the $T$-matrix. 
We take here the following representation
\be
T^{-1}=K^{-1}-diag( \bar{J}_1(s),\bar{J}_2(s)), 
\en
where $K$ must be real and symmetric and the functions 
$\bar{J}_i(s)$, $i=1,2$ read:  
\be
\bar{J}_i(s)={s\over\pi}\int_{s_i}^\infty {ds'\over s'(s'-s)}
{2q_i(s')\over\sqrt{s'}},
\en
with $s_1=(m_K+m_\pi)^2$ and $s_2=(m_K+m_{\eta'})^2$. 
Following the approach of Ref.~\cite{jop}, we use the following parametrization 
for the  $K$-matrix: 
\begin{equation}
\lbl{kmatparam}
K_{ij}(s)= {g_i g_j\over M_1^2- s } +{h_i h_j \over M_2^2 -s} 
+{a_{ij} + b_{ij} s \over 1 + (s/c)^2  },
\end{equation}
which includes two resonances and a background term. We determine the 
parameters by performing a fit over an energy range,
$1.25 \le m_{K\pi} \le 2.5\  {\rm GeV }$.
We used the experimental data
of LASS~\cite{aston88}, who measured the phase and the modulus of
the charged amplitude $K^+ \pi^- \to  K^+ \pi^- $ combined with the
earlier measurements of Estabrooks \textit{et al.}~\cite{estabrooks} of the
isospin $3/2$ component. The central values of the parameters determined
from the fit are as follows:
\begin{equation}
\lbl{fitres}
\begin{array}{lll}
M_1=   1.454, & g_1 =0.505, & g_2= 1.651,  \\
M_2=   1.988, & h_1 =0.784, & h_2= 1.144, \\
a_{11}= 2.371,  & a_{12}= 10.060, & a_{22}=-43.946, \\
b_{11}= -1.345, & b_{12}= -2.051, & b_{22}= 14.538 .
\end{array}
\end{equation}
All the above parameters are in units of GeV except for $a_{ij}$ which are 
dimensionless and $b_{ij}$ which are in units of ${\rm GeV}^{-2}$. 
The energy cutoff parameter $c$ is not fitted, it is set to $c=1$ 
${\rm GeV}^2$.
The number of data points is~$70$, the 
total $\chi^2$ is~$205$ 
and the number of parameters in this fit is~$12$. 
Figure~\ref{a0phi0} shows the comparison of the fit result with the
experimental data of Ref.~\cite{aston88}.
Using these results, when $t\to\infty$, 
the phases and inelasticity
parameters of the $S$-matrix satisfy:
$\delta_{K\pi}(\infty)=2\pi$, $\delta_{K\eta'}(\infty)=0$ 
 and $\eta_{K\pi}(\infty)=1$.

\subsection{Below the inelastic threshold}

\subsubsection{\boldmath $Chiral~ symmetry~ constraints~ on~ T_{12}$}

Chiral symmetry constrains scattering amplitudes which 
involve the $\eta'$
meson at low energy if one combines chiral symmetry with the large $N_c$ 
expansion~\cite{largeNc}. We will use here a systematic expansion scheme 
based on counting $1/N_c$ on the same footing as a chiral 
factor $p^2$~\cite{mou95,kaiserleutwyler}
\begin{equation}
\delta \equiv p^2 \sim {1\over N_c} .
\end{equation}
At order $\delta$ of this expansion the Lagrangian contains three independent
terms
\begin{equation}
\lbl{delta}
{\cal L}_{\delta}= {F_0^2\over 4}\left\{
\tr( D_\mu U D^\mu U^\dagger ) +\tr(\chi^\dagger U + U^\dagger \chi)\right\}
-{1\over2} M_0^2\, \phi_0^2, 
\end{equation} 
where $U$ is a unitary matrix which contains a nonet of pseudoscalar 
meson fields $\phi_0$,...,$\phi_8$.
At this order, $\eta-\eta'$ mixing involves one angle $\theta$
and its value may be determined such that the physical
$\eta$ mass is reproduced. This gives $\theta=-5.6^\circ$
and the prediction for the mass of the $\eta'$ meson is 
too large~\cite{etappprob} ($M_{\eta'}\simeq 1.6$ GeV). 
This problem is cured by going to the next order of this expansion. The
Lagrangian at order $\delta^2$ contains eight independent terms
\begin{eqnarray}
\lbl{delta2}
&& {\cal L}_{\delta^2}= L_2\tr(  D_\mu U^\dagger D_\nu U 
D^\mu U^\dagger D^\nu U)
+       (L_3+2L_2)\tr(  D_\mu U^\dagger D^\mu U D_\nu U^\dagger D^\nu U) 
\nonumber\\
&& + L_5\tr(D_\mu U^\dagger D^\mu U (\chi^\dagger U +U^\dagger\chi))
+ L_8\tr(\chi^\dagger U \chi^\dagger U + \chi U^\dagger \chi U^\dagger)
\nonumber\\
&& -iL_9\tr(F_{\mu\nu}^R D^\mu U D^\nu U^\dagger 
+ F_{\mu\nu}^L D^\mu U^\dagger D^\nu U)
+L_{10}\tr(U^\dagger F_{\mu\nu}^R U F^{\mu\nu\,L} )
\nonumber\\
&& +k_1 D_\mu\phi_0 D^\mu\phi_0 
 + ik_2 {F_0\over \sqrt6} 
{\phi_0}\,\tr(\chi^\dagger U - U^\dagger\chi)  .
\end{eqnarray}
This Lagrangian involves
the subset of the Gasser-Leutwyler~\cite{gl85} coupling constants $L_i$
which are of leading order in $N_c$
(the scale dependence shows up here at order $\delta^3$)
plus two additional couplings, $k_1$ and $k_2$. 
At order $\delta^2$, $\eta-\eta'$ mixing
involves two angles $\theta_0$ and $\theta_8$,
\begin{eqnarray}
&& \phi_8= {1\over\lambda_8} (\phi_\eta \cos\theta_8 
+ \phi_{\eta'} \sin\theta_8 ),
\nonumber\\
&& \phi_0= {1\over\lambda_0} (-\phi_\eta \sin\theta_0 
+ \phi_{\eta'} \cos\theta_0).
\end{eqnarray}
The factors $\lambda_0$ and
$\lambda_8$ can be expressed in terms of $L_5$,
and the angles $\theta_0$, $\theta_8$ can be expressed in terms of the chiral
couplings $L_5$, $L_8$ and the physical meson masses 
$m_\eta$, $m_{\eta'}$~\cite{mou95}. 
Using the Lagrangian's~\rf{delta} and \rf{delta2} a small calculation yields
the scattering amplitude $K\pi \to K\eta' $. It can be written in the
following form:
\be\lbl{TKpiKetap}
 T_{K \pi,K \eta'}(s,t,u)= 
 \sin\theta_8\, T_{8}(s,t,u) + \cos\theta_0\, T_{0}(s,t,u) 
+\sin\theta\,   T_{s}(s,t,u) + \cos\theta\,   T_{c}(s,t,u),
\en
with
\begin{eqnarray}
&& T_{8}(s,t,u)= -{\sqrt3\over36 \fpid} (-9t+8\mkd+\mpid+3\metaxd),
\nonumber\\
&& T_{0}(s,t,u)= {\sqrt6\over18\fpid} (2\mkd+\mpid),
\end{eqnarray}
and
\begin{eqnarray}
&& T_s(s,t,u)=L_3\, {-1 \over\sqrt3 \fpiq }\,[\,-2(t-2\mkd)(t-\metaxd-\mpid)
\nonumber\\
&&+(s-\mpid-\mkd)(s-\metaxd-\mkd) + (u-\mpid-\mkd)(u-\metaxd-\mkd)\,]
\nonumber\\
&&\quad + L_5\,{4\sqrt3\over27\fpiq}(\mkd-\mpid)(8\mkd+\mpid-3\metaxd)
\nonumber\\
&&\quad +L_8\,{-16\sqrt3\over 9\fpiq}(\mkd-\mpid)(2\mkd+\mpid)  ,
\end{eqnarray}
\begin{eqnarray}
&&T_c(s,t,u)= (L_3+2L_2)\,{2\sqrt6\over 3\fpiq}
[\,(t-2\mkd)(t-\metaxd-\mpid)
\nonumber\\
&&\quad+(s-\mpid-\mkd)(s-\metaxd-\mkd) +(u-\mpid-\mkd)(u-\metaxd-\mkd) ]
\nonumber\\
&&\quad
+L_5\,{-2\sqrt6\over27\fpiq}\, [\, 9 t (\mkd-\mpid)+ 31\mkd\mpid+3\mkd\metaxd
+4\mkq +6\metaxd\mpid-8\mpiq] 
\nonumber\\
&&\quad
+L_8\,{8\sqrt6\over9\fpiq} (8\mkd\mpid+2\mkq-\mpiq)
+\tilde{k}_2\,{\sqrt6(2\mkd+\mpid)\over9\fpid },
\end{eqnarray} 
with $\tilde{k}_2=k_2-k_1/2$.
Projecting eq.~\rf{TKpiKetap} 
on its $l=0$ partial wave gives $T_{12}(s)$. Its
value in numerical form at the $K\pi $ threshold $s_1$ is  at order  $(\delta+ \delta^2)$:
\begin{eqnarray}
&&T_{12}(s_1) =
0.32\sin\theta_8 +0.28\cos\theta_0 +L_2\, (-470.5 \cos\theta )
\nonumber\\
&& +L_3\,(-156.8\cos\theta +14.5\sin\theta)
+L_5\,(-229.1\cos\theta-21.5\sin\theta) 
\nonumber\\
&& +L_8\,(161.4\cos\theta-169.3\sin\theta)
+ \tilde{k}_2\,(0.56\cos\theta ).
\end{eqnarray}
The values of the couplings $L_5$, $L_8$ can be determined 
from the ratio of the decay constants $f_K/ f_\pi$ and the ratio of the quark 
masses $2m_s/(m_u+m_d)$ using the $\delta$ expansion up to order $\delta^2$. 
Using, for instance, the central values 
obtained from lattice QCD by the MILC collaboration~\cite{milc}
yields:
$L_5\simeq 1.97\times 10^{-3}$ and $L_8\simeq 0.87\times 10^{-3}$,
while for the mixing angles one obtains
$\theta_0\simeq -18.9^\circ$ and  $\theta_8\simeq -3.03^\circ$.
Fitting the $\eta$ and  $\eta'$ masses in the $\delta$ expansion gives
$\tilde{k}_2\simeq 0.12$~\cite{mou95}.
Finally, we need the values of $L_2$ and $L_3$. In the ordinary chiral
expansion, $L^r_2(\mu)$, $L^r_3(\mu)$ can be obtained either from
sum rules based on $\pi\pi$ scattering~\cite{gl85} or based on  
$K \pi$ scattering~\cite{abm,roypik} or from data on $K_{l4}$ decay 
form factors~\cite{riggenbach,bijkl4}. For illustration, let us adopt
the values from~\cite{roypik} and identify $L_2$, $L_3$ with
$L^r_2(\mu)$, $L^r_3(\mu)$ at $\mu=m_\rho$. This gives:
$L_2\simeq 1.3\times 10^{-3}$ and   $L_3\simeq -4.4\times 10^{-3}$ .
We can now deduce the value of 
the transition matrix element $T_{12}$. At leading order one finds:
$T_{12}(s_1)\simeq  0.25 $,
while including next-to-leading order corrections one obtains:
$T_{12}(s_1)\simeq  0.15 $.
Clearly, convergence is not very fast but we can expect the order of 
magnitude to be reasonable. 
This result will serve us in the construction of $T_{12}(s)$ 
in the unphysical region $s\le (m_K +m_{\eta'})^2$.

Finally, it is instructive to calculate the predictions for the values
of the scalar form factor components $F_1(0)$, $F_2(0)$ in this approach. 
A small calculation using the $\delta^2$ Lagrangian gives
\be\lbl{fszero}
 F_1(0)= 1, \ 
 F_2(0)= {\mkd-\metapd\over\mkd-\mpid} 
\left(\sin\theta_8-{8\sqrt2(\mkd-\mpid)\over3\fpid}L_5\,\cos\theta_0 \right)
\simeq 0.71 .
\en
The deviation of $ F_1(0)$ from 1 is proportional to $(m_s-\hat{m})^2$
according to the Ademollo-Gatto theorem~\cite{ademollogatto}.  
In the $\delta$ expansion approach, the deviation shows up at order 
$\delta^3$ because it is subleading in $N_c$. 
The value which we obtain for $ F_2(0)$ at order $\delta^2$ 
turns out to be  very similar to the one obtained in Ref.~\cite{jop} 
in a somewhat different
approach. The corrections to $ F_2(0)$
of order $\delta^3$, however, have no reason to be particularly small.
In fact, the value of $F_2(0)$ which we obtain from the solutions of the 
Muskhelishvili-Omn\`es equations solutions,
using chiral constraints on the component $F_1(t)$, is somewhat smaller than
in Eq.~\rf{fszero}: $F_2(0)\simeq 0.52$. An analogous result was obtained
in Ref.~\cite{jop}.

\subsubsection{\boldmath $Determination~ of ~T_{11}~ and~ T_{12}$}

Below the inelastic threshold, the two components $T_{11}$ and $T_{12}$
are needed in the unitarity equations for the scalar form factors.  
At first, let us look at
$T_{11}$. At low energies, $m_{K\pi} \lapprox 0.9$ GeV, 
$K \pi\to K \pi$ scattering 
(analogously to $\pi\pi \to \pi\pi$ scattering) is constrained by
Roy-Steiner equations~\cite{roysteiner} 
which result from combining dispersion relations
and crossing symmetry with elastic unitarity. We will use
the results obtained from a recent re-analysis of such equations~\cite{roypik}.
In the energy range  $0.9 \le m_{K\pi} \le 1.25$ GeV we also 
use the fit performed in that reference of the $K\pi$ elastic phase shift.

\begin{figure}[ht]
\centering
\includegraphics[width=10cm]{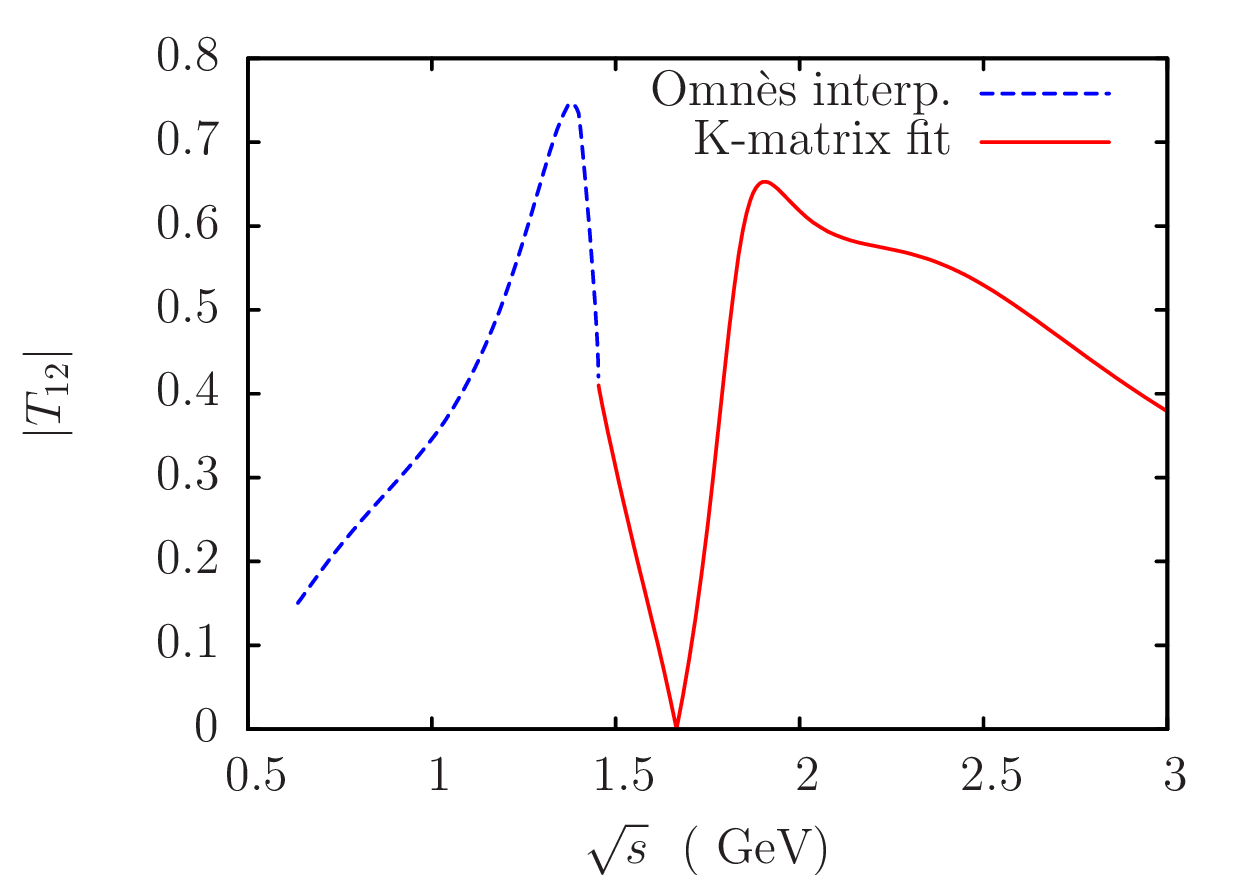}

\caption{Absolute value of the $T$-matrix element $T_{12}(s)$.
In the region $s \ge (m_K+m_{\eta'})$ it is obtained from the $K$-matrix fit,
and in the region $s \le (m_K+m_{\eta'})$ it is computed by a polynomial
approximation using the Omn\`es function as discussed in the text.\label{t0abs12}} 
\end{figure}

There remains to discuss the $K \pi \to K\eta' $ transition matrix element
$T_{12}$ in the unphysical region. For this purpose, we may use a simple
method which exploits the fact that the
phase, $\delta_{12}$, of $T_{12}$  is known for all values of the energy. 
Indeed, it is identical
with the  elastic phase below the inelastic threshold by Watson's theorem
and equal to $\delta_{K \pi}+\delta_{K\eta' }$ above because of two-channel 
unitarity. One can then compute the Omn\`es function:
\begin{equation}
\Omega_{12}(s)= \exp\left[{s\over\pi}
\int_{s_1}^\infty {ds'\over s'(s'-s)} \,
\delta_{12}(s')\right],
\end{equation}
and study $\psi(s)=\Omega_{12}^{-1}(s)\,T_{12}(s)$.
The function $\psi(s)$ has no right-hand cut
since $\im \psi(s)$ vanishes in the range  $(m_K+m_\pi)^2\le s \le\infty$.  
Therefore,
over a finite interval, we can approximate $\psi(s)$ by a polynomial. In 
practice, we use a polynomial of degree two and deduce the three parameters
of the polynomial from the known values of $\psi(s)$ at three points, 
$s=s_1,\ s_2$ and $s_3= (1.8)^2\ {\rm GeV}^2$.
The value at the $K\pi$ threshold $s_1$ is (approximately) known 
from the discussion above using the $\delta$
expansion, while the values at $s_2$ and $s_3$ are known from the
$K$-matrix fit. The result obtained in this manner
for the modulus of  $T_{12}$ is displayed in Fig.~\ref{t0abs12}.
This completes the determination of the three matrix elements 
$T_{ij}(s)$ in the energy region where they are needed in
Eqs.~\rf{dispfi} and~\rf{unitscal}.

\section{Determination of the decay  constants  $f_{K^*_0}$
and $ f_{K^*}$ in the complex pole approach \label{appendixfdecaypole}}

In this appendix, we quote the results for the decay constants
$f_{K_0^*}$ and $ f_{K^*}$ which are associated with matrix elements
involving the scalar meson $K^*_0(1430)$ and the vector meson $K^*(892)$
respectively, in the complex pole approach. 
For the decay constant $f_{K^*_0}$, 
let us use the definition proposed by Maltman~\cite{maltman99}:
\begin{equation}
\lbl{deffK*}
\braque{0\vert J^{su}(x)  \vert K^*_0(p)}= f_{K^*_0}\, m^2 _{K^*_0}
\,\exp(-ipx)
\end{equation}
with
$J^{su}(x)=\partial_\mu \bar{s}(x)\gamma^\mu u(x)$.
We introduce the two-point correlation function
associated with this current:
\begin{equation}
\Pi^{us}(t)=i\int d^4x \exp(ipx) \braque{0 \vert T [J^{su}(x)
(J^{su})^\dagger(0) ]\vert 0} .
\end{equation}
Using \rf{deffK*}, the contribution of the $K^*_0(1430)$  
to this correlation function, if it were a stable state, would be: 
\begin{equation}
\lbl{polePi}
\left.\Pi^{us}(t)\right\vert_{K^*_0(1430)} = {m^4_{K^*_0}\, f^2_{K^*_0}
\over m^2_{K^*_0}-t } .
\end{equation}
In reality, the $K^*_0(1430)$ is a resonance and it shows up as a pole
of $\Pi^{us}(t)$ on the second Riemann sheet. By analogy with Eq.~\rf{polePi},
we can identify the decay constant $f_{K^*_0} $
from the residue of the pole (which implies that it is a complex number).  
As before, in order to define the extension to the second sheet 
we consider the discontinuity of the function $\Pi^{us}(t)$ along 
the real axis:
\begin{equation}
\lbl{Pidisc}
\Pi^{su}(t+i\epsilon) - \Pi^{su}(t-i\epsilon)=-{3\over16\pi}
(m_K^2-m_\pi^2)^2 \sigma_{K \pi}(s+i\epsilon) f_0(s+i\epsilon)
f_0(s-i\epsilon)
\end{equation}
for $t$ real and lying in the range between the $K \pi$ and the $K \eta' $
thresholds. 
The factor 3 comes from summing over the two possible charge states
of the $K\pi$ system.
From Eq.~\rf{Pidisc}, one deduces that the extension to the 
second sheet must be defined as follows:
\begin{equation}
\lbl{PiII}
\Pi^{su}_{II}(t)= \Pi^{su}(t) + {3 \sigma_{K \pi}(t) (m_K^2-m_\pi^2)^2\,
(f_0(t))^2
\over 16\pi ( 1-2\sigma_{K \pi}(t) T_{11}^S(t))   } .
\end{equation}
We recognize again here the denominator function $D(t)$ which has a zero
at $t=t_0$.
We can identify the residue of the pole at $t=t_0$ with Eq.~\rf{polePi},  
replacing $m^2_{K^*_0}$ by $(t_0)^2$.
The following expression for $f_{K^*_0}$ results:
\begin{equation}
(f_{K^*_0})^2= -{ 3 (m_K^2-m_\pi^2)^2\over 16\pi (t_0)^2\alpha}
\sigma_{K \pi}(t_0) (f_0(t_0))^2 .
\end{equation}
Numerically, using the preceding results Eqs.~(\ref{t0}) and (\ref{f0t0}) we
obtain for the $ f_K/ f_\pi= 1.193$,
\begin{equation}
\label{fk0}
f_{K^*_0}= (31.3 +i\ 7.6)\ {\rm MeV} .
\end{equation}
The result is quasi real and comparable with the value obtained 
by Maltman~\cite{maltman99}, $f_{K^*_0}= 42.2$~MeV. 
 Varying $ f_K/ f_\pi$ one obtains:
\begin{eqnarray}
\lbl{fkfpivar1}
&& f_{K^*_0} =(36.7+i\ 7.5)\ {\rm MeV} \ \mbox{\rm if} \   {f_K\over  f_\pi}=1.203  \ \mbox{\rm and}\nonumber \\
&& f_{K^*_0} =(25.8+i\ 9.9)\ {\rm MeV}  \  \mbox{\rm if}\ { f_K\over  f_\pi}=1.183  .
\end{eqnarray}

Let us now present the analogous results for the vector form factor
$ f_1^{K\pi}(t)$ and the $K^*(892)$ resonance.
We want now to identify the decay constant $f_{K^*}$ associated with 
the vector meson $K^*(892)$, which may be defined as
\begin{equation}
\lbl{FKstdef}
\braque{0\vert j_\mu^{su}(x)\vert K^{*+}(p) }=  f_{K^*} \exp(-ipx)
\end{equation}
with $j_\mu^{su}(x)=\bar{s}(x)\gamma_\mu u(x)$.
For this purpose, we investigate the correlation function
\begin{equation}
\Pi_{\mu\nu}(q)= i\int d^4x {\rm e}^{iqx} \braque{ 0\vert T (
j_\mu ^{su}(x) j_\nu ^{us}(0) )\vert 0}
= (q_\mu q_\nu -q^2 g_{\mu\nu}) \Pi_1(q^2) + q_\mu q_\nu \Pi_0(q^2) .
\end{equation}
The discontinuity along the $K \pi$ elastic cut of $\Pi_1(q^2)$ reads:
\begin{equation}
\Pi_1(t+i\epsilon)-\Pi_1(t-i\epsilon) = -{1\over 4\pi t}
\sigma_{K \pi}(t+i\epsilon) q^2_{K \pi}(t+i\epsilon) 
f_1^{K\pi}(t+i\epsilon) f_1^{K\pi}(t-i\epsilon),
\end{equation}
which allows, as before, to obtain the definition on the second sheet
\begin{equation}
\left(\Pi_1(t)\right)^{II}= \Pi_1(t) +
{\sigma_{K \pi}(t) \left( q_{K \pi}(t) f_1^{K\pi}(t)\right)^2\over
4\pi t (1-2\sigma_{K \pi}(t) q^2_{K \pi}(t) T_{11}^P(t) )}.
\end{equation}
One then identifies the pole in this expression with the one generated
by a stable $K^*$ using \rf{FKstdef}, 
which, finally, gives $ f_{K^*}$ in terms of the vector form factor 
$f_1^{K\pi}$ 
\begin{equation}
f^2_{K^*}= - {\sigma_{K \pi}(t_1^{pole}) \left( q_{K \pi}(t_1^{pole}) f_1^{K\pi}(t_1^{pole})
\right)^2 \over 4\pi t_1^{pole} \beta } .
\end{equation}
Numerically, inserting the values  for $t_1^{pole}$, $f_1^{K\pi}(t_1^{pole})$
and $\beta$ (see Eqs.~(\ref{t1fplus})) gives
\begin{equation}
\label{fdecayvec}
 f_{K^*}\approx (213.9 -i\ 13.6)\ {\rm MeV} .
\end{equation}
The modulus of this decay decay constant, 214.3, is close to the value $f_V=218$ MeV we use in the $P$-wave amplitudes (see Eqs.~(\ref{K-pi+Sampli}) and (\ref{K0Spi-Pampli})).

\section{Effective decay constants and  two-body amplitudes 
{\boldmath $B^-\to\bar K_0^{*0}(1430)\pi^-$} and
{\boldmath  $B^-\to \bar K^{*0}(892) \pi^-$}  \label{appendix2a3body}}

We shall discuss below the case of  $B^-\to\bar K_0^{*0}(1430)\pi^-$ decays. 
It can be easily generalized to decays of other $B$ mesons.
For the two body $B^-\to\bar K_0^{*0}(1430)\pi^-$ decay mode the branching $\mathcal{B}_{2S}$ can be written in terms of the two body amplitude $\mathcal{M}_{2S}^-$,
\begin{equation}
\label{B-2}
\mathcal{B}_{2S}=\vert\mathcal{M}_{2S}^-\vert^2\ \dfrac{\vert\mathbf{p}_{\bar K_0^*}\vert}{8\pi M_B^2\Gamma_{B^-}},
\end{equation}
where $\vert\mathbf{p}_{\bar K_0^*}\vert$ is the modulus of the $\bar K_0^*(1430)$ momentum in the $B^-$ rest frame:
\begin{equation}
\label{pK0}
\vert\mathbf{p}_{\bar K_0^*}\vert=\dfrac{1}{2M_{B^-}}
\sqrt{
\left[M_B^2-\left(m_{\bar K_0^*}+m_{\pi}\right)^2\right]
\left[M_B^2-\left(m_{\bar K_0^*}-m_{\pi}\right)^2\right]
} ,
\end{equation}
$m_{\bar K_0^*}$ being the $\bar K_0^*(1430)$ mass.
In the vicinity of $m_{K^-\pi^+}=m_{\bar K_0^*}$ the two body amplitude $\mathcal{M}_{2S}^-$ is related to the three-body one $\mathcal{M}_S^-$~[Eq.~(\ref{K-pi+Sampli})],
\begin{equation}
\label{Mcal}
\mathcal{M}_S^-=\mathcal{M}_{2S}^-\ \Gamma_{\bar K_0^{*0}\to K^-\pi^+}^S(m_{K^-\pi^+}) .
\end{equation}
The vertex function $\Gamma^S$ can be expressed in terms of the scalar form factor 
$f_0^{K^-\pi^+}(m_{K^-\pi^+}^2)$ defined in Eq.~(\ref{Ktopiff}):
\begin{equation}
\label{Gamma-K+S}
\Gamma_{\bar K_0^{*0}\to K^-\pi^+}^S(m_{K^-\pi^+})=
\dfrac{1}{f_{\bar K_0^*(1430)}}\ \dfrac{m_{K}^2-m_{\pi}^2}{m_{K^-\pi^+}^2}
f_0^{K^-\pi^+}(m_{K^-\pi^+}^2) .
\end{equation}
In this equation $f_{\bar K_0^*(1430)}$ represents the not very well known $\bar K_0^*(1430)$ decay constant (see Appendix~\ref{appendixfdecaypole} and also Ref.~\cite{Cheng:2005nb}).
If the $m_{K^-\pi^+}$ effective mass is far from the resonance mass $m_{\bar K^{*0}}$ then, as one can see from Eq.~(\ref{K-pi+Sampli}), the relations~(\ref{Mcal}) and (\ref{Gamma-K+S}) cannot be used anymore, in particular, close to the $K\pi$ threshold and for $m_{K\pi} \gg m_{\bar K^*_0(1430)}$.

Integration of $\dfrac{d\mathcal{B}^-}{dm_{K^-\pi^+}}$ [Eq.~(\ref{dB-bis})] over the $m_{K^-\pi^+}$ range from $m_{\mathrm{min}}$ to $m_{\mathrm{max}}$,  where the $\bar K_0^*(1430)$ dominates, gives for the process $B^-\to(K^-\pi^+)_S\pi^-$ the branching fraction
\begin{multline}
\label{B_3S}
\mathcal{B}_{3S}=\int_{m_{\mathrm{min}}}^{m_{\mathrm{max}}}\dfrac{d\mathcal{B}_S^-}{dm_{K^-\pi^+}}\ dm_{K^-\pi^+}= \\
\dfrac{\vert\mathcal{M}_{2S}^-\vert^2}{4(2\pi)^3M_B^3\Gamma_{B}^-}
\int_{m_{\mathrm{min}}}^{m_{\mathrm{max}}}dm_{K^-\pi^+}\ m_{K^-\pi^+}
\vert\mathbf{p}_{\pi^+}\vert\ \vert\mathbf{p}_{\pi^-}\vert\ 
\vert\Gamma_{\bar K_0^{*0}\to K^-\pi^+}^S(m_{K^-\pi^+})\vert^2 .
\end{multline}
The branching ratios $\mathcal{B}_{2S}$ and $\mathcal{B}_{3S}$ are simply related by
\begin{equation}
\label{B3B2}
\mathcal{B}_{3S}=b\mathcal{B}_{2S}
\end{equation}
where $b=\dfrac{2}{3}\ 0.93$ is the secondary branching fraction for the decay 
$\bar K_0^*(1430)\to K^-\pi^+$~\cite{pdg08}.
Using Eqs.~(\ref{B-2}), (\ref{Gamma-K+S}) and (\ref{B_3S}), one obtains for the modulus square of the effective decay constant $f^{eff}_{\bar K_0^*(1430)}$,
\begin{equation}
\label{fK0}
\vert f^{eff}_{\bar K_0^*(1430)}\vert^2=\dfrac{\left(m_{K}^2-m_{\pi}^2\right)^2}
{b(2\pi)^2M_B\vert\mathbf{p}_{\bar K_0^*}\vert}
\int_{m_{\mathrm{min}}}^{m_{\mathrm{max}}}dm_{K^-\pi^+}m_{K^-\pi^+}^{-3}
\vert\mathbf{p}_{\pi^+}\vert\ \vert\mathbf{p}_{\pi^-}\vert
\vert f_0^{K^-\pi^+}(m_{K^-\pi^+}^2)\vert^2 .
\end{equation}
The knowledge of the scalar form factor allows one to calculate this effective decay constant.
Integration over the range $m_{\mathrm{min}} \leqslant  m_{K\pi} \leqslant m_{\mathrm{max}}$ with $m_{\mathrm{min}}=1 $ GeV and $m_{\mathrm{max}}=1.76$ GeV gives $\vert f^{eff}_{\bar K_0^*(1430)} \vert=31$ MeV.
This value is close to that of the decay constant calculated using the pole part
of the scalar form factor,  $\vert f_{\bar K_0^*(1430)}\vert=32$ MeV [see Eq.~(\ref{fk0})].
This agreement is expected as, in the $m_{K\pi}$ range (1, 1.76) GeV, the $\bar K_0^*(1430)$ pole part dominates (see e.g. Fig.~\ref{Modulae}).

The two-body $B^-\to \bar K^{*0}(892) \pi^-$ decay amplitude, $\mathcal{M}_{2P}^-$, can be expressed in terms of the three-body one $\mathcal{M}_{P}^-$~[Eq.~(\ref{K-pi+Pampli})],
\begin{equation}
\label{McalP}
\mathcal{M}_P^-=2\ \mathcal{M}_{2P}^-\ \Gamma_{ K^*\to K^-\pi^+}^P(m_{K^-\pi^+}) .
\end{equation}
The vertex function $\Gamma^P$ is related to the vector form factor 
$f_1^{K^-\pi^+}(m_{K^-\pi^+}^2)$ defined in Eq.~(\ref{Ktopiff}):
\begin{equation}
\label{Gamma-K+P}
\Gamma_{ K^{*}\to K^-\pi^+}^P(m_{K^-\pi^+})=
\dfrac{1}{m_{K^*}f_{K^*}}
f_1^{K^-\pi^+}(m_{K^-\pi^+}^2),
\end{equation}
where $m_{K^*}$ and $f_{K^*}$ are $K^*(892)$ mass and the decay constant, respectively. 
The two-body branching fraction for the $B^-\to \bar K^{*0}(892) \pi^-$ decay is
\begin{equation}
\label{B-2P}
\mathcal{B}_{2P}=\vert\mathcal{M}_{2P}^-\vert^2\ \dfrac{\vert\mathbf{p}_{ K^*}\vert^3}{8\pi m_{K^*}^2\Gamma_{B^-}},
\end{equation}
where $\vert\mathbf{p}_{K^*}\vert$ is the modulus of the $K^*(892)$ momentum in the $B^-$ rest frame.
It can be calculated from Eq.~(\ref{pK0}) replacing $m_{\bar K^*_0}$ by $m_{K^*}$.
The three-body branching fraction $B_{3P}$ for $B^-\to\bar K^{*0}(892)\pi^-$, $\bar K^{*0}\to K^-\pi^+$ is obtained by integration of the $P$-wave part of the effective mass distribution [see Eq.~(\ref{dB-bis})] from $m^P_{\mathrm{min}}$ to $m^P_{\mathrm{max}}$,  covering the range where the $ K^*(892)$ resonance dominates, 
\begin{multline}
\label{B_3P}
\mathcal{B}_{3P}=\int_{m^P_{\mathrm{min}}}^{m^P_{\mathrm{max}}}\dfrac{d\mathcal{B}_P^-}{dm_{K^-\pi^+}}\ dm_{K^-\pi^+}= \\
\dfrac{\vert\mathcal{M}_{2P}^-\vert^2}{3(2\pi)^3M_B^3\Gamma_{B}^-}
\int_{m^P_{\mathrm{min}}}^{m^P_{\mathrm{max}}}dm_{K^-\pi^+}\ m_{K^-\pi^+}
\vert\mathbf{p}_{\pi^+}\vert^3\ \vert\mathbf{p}_{\pi^-}\vert^3\ 
\vert\Gamma_{ K^{*}\to K^-\pi^+}^P(m_{K^-\pi^+})\vert^2 .
\end{multline}
Now $\mathcal{B}_{3P}=(2/3) \mathcal{B}_{2P}$, where the factor 2/3 is the secondary branching fraction for the decay 
$\bar K^{*0}(982)\to K^-\pi^+$.
As previously for the case of the $K^*_0(1430)$, taking into account the limited range of 
 $ m_{K\pi}$ between $m^P_{\mathrm{min}}$ and $m^P_{\mathrm{max}}$ and using Eqs.~ (\ref{Gamma-K+P}), (\ref{B-2P}) and (\ref{B_3P}), one obtains the modulus square of the effective decay constant $f^{eff}_{K^*}$
\begin{equation}
\label{fK892}
{\vert f^{eff}_ {K^*}}\vert^2=
\dfrac{1}{2\pi^2\ M_B^3\vert\mathbf{p}_{K^*}\vert^3}
\int_{m^P_{\mathrm{min}}}^{m^P_{\mathrm{max}}}dm_{K^-\pi^+}\ m_{K^-\pi^+}
\vert\mathbf{p}_{\pi^+}\vert^3\ \vert\mathbf{p}_{\pi^-}\vert^3
\vert f_1^{K^-\pi^+}(m_{K^-\pi^+}^2)\vert^2 .
\end{equation}
Integration from 0.82 to 0.97 GeV, range where the $K^*(892)$  dominates, gives $\vert f^{eff}_{ K^{*}} \vert=194$ MeV.
This value compares well with the decay constant calculated from the pole part of the vector form factor,  $\vert f_{ K^{*}}\vert=214$ MeV [see Eq.~(\ref{fdecayvec}].
A larger range of integration will improve the agreement.
In the limit of infinite $M_B$ mass and of zero width $K^*(892)$, the effective decay constant equals $f_{K^*}$.


\begin{thebibliography}{99}
\bibitem{Biesiada2007}
J. Biesiada (for the BaBar Collaboration) to appear in Proceedings to the Lake Louise Winter Institute, February 19-24, 2007, arXiv:hep-ex/0705.1001, \textit{Charmless hadronic B decays at BaBar} ; 
W.~Gradl (from the BABAR Collaboration), Proceedings of the VIIIth International Workshop on Heavy Quarks and Leptons (HQL06), M\"unich, October 16-20,  2006, Eds. S.~Recksiegel, A.Hoang, S.~Paul, \texttt{http://hql06.physik.tu-muenchen.de}, arXiv:hep-ex/0701032, \textit{Charmless B decays}.

\bibitem{Beneke2006}
M. Beneke, Nucl. Phys. B (Proc. Suppl.) \textbf{170}, 57 (2007),
\textit{Hadronic B decays}.

\bibitem{3bodyworkshop}
Three-Body Charmless B Decays Workshop,
\texttt{http://lpnhe-babar.in2p3.fr/3BodyCharmlessWS/}, February 1-3, 2006, LPNHE, Paris.

 \bibitem{AubertPRD72} 
B. Aubert,  \textsl{et al.} (BaBar Collaboration),  Phys. Rev. D {\bf 72}, 072003 (2005), Erratum-ibid. D {\bf 74}, 099903 (2006), 
{\it Dalitz-plot analysis of the decays $B^{\pm} \to K^\pm \pi^\mp \pi^\pm$}.

\bibitem{Abe2005} 
 K. Abe  \textsl{et al.} (Belle Collaboration), 
 arXiv:hep-ex/0509001, \textit{Search for direct $CP$ violation in three body charmless $B^{\pm} \to K^\pm \pi^\pm \pi^\mp$ decays}.
 
 \bibitem{Garmash:2005rv}
  A.~Garmash {\it et al.} (Belle Collaboration), Phys. Rev. Lett. {\bf 96}, 251803 (2006),
  {\it Evidence for large direct $CP$ violation in $B^\pm \to \rho(770)^0 K^\pm$ from
  analysis of the three-body charmless $B^\pm \to K^\pm \pi^\pm \pi^\mp$ decay};
  A. Garmash {\it et al.} (Belle Collaboration), Phys. Rev. {\bf D 71}, 092003 (2005), \textit{Dalitz analysis of the three-body charmless decays $B^+ \to K^+ \pi^+ \pi^-$  and  $B^+ \to K^+ K^+ K^-$}. 
 
\bibitem {AubertPRD73} 
B. Aubert, \textsl{et al.} (BaBar Collaboration),  Phys. Rev. D {\bf 73}, 031101(R) (2006), 
 {\it Measurements of neutral $B$ decay branching fractions to  $K^0_S \pi^+ \pi^-$
final states and the charge asymmetry of $B^{0} \to K^{*+} \pi^-$}.

\bibitem{Abe0509047} 
K. Abe,  \textsl{et al.} (Belle Collaboration), 
arXiv:hep-ex/0509047, {\it  Dalitz analysis of the three-body charmless decay $B^{0} \to K^0_S \pi^+ \pi^-$}.

\bibitem{Garmash2007} 
A. Garmash,  \textsl{et al.} (Belle Collaboration),  Phys. Rev. D {\bf 75}, 012006 (2007), 
{\it  Dalitz analysis of the three-body charmless $B^{0} \to K^0 \pi^+ \pi^-$ decay.}

\bibitem{AubertLP2007}
B. Aubert, \textsl{et al.} (BaBar Collaboration), arXiv: 0708.2097 [hep-ex], 
{\it Time-dependent Dalitz plot analysis of $B^0 \to K_S \pi^+ \pi^-$}.

\bibitem{Aubert:2008bj}
B. Aubert, \textsl{et al.} (BaBar Collaboration),
Phys. Rev. D \textbf{78}, 012004 (2008),
 \textit{Evidence for direct $CP$ violation from Dalitz-plot analysis of $B^\pm \to K^\pm \pi^\mp \pi^\pm$.}
 
 \bibitem{Aubert:2007bs}
 B. Aubert, \textsl{et al.} (BaBar Collaboration),
 Phys. Rev. D \textbf{78}, 052005 (2008),
\textit{Dalitz plot analysis of the decay $B^0(\bar B^0)\to K^\pm\pi^\mp\pi^0$}.


\bibitem{bene03}
  M.~Beneke and M.~Neubert,
  Nucl. Phys. \textbf{B675}, 333 (2003), 
  \textit{QCD factorization for $B\to PP$ and $B\to PV$ decays}.
  
  \bibitem{BauerFPCP06}
W. Bauer, Proceedings of the IV International Conference on Flavor Physics and $CP$ Violation (FPCP 2006), April 9 - 12, 2006, Vancouver, \texttt{http://www.slac.stanford.edu/econf/C060409/index.html}, arXiv:hep-ph/0606018, and references given therein, \textit{Hadronic $B$ decays from SCET}.
 
  
  \bibitem {2bodyQCDF}
N.~de~Groot, W.~N.~Cottingham and I.~B.~Whittingham, 
Phys. Rev.  D \textbf{68}, 113005 (2003), 
\textit{Factorization fits and the unitarity triangle in charmless two-body $B$ decays.} ; 
R.~Aleksan, P.-F.~Giraud, V.~Mor\'enas, O.~P\`ene and A.~S.~Safir, 
Phys. Rev. D \textbf{67}, 094019 (2003),
\textit{Testing QCD factorization and charming penguin diagrams in charmless $B\to PV$.} ;
D. Du, H. Gong, J. Sun, D. Yang, G. Zhu,
Phys. Rev. D \textbf{65}, 074001 (2002),
\textit{ Phenomenological analysis of $B\to PP$ decays with QCD factorization};
D. Du, H. Gong, J. Sun, D. Yang, G. Zhu, Phys. Rev. D \textbf{65}, 094025 (2002), Erratum-ibid. D \textbf{66}, 079904 (2002),  
\textit{Phenomenological analysis of charmless decays  $B\to PV$
 with QCD factorization};
 O. Leitner, X-H. Guo, A.W. Thomas,
J. Phys. G: Nucl. Part. Phys. \textbf{31}, 199 (2005),
\textit{Direct $CP$ violation, branching ratios and form factors $B\to\pi,\ B\to K$ in $B$ decays.}


\bibitem {Beneke3body} 
M. Beneke in Ref.~\cite{3bodyworkshop}, \textit{Quasi two-body and three-body decays in the heavy quark expansion}.

\bibitem {fkll} 
A. Furman, R. Kami\'nski, L.~Le\'sniak and B.~Loiseau,  
Phys. Lett.  B {\bf 622}, 207 (2005),
\textit{Long-distance effects and final state interactions in $B \to \pi\pi K$ and $B \to K\bar K K$ decays}.

\bibitem{Meissner:2000bc}
  U.~G.~Mei\ss ner and J.~A.~Oller, 
  Nucl. Phys.  {\bf A679}, 671 (2001), 
  \textit{$J/\psi \to \phi \pi \pi (K \bar K)$ decays, chiral dynamics and OZI violation}.

  \bibitem{El-Bennich2006}
   B. El-Bennich, A. Furman, R. Kami\'nski, L.~Le\'sniak and B.~Loiseau,
   Phys. Rev. D \textbf{74}, 114009 (2006),
   \textit{Interference between $f_0(980)$ and $\rho(770)^0$ resonances in $B\to\pi^+\pi^-K$ decays}.

 \bibitem{brodskylepage}
G.~P.~Lepage and S. J.~Brodsky, Phys. Rev. D {\bf 22}
 2157 (1980),
\textit{ Exclusive processes in perturbative Quantum Chromodynamics}.
  
\bibitem{aston84}
  D.~Aston {\it et al.},
  Nucl.\ Phys.\  {\bf B247}, 261 (1984),
 \textit{Partial wave analysis of the $\bar K^0 \pi^+\pi^-$ system produced in $K-P$  interactions at 11 GeV/c}.


\bibitem{aston87}
  D.~Aston {\it et al.},
  Nucl.\ Phys.\  {\bf B292}, 693 (1987), 
\textit{The strange meson resonances observed in the reaction $K^-p \to \bar K^0 \pi^+\pi^- n$ at 11 GeV/c}.

\bibitem{aston88}
D.~Aston et al., 
Nucl. Phys. {\bf B296}, 493 (1988),
\textit{A study of $K^-\pi^+$ scattering in the reaction $K^-p\to K^-\pi^+ n$ at 11 GeV/c}.

\bibitem{Watson52}
K.M. Watson,
Phys. Rev. \textbf{88}, 1163 (1952),
\textit{The effect of final state interaction on reaction cross sections}.

\bibitem{dgl}
  J.~F.~Donoghue, J.~Gasser and H.~Leutwyler,
  Nucl.\ Phys.\ {\bf B343}, 341 (1990),
  \textit{The decay of a light Higgs boson}.

\bibitem{jop} 
  M.~Jamin, J.~A.~Oller and A.~Pich,
  Nucl.\ Phys.\ {\bf B622}, 279 (2002), 
\textit{Strangeness-changing scalar form factors}.

\bibitem{jopffa}
  M.~Jamin, A.~Pich and J.~Portoles,
  Phys.\ Lett.\ B {\bf 640}, 176 (2006),
  \textit{Spectral distribution for the decay $\tau \to \nu/\tau K\pi$}.

\bibitem {MoussallampiKSandPwave} 
B. Moussallam, Eur. Phys. J. C \textbf{53}, 401 (2008), 
 \textit{Analyticity constraints on the strangeness changing vector current and applications to $\tau\to K\pi \nu_\tau$, $\tau\to K\pi\pi \nu_\tau$.}

\bibitem {Cheng0704.1049} 
H-Y. Cheng, C-K. Chua and A. Soni, Phys. Rev D \textbf{76}, 094006 (2007), \textit{Charmless three-body decays of B mesons}.
 
   \bibitem{Ciuchini:1997hb}
  M.~Ciuchini, E.~Franco, G.~Martinelli and L.~Silvestrini,
  Nucl. Phys.  {\bf B501}, 271 (1997),
  \textit{Charming penguins in B decays};
  C.~W.~Bauer, D.~Pirjol, I.~Z.~Rothstein, and I.~W.~Stewart,
  Phys. Rev. D {\bf 70}, 054015 (2004),
  \textit{$B\to M_1M_2$: Factorization, charming penguins, strong phases, and polarization}. 

 \bibitem{fleischer08}
 R. Fleischer, S. Jager, D. Pirjol and J. Zupan, 
 Phys. Rev. D {\bf 78}, 111501(R) (2008), \textit{Benchmarks for the new-physics search through $CP$ violation in $B^0 \to \pi^0 K_S$}. 
   
\bibitem{Beneke:2001ev}
  M.~Beneke, G.~Buchalla, M.~Neubert and C.~T.~Sachrajda,  Nucl.\ Phys.\  {\bf B606}, 245 (2001), \textit{QCD factorization in $B \to \pi K, \pi \pi$ decays and extraction of Wolfenstein  parameters}.

\bibitem{Cheng:2005nb}
  H.~Y.~Cheng, C.~K.~Chua and K.~C.~Yang,
 Phys.\ Rev.\  D {\bf 73}, 014017  (2006),
  \textit{Charmless hadronic B decays involving scalar mesons: implications to  the  nature of light scalar mesons.}


\bibitem{Khodjamirian:2006st}
  A.~Khodjamirian, T.~Mannel and N.~Offen,
  \newblock Phys. Rev.  D {\bf 75}, 054013 (2007),
  \newblock {\it Form factors from light-cone sum rules with $B$-meson distribution amplitudes}.

\bibitem{Lu:2007sg}
  C.~D.~Lu, W.~Wang and Z.~T.~Wei, Phys. Rev. D {\bf 76} 014013 (2007),
  \newblock {\it Heavy-to-light form factors on the light cone}.
  
 \bibitem{Albertus:2005ud}
  C.~Albertus, J.~M.~Flynn, E.~Hernandez, J.~Nieves and J.~M.~Verde-Velasco,
  Phys.\ Rev.\  D {\bf 72}, 033002 (2005)
  \newblock {\it  Semileptonic $B \to \pi$ decays from an Omn\`es improved nonrelativistic
  constituent quark model}.


\bibitem{Ebert:2006nz}
  D.~Ebert, R.~N.~Faustov and V.~O.~Galkin, Phys. Rev. D {\bf 75}, 074008 (2007),  \newblock {\it New analysis of semileptonic B decays in the relativistic quark model}.

\bibitem{Ivanov:2000aj}
  M.~A.~Ivanov, J.~G.~Korner and P.~Santorelli,
  \newblock Phys. Rev.  D {\bf 63}, 074010 (2001),
  \newblock {\it The semileptonic decays of the $B_c$ meson}.
  
\bibitem{Faessler:2002ut}
  A.~Faessler, T.~Gutsche, M.~A.~Ivanov, J.~G.~Korner and V.~E.~Lyubovitskij,
  \newblock  Eur.\ Phys.\ J.\ direct C {\bf 4}, 18 (2002),
  \newblock {\it The exclusive rare decays $B \to K (K^*) \bar l l$ and  $B_c \to D (D^*) \bar ll$ in a relativistic quark model}.

\bibitem{Melikhov:2001zv}
D.~Melikhov,
\newblock Eur. Phys. J. direct C {\bf 2}, 1 (2002),
\newblock {\it Dispersion approach to quark-binding effects in weak decays of
  heavy mesons}.
  
\bibitem{Burford:1995fc}
  D.~R.~Burford, H.~D.~Duong, J.~M.~Flynn, J.~Nieves, B.~J.~Gough, N.~M.~Hazel and H.~P.~Shanahan
                  [UKQCD Collaboration],
  Nucl.\ Phys.\  B {\bf 447}, 425 (1995),
\newblock {\it Form-factors for B $\to$ pi lepton anti-lepton-neutrino and B $\to$ K* gamma decays on the lattice}.
  
  \bibitem{Bowler:1999xn}
  K.~C.~Bowler {\it et al.}  (UKQCD Collaboration),
  Phys.\ Lett.\  B {\bf 486}, 111 (2000),
 \newblock {\it Improved $B \to \pi l \nu_l$ form factors from the lattice}.

\bibitem{Ivanov:2007cw}
  M.~A.~Ivanov, J.~G.~K\"orner, S.~G.~Kovalenko and C.~D.~Roberts,
  \newblock Phys. Rev. D {\bf 76}, 034018 (2007),
  \newblock {\it B-~to light-meson transition form factors}.

\bibitem{estabrooks}
  P.~Estabrooks, R.~K.~Carnegie, A.~D.~Martin, W.~M.~Dunwoodie, T.~A.~Lasinski and D.~W.~G.~Leith,
  Nucl. Phys. {\bf B133}, 490 (1978),
 \textit{Study of $K \pi$ scattering using the reactions $K^\pm p \to K^\pm \pi^+ n$ and 
$K^\pm p \to K^\pm \pi^- \Delta^{++}$ at 13-GeV/c}.

\bibitem{aston88b}
  D.~Aston {\it et al.},
 Phys. Lett. B {\bf 201}, 169 (1988), \textit{Observation of the selective coupling of $K^*$ states to the $K^- \eta$ channel}.

 \bibitem{muskhelishvili}
N.~I.~Muskhelishvili, {\sl Singular integral equations}, P. Noordhof,
1953.
 
\bibitem{mou95}
  B.~Moussallam,
  Phys.\ Rev.\ D {\bf 51}, 4939 (1995),
  \textit{Chiral sum rules for parameters of the order six Lagrangian in the W-Z  sector and application to $\pi_0$, $\eta$, $\eta'$ decays}.
  
\bibitem{kaiserleutwyler}
  H.~Leutwyler,
  Nucl.\ Phys.\ Proc.\ Suppl.\  {\bf 64}, 223 (1998),
  \textit{On the 1/N-expansion in chiral perturbation theory};  R.~Kaiser and H.~Leutwyler,
Proc. Workshop "Nonperturbative methods in quantum field theory", NITP/CSSM, University of Adelaide, Australia, Feb. 2-13, 1998, Eds. A.W. Schreiber, A.G. Williams and A.W. Thomas (World Scientific, Singapore 1998),   arXiv:hep-ph/9806336, \textit{Pseudoscalar decay constants at large $N_c$}.

\bibitem{gl85}
  J.~Gasser and H.~Leutwyler,
  Nucl. Phys. {\bf B250}, 465 (1985),
  \textit{Chiral Perturbation Theory: expansions in the mass of the strange quark}.
  
\bibitem{gl85ff}
  J.~Gasser and H.~Leutwyler,
  Nucl. Phys. {\bf B250}, 517 (1985),
 \textit{ Low-energy expansion of meson form-factors}.

\bibitem{leutwyler-roos}
  H.~Leutwyler and M.~Roos,
  Z.\ Phys.\  C {\bf 25}, 91 (1984),
  \textit{Determination of the elements $V_{us}$ and $V_{ud}$ of the Kobayashi-Maskawa
  matrix}.

\bibitem{pdg08}
  C. Amsler {\it et al.} (Particle Data Group),
  Phys. Lett. B {\bf 667}, 1 (2008),
  \textit{Review of particle physics}.
  
  
\bibitem{L6}
  B.~Moussallam,
  Eur.\ Phys.\ J.\  C {\bf 14}, 111 (2000),
  \textit{$N_f$ dependence of the quark condensate from a chiral sum rule}.

\bibitem{taylorq}
J. R. Taylor, Scattering theory, Wiley, New York (1972).

\bibitem{duncanmueller} A.~Duncan and A. H.~Mueller, Phys. Rev. D {\bf 21}, 1636 
(1980), \textit{Asymptotic behavior of composite-particle form factors and the renormalization group}.



 \bibitem{Descotes06}  S. Descotes-Genon, B. Moussallam, Eur. Phys. J. {\bf C 48}, 553 (2006),
    \textit{The $K^*_0(800)$ scalar resonance from Roy-Steiner representations of $\pi K$ scattering}.
    
  \bibitem{Dunwoodie06} W. Dunwoodie in Ref.~\cite{3bodyworkshop}, \textit{Relevance of LASS results to $B$-factory analyses}.

\bibitem{Belle1108}
 J. Dalseno  {\it et al.} (Belle Collaboration),  arXiv: 0811.3665,
    \textit{Time-dependent Dalitz-plot measurement of $CP$ parameters in $B^0 \to K^0_S\pi^+\pi^-$  decays}.

\bibitem{Ball:2007rt}
  P.~Ball and G.~W.~Jones, JHEP {\bf 0703}, 069 (2007),  \textit{Twist-3 distribution amplitudes of $K^*$ and $\phi$ mesons}.


\bibitem{largeNc}
  E.~Witten,
  Nucl.\ Phys.\ {\bf B156}, 269 (1979),
  \textit{Current algebra theorems for the U(1) Goldstone Boson};
  G.~Veneziano,
  Nucl.\ Phys.\ {\bf B159}, 213 (1979),
  \textit{ U(1) without instantons}.

\bibitem{etappprob}
  H.~Georgi,
  Phys.\ Rev.\  D {\bf 49}, 1666 (1994), 
  \textit{Bound on $m_\eta / m_{\eta'}$ for large $N_C$}.

\bibitem{milc}
  C.~Bernard {\it et al.},
 Proceedings of the 5th International Workshop on Chiral Dynamics, Theory and Experiment Durham/Chapel Hill, North Carolina, USA, Edts. M.~W.~Ahmed, H.~Gao, B.~R.~Holstein,
World Scientific Pub Co Inc Published 2008/01,
  [arXiv:hep-lat/0611024],  
 \textit{Low energy constants from the MILC Collaboration}.

\bibitem{abm}
  B.~Ananthanarayan, P.~B\"uttiker and B.~Moussallam,
  Eur.\ Phys.\ J.\ C {\bf 22}, 133 (2001),
  \textit{$\pi K$ sum rules and the SU(3) chiral expansion}.

\bibitem{roypik}
  P.~B\"uttiker, S.~Descotes-Genon and B.~Moussallam,
  Eur.\ Phys.\ J.\ C {\bf 33}, 409 (2004),
  \textit{A new analysis of $\pi K$  scattering from Roy and Steiner type equations}.

\bibitem{riggenbach}
  C.~Riggenbach, J.~Gasser, J.~F.~Donoghue and B.~R.~Holstein,
  Phys.\ Rev.\ D {\bf 43}, 127 (1991),
\textit{Chiral symmetry and the large-$N_C$ limit in $K_{l4}$ decays}.
 
\bibitem{bijkl4}
  J.~Bijnens,
  Nucl. Phys. {\bf B337}, 635 (1990),
   \textit{$K_{l4}$ Decays and the low-energy expansion}.

\bibitem{ademollogatto}
  M.~Ademollo and R.~Gatto,
  Phys. Rev. Lett. {\bf 13},264 (1964),
\textit{Nonrenormalization theorem for the strangeness violating vector currents}.

\bibitem{roysteiner}
  F.~Steiner,
  Fortsch.\ Phys.\  {\bf 19}, 115 (1971),
 \textit{ Partial wave crossing relations for meson-baryon scattering};
  S.~M.~Roy,
  Phys.\ Lett.\ B {\bf 36}, 353 (1971),
  \textit{Exact integral equation for pion-pion scattering involving only physical region partial waves}.

\bibitem{maltman99}
  K.~Maltman,
  Phys.\ Lett.\ B {\bf 462}, 14 (1999),
   \textit{The $a_0(980)$, $a_0(1450)$ and $K^*_0(1430)$ scalar decay constants and the isovector scalar spectrum}.

\end{thebibliography}
\end{document}